\documentclass{book}
\usepackage{roboto}

\usepackage[english]{babel}
\usepackage[letterpaper,top=2cm,bottom=2cm,left=3cm,right=3cm,marginparwidth=1.75cm]{geometry}

\usepackage{pdflscape}
\usepackage{rotating}
\usepackage{longtable}
\usepackage{array}
\usepackage{float}
\usepackage{amsmath}
\usepackage{graphicx}
\usepackage{inconsolata}
\usepackage[colorlinks=true, allcolors=blue]{hyperref}
\usepackage{enumitem}
\usepackage{tikz} 
\usepackage{listings}
\usepackage{xcolor}
\usepackage[T1]{fontenc}
\usepackage{IEEEtrantools}  
\usepackage{epigraph}  
\usepackage{cite}

\definecolor{bgcolor}{rgb}{0.97,0.97,0.97}
\definecolor{codeblue}{rgb}{0.1,0.1,0.8}
\definecolor{codegreen}{rgb}{0,0.4,0}
\definecolor{codegray}{rgb}{0.4,0.4,0.4}
\definecolor{codepurple}{rgb}{0.5,0,0.5}
\definecolor{codered}{rgb}{0.6,0.2,0.2}
\definecolor{lightgray}{rgb}{0.9,0.9,0.9}
\definecolor{darkgray}{rgb}{0.6,0.6,0.6} 

\makeatletter
\renewcommand{\paragraph}{%
  \@startsection{paragraph}{4}{\z@}{1ex}{-1em}{\normalfont\normalsize\bfseries\color{gray}}}
\makeatother

\lstdefinestyle{python}{
    language=Python,
    basicstyle=\ttfamily\small\color{black}\usefont{T1}{zi4}{m}{n},  
    keywordstyle=\bfseries\color{codeblue},  
    stringstyle=\color{codegreen},  
    commentstyle=\slshape\color{codegray},  
    showstringspaces=false,
    numbers=left,
    numberstyle=\tiny\color{codegray},  
    stepnumber=1,
    numbersep=8pt,
    frame=single,
    rulecolor=\color{darkgray},  
    breaklines=true,
    backgroundcolor=\color{bgcolor},
    tabsize=4,
    captionpos=b,
    morekeywords={self}, 
}

\lstdefinestyle{cmd}{
    language=bash,
    basicstyle=\ttfamily\small\color{black}\usefont{T1}{zi4}{m}{n},  
    keywordstyle=\bfseries\color{blue},
    stringstyle=\color{codegreen},
    commentstyle=\itshape\color{gray},
    showstringspaces=false,
    numbers=none,
    frame=single,
    rulecolor=\color{darkgray},  
    breaklines=true,
    backgroundcolor=\color{bgcolor},
    tabsize=4,
    captionpos=b,
}

\lstdefinelanguage{Solidity}{
    keywordstyle=\bfseries\color{codeblue},  
    commentstyle=\itshape\color{darkgray},   
    stringstyle=\color{codegreen},           
    keywords={pragma, solidity, contract, public, private, function, returns, event, emit, require, msg, address, uint, uint256, bool, mapping, struct}, 
    sensitive=true,                      
    morecomment=[l]{//},                 
    morecomment=[s]{/*}{*/},             
    morestring=[b]",                     
}

\lstdefinestyle{solidity}{
    language=Solidity,
    basicstyle=\ttfamily\small\color{black},  
    keywordstyle=\bfseries\color{codeblue},  
    stringstyle=\color{codegreen},           
    commentstyle=\slshape\color{gray},   
    showstringspaces=false,              
    numbers=left,                        
    numberstyle=\tiny\color{darkgray},       
    stepnumber=1,
    numbersep=8pt,
    frame=single,                        
    rulecolor=\color{darkgray},          
    breaklines=true,                     
    backgroundcolor=\color{white},       
    tabsize=4,                           
    captionpos=b,                        
}

\title{Mastering AI: Big Data, Deep Learning, and the Evolution of Large Language Models - Blockchain and Applications}

\author{
    Pohsun Feng\textsuperscript{*†} \\ 
    \textit{National Taiwan Normal University} \\
    41075018h@ntnu.edu.tw
    \and
    Ziqian Bi\textsuperscript{*†} \\ 
    \textit{Indiana University} \\
    bizi@iu.edu
    \and
    Lawrence K.Q. Yan \\ 
    \textit{Hong Kong University of Science and Technology} \\
    kqyan@connect.ust.hk
    \and
    Yizhu Wen \\ 
    \textit{University of Hawaii} \\
    yizhuw@hawaii.edu
    \and
    Benji Peng \\ 
    \textit{AppCubic} \\
    benji@appcubic.com
    \and
    Junyu Liu \\ 
    \textit{Kyoto University} \\
    liu.junyu.82w@st.kyoto-u.ac.jp
    \and
    Caitlyn Heqi Yin \\ 
    \textit{University of Wisconsin-Madison} \\
    hyin66@wisc.edu
    \and
    Tianyang Wang \\ 
    \textit{Xi’an Jiaotong-Liverpool University} \\
    Tianyang.Wang21@student.xjtlu.edu.cn
    \and
    Keyu Chen \\ 
    \textit{Georgia Institute of Technology} \\
    kchen637@gatech.edu
    \and
    Sen Zhang \\ 
    \textit{Rutgers University} \\
    sen.z@rutgers.edu
    \and
    Ming Li \\ 
    \textit{Georgia Institute of Technology} \\
    mli694@gatech.edu
    \and
    Jiawei Xu \\ 
    \textit{Purdue University} \\
    xu1644@purdue.edu
    \and
    Ming Liu \\ 
    \textit{Purdue University} \\
    liu3183@purdue.edu
    \and
    Xuanhe Pan \\ 
    \textit{University of Wisconsin-Madison} \\
    xpan73@wisc.edu
    \and
    Jinlang Wang \\ 
    \textit{University of Wisconsin-Madison} \\
    jinlang.wang@wisc.edu
    \and
    Xinyuan Song \\ 
    \textit{Emory University} \\
    xinyuan.song@emory.edu
    \and
    Qian Niu \\ 
    \textit{Kyoto University} \\
    niu.qian.f44@kyoto-u.ac.jp
}

\date{} 

\begin{document}

\maketitle

\begingroup
\renewcommand\thefootnote{}\footnote{
    \textsuperscript{*} Equal contribution \\
    \textsuperscript{$\dagger$} Corresponding author
}
\addtocounter{footnote}{0}
\endgroup

\epigraph{"It may be that our role on this planet is not to worship God but to create him."}{\textit{Arthur C. Clarke}}

\tableofcontents  

\part{Fundamentals of Cryptography}
\chapter{Overview of Cryptography}

\section{The Origins of Cryptography}

\subsection{Cryptography in Ancient Egypt and Ancient Greece}
Cryptography has its roots deep in ancient history, with various early civilizations employing basic cryptographic methods to protect their messages \cite{subramani2023review,aumasson2024serious,thabit2023cryptography,pillai2024analyzing,zulkifli2007evolution}. These techniques were essential for military, political, and diplomatic communications, where secrecy could mean the difference between success and failure.

In ancient Egypt, one of the earliest known examples of cryptography was found in the tombs of pharaohs \cite{polis2023guide}. Egyptian scribes used a method of encryption in which non-standard hieroglyphs were substituted for more common ones. This wasn't designed for secrecy in the way we understand cryptography today but may have been used to obscure the message from the untrained.

The ancient Greeks took cryptography further, especially with the development of techniques like the \textbf{Scytale}. The Scytale was a tool used by the Spartans in military communication\cite{guerra2024embryonic}. It involved wrapping a strip of parchment around a cylinder of a specific diameter. The message was written across the parchment, and when unwrapped, it would appear as random letters. Only by wrapping it around another cylinder of the same diameter could the original message be deciphered, making it one of the earliest examples of transposition ciphers.

Another early technique was \textbf{Steganography}, the practice of hiding a message within another, more benign message or medium\cite{rustad2023digital,jamil1999steganography}. One of the famous examples is from Herodotus, who tells of a secret message being written underneath the wax of a wax tablet.

\subsection{Caesar Cipher and Classical Encryption}
One of the most famous classical encryption techniques is the \textbf{Caesar Cipher} \cite{ardiansyah2023implementasi,noviyanti2analysis}, named after Julius Caesar, who used this method to protect his military communications. The Caesar cipher is a type of substitution cipher where each letter in the plaintext is shifted by a fixed number of positions down the alphabet. For example, with a shift of 3, \textit{A} would become \textit{D}, \textit{B} would become \textit{E}, and so on. 

\noindent The encryption process can be defined as:

\[
E(x) = (x + n) \mod 26
\]

Where:
\begin{itemize}
    \item \(x\) is the position of the letter in the alphabet.
    \item \(n\) is the number of positions to shift.
    \item 26 represents the total number of letters in the English alphabet.
\end{itemize}

\noindent The decryption process simply reverses the shift:

\[
D(x) = (x - n) \mod 26
\]

While the Caesar cipher is easy to implement, its security is extremely weak by modern standards because it only has 25 possible shifts (excluding no shift). An attacker could easily try all 25 possibilities, making it vulnerable to \textit{brute-force} attacks.

\textbf{Example of Caesar Cipher in Python:}

\begin{lstlisting}[style=python]
def caesar_cipher_encrypt(text, shift):
    result = ""
    for char in text:
        if char.isalpha():
            shift_base = 65 if char.isupper() else 97
            result += chr((ord(char) - shift_base + shift) % 26 + shift_base)
        else:
            result += char
    return result

def caesar_cipher_decrypt(text, shift):
    return caesar_cipher_encrypt(text, -shift)

# Example usage
plaintext = "HELLO WORLD"
shift = 3
encrypted = caesar_cipher_encrypt(plaintext, shift)
decrypted = caesar_cipher_decrypt(encrypted, shift)

print(f"Plaintext: {plaintext}")
print(f"Encrypted: {encrypted}")
print(f"Decrypted: {decrypted}")
\end{lstlisting}

While the Caesar cipher was effective in ancient times, modern 
cryptographic techniques are far more sophisticated, relying on complex mathematical structures and computational difficulty to achieve security.

\section{Development of Modern Cryptography}

\subsection{Shannon's Communication Theory}
Modern cryptography began to take shape with the advent of computers and the formalization of communication theory. One of the foundational figures in this field is \textbf{Claude Shannon}, who is often considered the father of modern cryptography.

In 1949, Shannon published a seminal paper titled "\textit{Communication Theory of Secrecy Systems}" \cite{shannon1949communication}in which he applied the concepts of information theory to cryptography. Shannon introduced several key concepts that are still vital to modern cryptographic algorithms, such as \textbf{entropy}\cite{simion2020entropy}, \textbf{confusion}\cite{hu2021quantum}, and \textbf{diffusion}\cite{hu2021quantum}:

\begin{itemize}
    \item \textbf{Entropy:} In cryptography, entropy refers to the uncertainty or randomness in a system\cite{simion2020entropy}. The higher the entropy, the more secure a cryptographic system is. Shannon demonstrated that systems with low entropy are easier to predict and therefore more vulnerable to attacks.
    
    \item \textbf{Confusion:} Confusion is a technique used to make the relationship between the key and the ciphertext as complex as possible\cite{hu2021quantum}. This is done to prevent an attacker from finding patterns that could lead to determining the key. Substitution techniques, such as those used in substitution ciphers, are a way to introduce confusion.
    
    \item \textbf{Diffusion:} Diffusion spreads the influence of each plaintext digit over many ciphertext digits, making it harder to deduce the plaintext by examining the ciphertext\cite{hu2021quantum}. Transposition ciphers are an example of diffusion in action.
\end{itemize}

Shannon's work laid the groundwork for understanding the fundamental properties that any secure encryption system should have. His theories directly influence the design of modern encryption algorithms, such as \textbf{AES} (Advanced Encryption Standard)\cite{smid2021development} and \textbf{RSA} (Rivest–Shamir–Adleman)\cite{cohen2011architecture}, which employ high levels of confusion and diffusion to ensure security.

\subsection{Security Definitions in Cryptography}
In cryptography, the goals of securing communication or data can be summarized using three core principles: \textbf{Confidentiality}, \textbf{Integrity}, and \textbf{Authenticity}. These are often referred to as the CIA triad\cite{khansa2014assessing}.

\begin{itemize}
    \item \textbf{Confidentiality:} Ensuring that only authorized parties can access the information. Encryption is one of the main tools used to achieve confidentiality. For instance, when you visit a secure website, your browser and the server use encryption protocols like \textbf{TLS} (Transport Layer Security) to keep the data confidential.
    
    \item \textbf{Integrity:} Ensuring that the data has not been tampered with or altered during transmission or storage. One common technique used to ensure integrity is the use of cryptographic hash functions. A hash function produces a fixed-size output from an arbitrary-length input, and even the slightest change in the input will produce a significantly different output. This property is used in many systems to verify data integrity.
    
    \item \textbf{Authenticity:} Ensuring that the parties involved in the communication are who they claim to be. Authentication can be achieved through techniques like digital signatures, where a sender signs a message with their private key, and the recipient verifies the signature using the sender's public key.
\end{itemize}

Cryptographic security models, such as \textbf{Kerckhoffs's Principle}\cite{ke2019generative}, emphasize that the security of a system should rely on the secrecy of the key rather than the secrecy of the algorithm itself. This means that cryptographic algorithms should be open to public scrutiny, but the key must remain secret.

In designing cryptographic algorithms, these principles guide how we approach problems like key management, authentication protocols, and secure communication channels. For instance, modern encryption methods like \textbf{AES}\cite{inam2023new} use both confusion and diffusion to ensure confidentiality, while digital signatures ensure authenticity, and hash functions help maintain integrity.

\section{The Role of Cryptography in Blockchain}
Cryptography plays a pivotal role in blockchain technology, ensuring that the system remains secure, trustless, and tamper-resistant. In this section, we will explore how cryptographic techniques such as encryption, hashing, and digital signatures ensure data confidentiality, integrity, and trustlessness, eliminating the need for intermediaries.

\subsection{Ensuring Data Confidentiality and Integrity}
One of the key benefits of blockchain technology is the ability to ensure that data remains both confidential and secure from unauthorized changes\cite{lin2024symmetry}. Cryptographic techniques, specifically symmetric and asymmetric encryption, are employed to safeguard sensitive information.

\subsubsection{Symmetric Encryption}
Symmetric encryption involves using the same key for both encryption and decryption\cite{sudeva2017symmetric}. This method is highly efficient and suitable for encrypting large amounts of data, making it ideal for protecting user data within a blockchain network. However, the challenge with symmetric encryption is securely sharing the secret key, as anyone who has the key can decrypt the data.

\subsubsection{Asymmetric Encryption}
Asymmetric encryption, also known as public-key cryptography, uses a pair of keys—a public key and a private key\cite{mohamad2021research}. The public key is shared openly and used for encryption, while the private key is kept secret and used for decryption. This method provides a higher level of security compared to symmetric encryption and is commonly used in blockchain systems to protect users' identities and ensure that only authorized parties can access specific information.

For instance, in Ethereum, a user’s public key can be used to encrypt a transaction, but only the holder of the corresponding private key can decrypt it and prove ownership. This ensures confidentiality while maintaining data accessibility for authorized users.

\subsubsection{Ensuring Integrity}
While encryption protects the confidentiality of data, cryptography also ensures that the data remains unaltered. Integrity in blockchain is often achieved through cryptographic hash functions\cite{kuznetsov2021performance}, which we will cover in more detail in a later section. For now, consider that any modification to a block in the blockchain would change the hash of the block, making tampering evident.

\subsection{Achieving Trustlessness: Cryptography as the Foundation}
The concept of \textit{trustlessness} refers to the elimination of the need for trusted third parties (intermediaries) in a transaction\cite{bruun2020infrastructures}. Blockchain achieves this through cryptography, where the rules of the system and trust are embedded in mathematical principles, not reliant on human intervention\cite{kuznetsov2021performance}.

\subsubsection{Public Key Cryptography}
In blockchain, public key cryptography is the foundation of user identity\cite{tewari2019blockchain}. Each participant in the network has a cryptographic key pair—a public key that can be shared and a private key that is kept secret. 

When Alice wants to send cryptocurrency to Bob, she broadcasts a transaction, which includes her public key and a digital signature created using her private key. Nodes in the network can use Alice’s public key to verify that the transaction was indeed signed by her private key, thus proving authenticity without needing a central authority to validate it.

\begin{lstlisting}[style=python]
# Example of generating a public/private key pair using Python's cryptography library
from cryptography.hazmat.primitives.asymmetric import rsa

# Generate a private key
private_key = rsa.generate_private_key(
    public_exponent=65537,
    key_size=2048
)

# Get the public key from the private key
public_key = private_key.public_key()
\end{lstlisting}

\subsubsection{Digital Signatures}
Digital signatures are another crucial cryptographic tool used in blockchain to ensure trustlessness\cite{fang2020digital}. When a user signs a transaction with their private key, it generates a digital signature that can be verified using their public key. This mechanism allows the blockchain to verify the authenticity of transactions without intermediaries.

For example, when Bob receives a signed transaction from Alice, he can verify the digital signature using Alice's public key, ensuring that the transaction is legitimate and came from Alice, not someone pretending to be her.

\begin{lstlisting}[style=python]
# Example of signing a message with a private key
from cryptography.hazmat.primitives import hashes
from cryptography.hazmat.primitives.asymmetric import padding

message = b"Blockchain transaction"

# Sign the message
signature = private_key.sign(
    message,
    padding.PSS(
        mgf=padding.MGF1(hashes.SHA256()),
        salt_length=padding.PSS.MAX_LENGTH
    ),
    hashes.SHA256()
)

# The signature can be verified using the public key
public_key.verify(
    signature,
    message,
    padding.PSS(
        mgf=padding.MGF1(hashes.SHA256()),
        salt_length=padding.PSS.MAX_LENGTH
    ),
    hashes.SHA256()
)
\end{lstlisting}

By using digital signatures, blockchain transactions become tamper-proof and self-verifiable, contributing to the overall trustless nature of the system.

\subsection{The Different Roles of Encryption and Hashing in Blockchain}
Cryptography is central to blockchain, but it serves different purposes based on the technique used. Specifically, encryption and hashing play distinct yet complementary roles in ensuring both confidentiality and integrity in blockchain systems.

\subsubsection{Encryption for Confidentiality}
Encryption, as discussed earlier, is primarily used to ensure that data remains confidential and can only be accessed by authorized parties\cite{zhang2018ensuring}. In blockchain, encryption is mainly employed to protect user data and communication between nodes. 

For example, in private blockchains where sensitive business information is shared, symmetric and asymmetric encryption can be used to ensure that only authorized participants can view and modify data.

\subsubsection{Hashing for Integrity}
In contrast to encryption, hashing does not aim to protect confidentiality but rather ensures data integrity\cite{mohammed2020maintaining}. A hash function takes an input (or "message") and returns a fixed-size string of bytes\cite{pappu2002physical}. Even a small change in the input drastically changes the output hash, making it easy to detect tampering. 

Hashing is widely used in blockchain for:
\begin{itemize}
    \item \textbf{Block identification}: Each block in a blockchain contains a hash of the previous block, creating a chain of blocks that ensures any alteration would break the chain\cite{monrat2019survey}.
    \item \textbf{Proof of Work (PoW)}: In Bitcoin, miners compete to solve a cryptographic puzzle, which involves finding a hash value below a certain threshold\cite{cojocaru2023quantum}. This process secures the network and prevents double-spending.
    \item \textbf{Transaction integrity}: Each transaction is hashed and included in a block, making it immutable once added to the blockchain\cite{patel2021review}.
\end{itemize}

In Bitcoin, for example, the SHA-256 hash function is used extensively. The hash of each block links to the previous block, forming the blockchain. If anyone tries to tamper with a block, its hash will change, invalidating the subsequent blocks.

\begin{lstlisting}[style=python]
import hashlib

# Hash a message using SHA-256
message = b"Blockchain Data"
hash_object = hashlib.sha256(message)
hex_dig = hash_object.hexdigest()

print(f"SHA-256 Hash: {hex_dig}")
\end{lstlisting}

\subsubsection{Proof of Work (PoW)}
Proof of Work is a consensus algorithm that relies on hashing to secure the blockchain\cite{cojocaru2023quantum}. In PoW-based blockchains like Bitcoin, miners compete to solve a complex mathematical problem that involves finding a hash value. Once a miner finds the correct hash, they can add the new block to the blockchain, and the process starts again.

The puzzle is computationally expensive to solve but easy for others to verify, making it an effective mechanism to secure the network without relying on trust in any single party.

\begin{lstlisting}[style=python]
# Example of a simple proof of work
import hashlib

def proof_of_work(last_proof):
    proof = 0
    while valid_proof(last_proof, proof) is False:
        proof += 1
    return proof

def valid_proof(last_proof, proof):
    guess = f'{last_proof}{proof}'.encode()
    guess_hash = hashlib.sha256(guess).hexdigest()
    return guess_hash[:4] == "0000"

last_proof = 100
new_proof = proof_of_work(last_proof)
print(f"Proof of Work: {new_proof}")
\end{lstlisting}

Thus, encryption and hashing work together in blockchain: encryption ensures confidentiality, while hashing ensures integrity and security through consensus mechanisms like Proof of Work.
\chapter{Symmetric Encryption Algorithms}

    \section{Definition and Characteristics of Symmetric Encryption}
    
    Symmetric encryption is a cryptographic  method in which the same key is used for both encryption and decryption\cite{Nielson2019SymmetricET}. It is one of the most fundamental and widely used encryption techniques due to its simplicity and efficiency. In this type of encryption, the sender encrypts the plaintext using a key, and the receiver decrypts the ciphertext with the same key\cite{Hellwig2020BlockchainCP}. This method ensures confidentiality, as long as the key remains secret between the communicating parties.

    The main characteristics of symmetric encryption include\cite{Gharat2014OverviewOS}:
    \begin{itemize}
        \item \textbf{Single key}: The same key is used for both encryption and decryption.
        \item \textbf{Speed}: Symmetric encryption algorithms are generally faster than asymmetric encryption algorithms, making them suitable for encrypting large amounts of data.
        \item \textbf{Simplicity}: The algorithms are generally easier to implement compared to asymmetric encryption.
        \item \textbf{Confidentiality}: The security of symmetric encryption relies heavily on the secrecy of the encryption key.
    \end{itemize}
    
    \subsection{The Key Sharing Problem}
    
    One of the major challenges in symmetric encryption is the secure distribution of the encryption key\cite{yashaswini2015key}. Since the same key is used for both encryption and decryption, it must be shared securely between the communicating parties. This is known as the \textit{key sharing problem}.

    If the key is intercepted during transmission, the security of the encrypted data is compromised. Therefore, securely exchanging the encryption key between the sender and receiver is a critical issue\cite{}.

    Some methods to address the key sharing problem include:
    \begin{itemize}
        \item \textbf{Key Exchange Protocols}: Protocols such as the Diffie-Hellman key exchange\cite{Canetti2001AnalysisOK} allow two parties to securely share a secret key over an unsecured communication channel without revealing the key to any third party.
        \item \textbf{Pre-shared Keys}: In some cases, the key may be shared beforehand, for instance, physically exchanging a USB drive containing the key. However, this method is not practical for large-scale communication.
        \item \textbf{Hybrid Encryption}: A common approach is to use asymmetric encryption to securely exchange the symmetric key\cite{Simmons1979SymmetricAA}. For example, in SSL/TLS protocols\cite{2021RSAKS}, asymmetric encryption is used to exchange a session key, which is then used for faster symmetric encryption during the communication.
    \end{itemize}
    
    \subsection{Security Analysis of Symmetric Encryption}
    
    Symmetric encryption provides several security benefits, but it is not without vulnerabilities. Let's analyze both its strengths and weaknesses in modern cryptographic contexts.

    \subsubsection{Strengths}
    \begin{itemize}
        \item \textbf{Efficiency}: Symmetric encryption is computationally efficient, making it ideal for encrypting large datasets or streaming data\cite{Raigoza2016EvaluatingPO}.
        \item \textbf{Simplicity}: The implementation of symmetric encryption algorithms is generally more straightforward than asymmetric encryption\cite{AlShabi2019ASO}.
        \item \textbf{Low Computational Cost}: Symmetric encryption is less resource-intensive compared to its asymmetric counterpart, which makes it suitable for devices with limited computational power\cite{nour2024securing}.
    \end{itemize}
    
    \subsubsection{Weaknesses}
    \begin{itemize}
        \item \textbf{Brute-force Attacks}: Symmetric encryption is vulnerable to brute-force attacks, where an attacker systematically attempts every possible key\cite{Marinakis2013MinimumKL}. To mitigate this, modern symmetric algorithms use sufficiently long key sizes (e.g., AES with a 256-bit key)\cite{Blaze1996MinimalKL}.
        \item \textbf{Key Distribution}: As mentioned earlier, securely distributing the key remains a critical challenge\cite{yashaswini2015key}.
        \item \textbf{Key Management}: Managing and securely storing keys, especially in large-scale systems with multiple parties, can be difficult. If keys are not securely stored, the encryption becomes vulnerable to key exposure\cite{Damera2012SecureSO}.
    \end{itemize}
    
    \section{Stream Ciphers}
    
    Stream ciphers are a type of symmetric encryption that encrypt data one bit or byte at a time, as opposed to block ciphers, which encrypt data in fixed-size blocks\cite{cryptoeprint:2004/094}. This allows stream ciphers to be highly efficient and low latency, making them ideal for applications like real-time communications.

    \subsection{Encryption and Decryption Process of Stream Ciphers}
    
    The encryption process in stream ciphers involves generating a keystream, a pseudo-random sequence of bits, which is XORed with the plaintext to produce ciphertext\cite{Gorbenko2017TheRO}. Similarly, the decryption process XORs the ciphertext with the same keystream to retrieve the original plaintext.

    Let’s illustrate this with a simple Python example:
    
    \begin{lstlisting}[style=python]
    # Simple stream cipher example in Python
    def xor_bytes(data, key):
        return bytes([b ^ key[i % len(key)] for i, b in enumerate(data)])

    # Example of encryption
    plaintext = b"Hello World!"
    key = b"mysecretkey"
    ciphertext = xor_bytes(plaintext, key)
    print("Ciphertext:", ciphertext)

    # Example of decryption
    decrypted = xor_bytes(ciphertext, key)
    print("Decrypted:", decrypted)
    \end{lstlisting}
    
    In this example, the XOR operation is performed byte-by-byte between the plaintext and the key to generate the ciphertext. The same operation is applied to the ciphertext with the same key to recover the original plaintext.

    \subsection{The RC4 Algorithm and Its Applications}
    
    RC4 is one of the most well-known stream ciphers\cite{Jindal2015ASO}. It was widely used in protocols such as WEP (Wired Equivalent Privacy)\cite{Stoic2012RC4SC} and SSL/TLS\cite{Khine2009ANV}. RC4 operates by initializing a key-scheduling algorithm (KSA)\cite{ohigashi2008new} to create a keystream from a secret key. This keystream is XORed with the plaintext to produce ciphertext.

    However, RC4 has been found to have several vulnerabilities, particularly due to biases in the keystream\cite{Paterson2014BigBH}. Attackers can exploit these biases to recover parts of the key or plaintext, making RC4 insecure for many applications.

    Below is a brief explanation of how RC4 works:
    
    \begin{itemize}
        \item \textbf{Key-scheduling Algorithm (KSA)}: The KSA initializes the internal state of the cipher using the provided key.
        \item \textbf{Pseudo-random Generation Algorithm (PRGA)}: The PRGA generates the keystream, which is XORed with the plaintext\cite{Maity2017AnEL}.
    \end{itemize}
    
    Despite its widespread historical use, RC4 has been deprecated in modern security protocols due to its vulnerabilities.

    \subsection{Advantages and Limitations of Stream Ciphers}
    
    Stream ciphers offer several advantages, but they also come with limitations. Let's discuss both in detail.

    \subsubsection{Advantages}
    \begin{itemize}
        \item \textbf{Low Latency}: Since stream ciphers encrypt data one bit or byte at a time, they are well-suited for environments where low latency is essential, such as real-time communication\cite{Venkatesulu2016ASC}.
        \item \textbf{Efficiency}: Stream ciphers are typically faster than block ciphers for encrypting small amounts of data or streaming data\cite{Sharif2010PerformanceAO}.
        \item \textbf{Simplicity in Hardware}: Stream ciphers can be implemented efficiently in hardware, making them suitable for embedded systems and constrained environments\cite{Llise2005EfficientIO}.
    \end{itemize}
    
    \subsubsection{Limitations}
    \begin{itemize}
        \item \textbf{Key Reuse Vulnerability}: One of the biggest weaknesses of stream ciphers is the susceptibility to attacks if the same key is used more than once. Reusing the same key with different data can lead to key recovery attacks\cite{Stone2020RethinkingTW}.
        \item \textbf{Known Attacks}: Stream ciphers like RC4 have known vulnerabilities\cite{Paterson2014BigBH}. For example, biases in the RC4 keystream can be exploited to break its encryption.
        \item \textbf{Synchronization Requirements}: In some stream cipher implementations, both sender and receiver must remain synchronized for the decryption process to work correctly\cite{Armknecht2004ExtendingTR}. Loss of synchronization can result in data corruption.
    \end{itemize}

\section{Block Ciphers}
    \subsection{Fundamentals of Block Cipher Encryption}
    Block ciphers are encryption methods that operate on fixed-size blocks of data\cite{Singh2014BlockCI}. In block ciphers, the input data is divided into blocks of a specified length (typically 64 or 128 bits), and each block is encrypted separately using a symmetric key. This is in contrast to stream ciphers, which encrypt data bit by bit or byte by byte in a continuous stream.

    The basic operation of a block cipher involves taking a block of plaintext and a key as input and producing a block of ciphertext of the same size as the output. This process is highly deterministic, meaning that for a given input block and key, the output ciphertext will always be the same. However, the challenge in block cipher design lies in ensuring that the ciphertext is indistinguishable from random data.

    \subsubsection{Comparison to Stream Ciphers:}
    Block ciphers are generally more secure than stream ciphers when it comes to applications requiring high levels of confidentiality, particularly in environments where the same data might be encrypted multiple times. However, block ciphers are less efficient for real-time encryption, such as video streaming or voice transmission, where stream ciphers are typically more performant.

    \subsection{Block Cipher Modes: ECB, CBC, CFB, OFB, CTR}
    Block ciphers can be operated in various modes to achieve different cryptographic goals, such as confidentiality, integrity, or authentication. The mode of operation defines how each block of plaintext is processed and how it interacts with other blocks. Here, we will introduce five common block cipher modes:

    \subsubsection{Electronic Codebook (ECB):}
    In ECB mode\cite{Rachman2010PERBANDINGANMC}, each block of plaintext is encrypted independently of other blocks. While this mode is simple and fast, it has a significant weakness: identical plaintext blocks produce identical ciphertext blocks. This can reveal patterns in the plaintext, making ECB unsuitable for encrypting large datasets or images.
    
    \subsubsection{Cipher Block Chaining (CBC):}
    CBC mode addresses the weaknesses of ECB by chaining the blocks together\cite{Vaidehi2014DesignAA}. In this mode, each plaintext block is XORed with the previous ciphertext block before encryption. To ensure the first block is encrypted securely, an initialization vector (IV) is used. CBC is more secure than ECB but requires sequential encryption, making parallel processing difficult.
    
    \subsubsection{Cipher Feedback (CFB):}
    CFB operates similarly to CBC, but it can process smaller units than the block size, such as bits or bytes, making it more flexible for streaming data\cite{Gope2015IntegrityAwarePC}. In CFB, the previous ciphertext block is encrypted, and the result is XORed with the plaintext to produce the current ciphertext block. It also requires an IV but allows error propagation.

    \subsubsection{Output Feedback (OFB):}
    OFB mode is similar to CFB but avoids error propagation\cite{Sung2001ConcreteSA}. It generates a keystream by repeatedly encrypting the IV, which is XORed with the plaintext to produce the ciphertext. Since the plaintext never directly influences the ciphertext, a transmission error in one block does not affect subsequent blocks.

    \subsubsection{Counter (CTR):}
    In CTR mode, instead of chaining blocks, a counter is used, which is incremented for each block\cite{Rogaway2012CommentsTN}. The counter is encrypted, and the result is XORed with the plaintext to produce the ciphertext. CTR mode supports parallel processing of blocks and is efficient for applications where fast encryption is needed.

\section{Single-Key Encryption Algorithms}
    \subsection{Caesar Cipher}
    \subsubsection{Historical Background of Caesar Cipher}
    The Caesar cipher\cite{Allen2017TheCC} is one of the simplest and oldest encryption algorithms. It is a substitution cipher named after Julius Caesar, who reportedly used it to protect military communications. The Caesar cipher works by shifting each letter in the plaintext by a fixed number of positions in the alphabet. For example, with a shift of 3, A becomes D, B becomes E, and so on. Once the end of the alphabet is reached, the cipher wraps around to the beginning. 

    Caesar used this cipher with a shift of 3 to send encrypted messages, and although it was effective for simple military communication in his time, it is very insecure by modern standards.

    \subsubsection{Encryption and Decryption Process of Caesar Cipher}
    The Caesar cipher is a symmetric encryption algorithm, meaning the same key (shift value) is used for both encryption and decryption. The algorithm for encryption and decryption can be described as follows:

    \paragraph{Encryption:}
    For each letter in the plaintext:
    \begin{itemize}
        \item Find the corresponding letter in the alphabet after shifting by a certain number of positions.
        \item Replace the original letter with the new letter.
    \end{itemize}

    \paragraph{Decryption:}
    For each letter in the ciphertext:
    \begin{itemize}
        \item Find the corresponding letter in the alphabet by shifting in the opposite direction by the same number of positions.
        \item Replace the original letter with the new letter.
    \end{itemize}

    Below is a Python implementation of the Caesar cipher.

\begin{lstlisting}[style=python]
def caesar_encrypt(plaintext, shift):
    encrypted_text = ""
    for char in plaintext:
        if char.isalpha():
            shift_base = 65 if char.isupper() else 97
            encrypted_text += chr((ord(char) - shift_base + shift) % 26 + shift_base)
        else:
            encrypted_text += char
    return encrypted_text

def caesar_decrypt(ciphertext, shift):
    decrypted_text = ""
    for char in ciphertext:
        if char.isalpha():
            shift_base = 65 if char.isupper() else 97
            decrypted_text += chr((ord(char) - shift_base - shift) % 26 + shift_base)
        else:
            decrypted_text += char
    return decrypted_text

# Example usage:
plaintext = "HELLO WORLD"
shift = 3
ciphertext = caesar_encrypt(plaintext, shift)
print(f"Encrypted: {ciphertext}")
decrypted = caesar_decrypt(ciphertext, shift)
print(f"Decrypted: {decrypted}")
\end{lstlisting}

    \subsubsection{Security Analysis of Caesar Cipher}
    The Caesar cipher is highly insecure by today's cryptographic standards for several reasons:
    \begin{itemize}
        \item \textbf{Limited key space:} Since there are only 26 possible shifts (for the English alphabet), the Caesar cipher is vulnerable to brute-force attacks. An attacker could try all 26 possible shifts and easily determine the correct one.
        \item \textbf{Vulnerability to frequency analysis:} In languages like English, certain letters appear more frequently than others (e.g., E, T, A). By analyzing the frequency of letters in the ciphertext, an attacker can make educated guesses about the plaintext.
    \end{itemize}
    
    While the Caesar cipher may be useful for educational purposes, it offers no real security in modern applications. More advanced encryption techniques are necessary for securing sensitive information.

\subsection{Vigenère Cipher}

\subsubsection{Principle of Polyalphabetic Substitution}
The Vigenère cipher is a polyalphabetic substitution cipher\cite{RubinsteinSalzedo2018TheVC}, which means it uses multiple substitution alphabets to encode the plaintext. In contrast to a simple substitution cipher, where each letter in the plaintext is always replaced by the same letter in the ciphertext, the Vigenère cipher varies the substitution depending on the position of the letter and the key. 

The basic idea of polyalphabetic substitution is to use different Caesar shifts for each letter in the message, with the shifts determined by a repeating keyword. This significantly increases the complexity of the cipher, making it much harder to break compared to simple substitution ciphers, as the same letter in the plaintext can be encoded as different letters in the ciphertext depending on the position of the key.

For example, consider the plaintext "HELLO" and the key "KEY". In a Vigenère cipher:
\begin{itemize}
    \item The first letter 'H' is shifted by 'K' (which represents a shift of 10), resulting in 'R'.
    \item The second letter 'E' is shifted by 'E' (which represents a shift of 4), resulting in 'I'.
    \item The third letter 'L' is shifted by 'Y' (which represents a shift of 24), resulting in 'J'.
    \item The fourth letter 'L' is shifted by 'K' (starting the key over), resulting in 'V'.
    \item The fifth letter 'O' is shifted by 'E', resulting in 'S'.
\end{itemize}
Thus, the ciphertext becomes "RIJVS". As this example illustrates, the Vigenère cipher distributes the frequency of each letter across the ciphertext, making frequency analysis, which is effective against simple substitution ciphers, much less useful.

\subsubsection{Encryption and Decryption Process of Vigenère Cipher}
To encrypt a message using the Vigenère cipher, we use the following steps:

\begin{enumerate}
    \item \textbf{Choose a key:} The key is a repeated string of characters (e.g., "KEY").
    \item \textbf{Align the key with the plaintext:} Repeat the key as necessary to match the length of the plaintext.
    \item \textbf{Convert letters to numbers:} Convert each letter of the plaintext and key to its corresponding position in the alphabet (A=0, B=1, \dots, Z=25).
    \item \textbf{Add the key values to the plaintext values:} For each letter, add the corresponding key value to the plaintext value (mod 26), resulting in the ciphertext.
    \item \textbf{Convert the numbers back to letters:} Convert the resulting numbers back to letters to form the ciphertext.
\end{enumerate}

To decrypt, the same process is used, but instead of adding the key values, we subtract them.

\paragraph{Python Implementation}
Below is an implementation of the Vigenère cipher in Python.

\begin{lstlisting}[style=python]
def vigenere_encrypt(plaintext, key):
    key = key.upper()
    ciphertext = []
    for i, letter in enumerate(plaintext):
        if letter.isalpha():
            shift = ord(key[i % len(key)]) - ord('A')
            encrypted_char = chr((ord(letter.upper()) - ord('A') + shift) % 26 + ord('A'))
            ciphertext.append(encrypted_char)
        else:
            ciphertext.append(letter)
    return ''.join(ciphertext)

def vigenere_decrypt(ciphertext, key):
    key = key.upper()
    plaintext = []
    for i, letter in enumerate(ciphertext):
        if letter.isalpha():
            shift = ord(key[i % len(key)]) - ord('A')
            decrypted_char = chr((ord(letter.upper()) - ord('A') - shift) % 26 + ord('A'))
            plaintext.append(decrypted_char)
        else:
            plaintext.append(letter)
    return ''.join(plaintext)

# Example usage
plaintext = "HELLO"
key = "KEY"
ciphertext = vigenere_encrypt(plaintext, key)
print("Ciphertext:", ciphertext)

decrypted_text = vigenere_decrypt(ciphertext, key)
print("Decrypted Text:", decrypted_text)
\end{lstlisting}

This code demonstrates both the encryption and decryption processes of the Vigenère cipher. You can adjust the \texttt{plaintext} and \texttt{key} variables to test the cipher with different inputs.

\subsubsection{Kasiski Examination and Cryptanalysis}
The Kasiski examination\cite{Hananto2019AnalyzingTK} is a method used to break the Vigenère cipher by identifying repeated sequences in the ciphertext. Since the cipher relies on a repeating keyword, patterns can emerge when the key is repeated, especially if parts of the plaintext contain repeated words or phrases.

The basic process of Kasiski examination is as follows:
\begin{enumerate}
    \item \textbf{Identify repeated sequences}: Search the ciphertext for repeated sequences of characters. These repeated sequences likely correspond to parts of the plaintext that were encrypted using the same key segment.
    \item \textbf{Calculate the distance between repetitions}: For each repeated sequence, calculate the distance (in terms of character positions) between occurrences.
    \item \textbf{Determine the key length}: The greatest common divisor (GCD) of these distances is likely to be the length of the key, as repeated segments of the key are responsible for encrypting the repeated segments of the plaintext.
    \item \textbf{Perform frequency analysis}: Once the key length is known, break the ciphertext into columns where each column corresponds to a letter of the key. Each column can be analyzed independently using frequency analysis, as each column represents a simple Caesar cipher.
\end{enumerate}

By exploiting these repeated sequences, the Kasiski examination significantly reduces the complexity of cryptanalyzing a Vigenère cipher, especially when the key is relatively short.

\subsection{Data Encryption Standard (DES)}

\subsubsection{Basic Structure of the DES Algorithm}
The Data Encryption Standard (DES) is a symmetric-key block cipher that was widely used for data encryption in the late 20th century\cite{Landau2000StandingTT}. DES operates on blocks of 64 bits of data and uses a 56-bit key. The core of DES is a \textit{Feistel network}\cite{Kumari2023AnRI}, which involves splitting the data into two halves and applying a series of encryption rounds.

\paragraph{Feistel Network}
In each round of DES, the right half of the data block is expanded, permuted, and combined with a subkey using XOR operations. This result is then processed by a set of substitution boxes (S-boxes) to introduce non-linearity. The modified right half is then swapped with the left half for the next round. After 16 rounds, the two halves are recombined, and a final permutation is applied to produce the encrypted output.

\begin{itemize}
    \item \textbf{Input}: A 64-bit block of plaintext.
    \item \textbf{Key}: A 56-bit encryption key.
    \item \textbf{Output}: A 64-bit block of ciphertext.
\end{itemize}

\subsubsection{Encryption Rounds and Subkey Generation}
DES performs 16 rounds of encryption, with each round using a different subkey derived from the original key. The steps in each round are as follows:
\begin{enumerate}
    \item \textbf{Expansion}: The 32-bit right half of the data is expanded to 48 bits using an expansion permutation.
    \item \textbf{Key mixing}: The expanded right half is XORed with a 48-bit subkey.
    \item \textbf{Substitution}: The result is passed through 8 S-boxes, each of which takes a 6-bit input and produces a 4-bit output.
    \item \textbf{Permutation}: The 32-bit output from the S-boxes is permuted using a predefined permutation table.
    \item \textbf{Swap}: The left and right halves of the data are swapped before proceeding to the next round.
\end{enumerate}

Subkey generation involves applying a permutation to the 56-bit key, splitting it into two halves, and then rotating the halves before applying another permutation to produce a 48-bit subkey for each round.

\subsubsection{Security Issues of DES}
DES, once considered secure, has significant vulnerabilities by modern standards. The primary issue is its relatively short key length of 56 bits, which makes it susceptible to \textit{brute-force attacks}. With the advances in computing power, it is now feasible to try all possible 56-bit keys in a reasonable amount of time.

Furthermore, DES is vulnerable to \textit{differential cryptanalysis} and \textit{linear cryptanalysis}, which exploit specific patterns in the encryption process. Due to these weaknesses, DES was replaced by more secure algorithms, such as \textit{Triple DES (3DES)}\cite{Venigalla2012IMPLEMENTATIONOT} and \textit{Advanced Encryption Standard (AES)}\cite{Selent2010ADVANCEDES}.

\subsection{Triple DES (3DES)}
    \subsubsection{Operation Process of 3DES}
    Triple DES (3DES)\cite{Chen2021AccountingDE} enhances the security of the original Data Encryption Standard (DES) by applying the DES algorithm three times with three different keys. The basic operation process is as follows:
    
\begin{enumerate}
    \item \textbf{Encryption (E):} The data is first encrypted with the first key ($K_1$) using DES.
    \item \textbf{Decryption (D):} The result from step 1 is decrypted using DES with the second key ($K_2$).
    \item \textbf{Encryption (E):} The output from step 2 is encrypted again with the third key ($K_3$).
\end{enumerate}

    Therefore, the encryption process can be represented as:
    \[
    \text{Ciphertext} = E_{K_3}(D_{K_2}(E_{K_1}(\text{Plaintext})))
    \]
    
    Similarly, the decryption process is the reverse of this:
    \[
    \text{Plaintext} = D_{K_1}(E_{K_2}(D_{K_3}(\text{Ciphertext})))
    \]

    This triple application of DES significantly improves the security by increasing the effective key length from 56 bits in DES to 168 bits in 3DES (three 56-bit keys). 

    \textbf{Example:} If we want to encrypt a 64-bit block of data using 3DES:
    
    \begin{itemize}
        \item First, we apply the DES algorithm using key $K_1$.
        \item Then, we decrypt the result using key $K_2$.
        \item Finally, we encrypt the intermediate output using key $K_3$.
    \end{itemize}

    \subsubsection{Applications and Security of 3DES}
    3DES has been widely used in industries that require a high level of security, such as in financial systems and government institutions. For example, it has been used in \textit{payment gateways}, \textit{automated teller machines (ATMs)}, and \textit{electronic financial transactions}.

    The security of 3DES is much stronger than DES due to its triple encryption. However, it still has some drawbacks:

    \begin{itemize}
        \item \textbf{Performance:} 3DES is relatively slow compared to modern encryption algorithms like AES. This is because it applies the DES algorithm three times, which increases the computational overhead.
        \item \textbf{Key Length:} While 168-bit keys provide better security than DES, 3DES can still be vulnerable to certain attacks, such as the \textit{meet-in-the-middle attack}, which reduces its effective security to about 112 bits.
    \end{itemize}
    
    Because of these performance limitations and the fact that stronger algorithms have been developed, 3DES is being phased out and replaced by the Advanced Encryption Standard (AES) in many applications.

\subsection{Advanced Encryption Standard (AES)}
    \subsubsection{Design Principles of AES}
    The Advanced Encryption Standard (AES)\cite{Selent2010ADVANCEDES} is a symmetric block cipher that was developed as a replacement for DES and 3DES. AES was selected as the encryption standard by the U.S. government in 2001\cite{Rao2017ASO}. Its design is based on a structure known as the \textit{Substitution-Permutation Network (SPN)}\cite{Li2011AND}, which is resistant to several forms of cryptanalysis, including \textit{linear} and \textit{differential cryptanalysis}.

    Key design principles of AES include:
    
    \begin{itemize}
        \item \textbf{Substitution:} AES uses a non-linear substitution step (called the S-box)\cite{Scripcariu2013OnTS} that replaces each byte with another byte according to a fixed lookup table. This helps introduce confusion into the data, making it harder for attackers to find relationships between the plaintext and ciphertext.
    
        \item \textbf{Permutation:} AES also includes a permutation step (known as the ShiftRows and MixColumns steps)\cite{Mahrousa2020ANM}, which re-arranges the data to further complicate the relationships between the plaintext and ciphertext. This adds diffusion, spreading the influence of each byte of the plaintext across the entire ciphertext.

        \item \textbf{Key Schedule:} AES employs a key schedule to expand the encryption key into multiple round keys, which are used in each round of encryption.
    \end{itemize}

    \subsubsection{AES Block Sizes and Key Lengths}
    AES operates on fixed-size blocks of 128 bits. It supports three key lengths: 128 bits, 192 bits, and 256 bits. These key lengths affect both the number of rounds of encryption and the overall security of the algorithm:
    
    \begin{itemize}
        \item \textbf{AES-128:} Uses a 128-bit key and 10 rounds of encryption.
        \item \textbf{AES-192:} Uses a 192-bit key and 12 rounds of encryption.
        \item \textbf{AES-256:} Uses a 256-bit key and 14 rounds of encryption.
    \end{itemize}

    A longer key length results in a higher level of security. For example, AES-256 is much more resistant to brute-force attacks than AES-128 due to its longer key length, making it suitable for highly sensitive data.

    \textbf{Example:} Suppose you are securing a sensitive financial transaction. You might choose AES-256 to ensure maximum security, while for encrypting less sensitive data like log files, AES-128 could be used for its faster performance.

    \subsubsection{Encryption Process and Security of AES}
    The AES encryption process involves multiple steps applied to the data block in several rounds. Here's a step-by-step breakdown:

    \begin{itemize}
        \item \textbf{Key Expansion:} The original key is expanded into a set of round keys, which are derived from the initial encryption key.
    
        \item \textbf{Initial Round:} The first step is an \textit{AddRoundKey} operation where the plaintext block is XORed with the first round key.

        \item \textbf{Main Rounds (9, 11, or 13 rounds depending on the key size):}
        \begin{itemize}
            \item \textit{SubBytes:} Each byte of the block is substituted with a new byte from the S-box.
            \item \textit{ShiftRows:} The rows of the state matrix are shifted to the left by different offsets to introduce diffusion.
            \item \textit{MixColumns:} The columns of the matrix are mixed by applying a matrix multiplication. This further diffuses the data.
            \item \textit{AddRoundKey:} The result is XORed with a round key derived from the original key.
        \end{itemize}

        \item \textbf{Final Round:} The last round omits the MixColumns step and only applies SubBytes, ShiftRows, and AddRoundKey.
    \end{itemize}

    \textbf{Example:} Let's say we are encrypting a block of text. Using AES-128, the process would begin by XORing the plaintext with the first round key, then performing SubBytes, ShiftRows, MixColumns, and AddRoundKey for 9 rounds, and finally applying one last round of SubBytes, ShiftRows, and AddRoundKey.

    AES is currently considered the gold standard for encryption due to its strong resistance to known cryptographic attacks and its efficiency. It is widely used in various fields, including \textit{secure communication}, \textit{file encryption}, \textit{wireless security protocols (such as WPA2)}, and \textit{blockchain technology}.

\chapter{Asymmetric Encryption Algorithms}

\section{Basic Concepts of Asymmetric Encryption}

Asymmetric encryption, also known as public-key encryption, is a cryptographic method that uses a pair of keys to encrypt and decrypt data\cite{Simmons1979SymmetricAA}. This approach eliminates the need for both parties to share a secret key ahead of time, making communication more secure.

\subsection{Definition of Public and Private Keys}

In asymmetric encryption, two keys are generated: a public key and a private key. These keys are mathematically related, but it is computationally infeasible to derive the private key from the public key.

\begin{itemize}
    \item \textbf{Public Key:} This key is shared openly and can be distributed to anyone. It is used to encrypt messages or verify signatures.
    \item \textbf{Private Key:} This key must remain confidential to the owner. It is used to decrypt messages encrypted with the corresponding public key or to sign data to ensure authenticity.
\end{itemize}

The security of asymmetric encryption is based on the fact that while the public key can be freely distributed, only the private key can decrypt the data, ensuring that sensitive information remains secure.

\subsection{Public Key Encryption and Private Key Decryption}

In public-key cryptography, the public key is used for encrypting a message, and the private key is used to decrypt it. Let's walk through an example to clarify this concept.

\textbf{Example:} Alice wants to send a secret message to Bob. Here's how they do it:

\begin{itemize}
    \item Bob shares his public key with Alice.
    \item Alice uses Bob's public key to encrypt a message: ``Hello, Bob!''.
    \item Alice sends the encrypted message to Bob.
    \item Bob uses his private key to decrypt the message and read the original content.
\end{itemize}

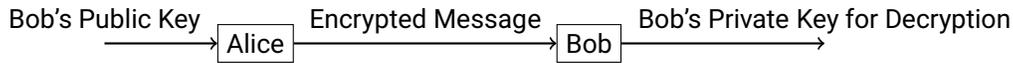
\begin{center}
\begin{tikzpicture}
  \node (alice) [rectangle, draw] {Alice};
  \node (bob) [rectangle, draw, right=2cm] at (2,0) {Bob}; 
  \draw[->, thick] (alice.east) -- node[above] {Encrypted Message} (bob.west);
  \draw[<-, thick] (alice.west) -- ++(-1.5cm, 0) node[above] {Bob's Public Key};
  \draw[->, thick] (bob.east) -- ++(2.7cm, 0cm
) node[above] {Bob's Private Key for Decryption};
\end{tikzpicture}
\end{center}

In this way, only Bob can decrypt the message because only he possesses the private key. Even if someone intercepts the encrypted message, they cannot read it without Bob's private key.

\subsection{Advantages and Challenges of Asymmetric Encryption}

\textbf{Advantages:}
\begin{itemize}
    \item \textbf{No Need for Pre-shared Keys:} Asymmetric encryption eliminates the need to exchange a secret key over insecure channels.
    \item \textbf{Scalability:} Public keys can be distributed to many users, making this approach scalable for systems where many parties are involved.
    \item \textbf{Authentication:} Asymmetric encryption provides an authentication mechanism via digital signatures, where the private key is used to sign data, and the public key verifies it.
\end{itemize}

\textbf{Challenges:}
\begin{itemize}
    \item \textbf{Computationally Intensive:} Asymmetric encryption is significantly slower compared to symmetric encryption due to the complex mathematical operations involved.
    \item \textbf{Key Management:} Even though public keys can be freely distributed, private keys must be carefully managed and kept secure. Loss of a private key could lead to data being irrecoverable.
\end{itemize}

\section{RSA Algorithm}

The RSA algorithm\cite{Zhong2022AnOO} is one of the most widely used asymmetric encryption algorithms. It is based on the difficulty of factoring large numbers, which ensures the security of the cryptosystem.

\subsection{Working Principle of RSA Algorithm}

RSA works by generating two large prime numbers and using them to compute a pair of keys: one for encryption (public key) and one for decryption (private key). The mathematical foundation of RSA is based on modular exponentiation and the difficulty of prime factorization.

The main steps in RSA are:
\begin{enumerate}
    \item Choose two large prime numbers, \( p \) and \( q \).
    \item Compute \( n = p \times q \) (the modulus).
    \item Calculate \( \phi(n) = (p - 1) \times (q - 1) \) (Euler's totient function).
    \item Select an integer \( e \) such that \( 1 < e < \phi(n) \) and \( e \) is coprime with \( \phi(n) \). This becomes the public key exponent.
    \item Compute \( d \), the modular multiplicative inverse of \( e \), such that \( e \times d \equiv 1 \, (\text{mod} \, \phi(n)) \). This is the private key exponent.
\end{enumerate}

\subsection{Large Prime Numbers and Key Generation}

The security of RSA relies on the difficulty of factoring large numbers. The two large prime numbers, \( p \) and \( q \), are chosen such that their product, \( n \), is hard to factor back into \( p \) and \( q \). Key generation in RSA follows this process:

\begin{lstlisting}[style=python]
from sympy import randprime, mod_inverse

# Step 1: Generate large prime numbers
p = randprime(10**10, 10**11)
q = randprime(10**10, 10**11)

# Step 2: Calculate n and Euler's totient function
n = p * q
phi_n = (p - 1) * (q - 1)

# Step 3: Choose public exponent e (commonly 65537)
e = 65537

# Step 4: Calculate private key d
d = mod_inverse(e, phi_n)

print(f"Public Key: (n={n}, e={e})")
print(f"Private Key: (n={n}, d={d})")
\end{lstlisting}

\subsection{Encryption and Decryption Process of RSA}

In RSA, encryption is performed using the public key and modular exponentiation. Decryption is performed using the private key. Let's look at how this works in practice.

\textbf{Encryption:} To encrypt a message \( m \), compute the ciphertext \( c \) as follows:

\[
c = m^e \mod n
\]

\textbf{Decryption:} To decrypt the ciphertext \( c \), compute the original message \( m \) as follows:

\[
m = c^d \mod n
\]

\begin{lstlisting}[style=python]
# Encryption
message = 123456789  # Example message to encrypt
ciphertext = pow(message, e, n)
print(f"Encrypted message: {ciphertext}")

# Decryption
decrypted_message = pow(ciphertext, d, n)
print(f"Decrypted message: {decrypted_message}")
\end{lstlisting}

In this example, the message is encrypted using the public key (exponent \( e \) and modulus \( n \)) and decrypted using the private key (exponent \( d \) and modulus \( n \)).

\subsection{Applications and Security Analysis of RSA}

\textbf{Applications:}
\begin{itemize}
    \item \textbf{Digital Signatures:} RSA is widely used for creating digital signatures, which ensure that a message or document is authentic and unaltered.
    \item \textbf{Secure Communication:} RSA is used in establishing secure communication channels, such as in SSL/TLS protocols for web security.
    \item \textbf{Email Encryption:} RSA is commonly implemented in secure email systems like PGP (Pretty Good Privacy) to encrypt messages.
\end{itemize}

\textbf{Security Analysis:}
\begin{itemize}
    \item \textbf{Strength of RSA:} The strength of RSA comes from the difficulty of factoring large numbers. The larger the prime numbers used to generate the keys, the more secure the RSA system.
    \item \textbf{Factorization Attacks:} As of now, no efficient algorithm exists for factoring large integers, but advances in factorization techniques (e.g., quantum computing) pose a potential future threat.
    \item \textbf{Quantum Computing:} Shor's algorithm, which can run on quantum computers, can efficiently factor large integers, which would break RSA. Researchers are working on quantum-resistant algorithms to address this.
\end{itemize}

\section{Elliptic Curve Cryptography (ECC)}

\subsection{Mathematical Foundation of ECC}
Elliptic Curve Cryptography (ECC)\cite{Koppl2021ApplicationOC} is based on the algebraic structure of elliptic curves over finite fields. An elliptic curve is represented by an equation of the form:

\[
y^2 = x^3 + ax + b
\]

where \(a\) and \(b\) are constants that define the shape of the curve, and the curve must satisfy the condition:

\[
4a^3 + 27b^2 \neq 0
\]

This ensures that the curve has no cusps or self-intersections. The set of points on this curve, along with a special "point at infinity," form a group, which is used in cryptography.

One of the most important properties of elliptic curves is the difficulty of the Elliptic Curve Discrete Logarithm Problem (ECDLP)\cite{Sadkhan2021DevelopmentOS}. Given two points \(P\) and \(Q\) on an elliptic curve, where \(Q = kP\) (i.e., \(Q\) is the result of adding \(P\) to itself \(k\) times), finding \(k\) is computationally hard. This problem provides the security foundation for ECC, similar to how the difficulty of factoring large numbers provides the security for RSA.

For example, on an elliptic curve over a finite field, consider the following operation:

\[
P + P = Q
\]

Here, finding the value of \(k\) such that \(Q = kP\) is analogous to solving the discrete logarithm problem in ECC.

\subsection{Generation and Encryption Using Elliptic Curves}
ECC uses the properties of elliptic curves for key generation, encryption, and decryption. A key pair consists of a private key, which is a randomly selected integer, and a public key, which is a point on the elliptic curve.

Let’s walk through key generation and encryption:

\textbf{Key Generation:}
\begin{itemize}
    \item Choose a random integer \(d\), which will be the private key.
    \item Calculate the public key \(Q = dG\), where \(G\) is a publicly known base point on the elliptic curve, and \(dG\) means adding \(G\) to itself \(d\) times.
\end{itemize}

\textbf{Encryption:}
To encrypt a message \(M\) using the recipient’s public key \(Q\):
\begin{itemize}
    \item Represent \(M\) as a point on the elliptic curve \(P_M\).
    \item Choose a random integer \(k\).
    \item Compute the ciphertext as the pair of points \((kG, P_M + kQ)\).
\end{itemize}

\textbf{Decryption:}
To decrypt the ciphertext \((C_1, C_2)\), where \(C_1 = kG\) and \(C_2 = P_M + kQ\):
\begin{itemize}
    \item Use the private key \(d\) to compute \(dC_1 = dkG = kQ\).
    \item Subtract \(kQ\) from \(C_2\): \(P_M = C_2 - kQ\).
\end{itemize}

This process ensures that only the recipient, who knows the private key \(d\), can decrypt the message.

\textbf{Python Example for ECC Key Generation:}
\begin{lstlisting}[style=python]
from tinyec import registry
import secrets

curve = registry.get_curve('brainpoolP256r1')

# Private key
privKey = secrets.randbelow(curve.field.n)

# Public key
pubKey = privKey * curve.g

print("Private key:", privKey)
print("Public key:", pubKey)
\end{lstlisting}

\subsection{Advantages of ECC: Efficiency and Security}
ECC offers several significant advantages over traditional cryptographic algorithms such as RSA:

\begin{itemize}
    \item \textbf{Smaller Key Sizes:} ECC provides the same level of security as RSA but with much smaller key sizes. For instance, a 256-bit key in ECC provides roughly the same security as a 3072-bit key in RSA.
    \item \textbf{Faster Computations:} Due to the smaller key sizes, the mathematical operations in ECC are faster and require less computational power. This is particularly important for environments with limited resources, such as mobile devices.
    \item \textbf{Lower Resource Consumption:} ECC's efficiency leads to reduced memory, bandwidth, and energy consumption, making it ideal for devices like smartphones, IoT devices, and embedded systems.
\end{itemize}

\textbf{Comparison between ECC and RSA:} \\

\begin{center}
\begin{tabular}{|c|c|c|}
    \hline
    \textbf{Security Level (bits)} & \textbf{ECC Key Size} & \textbf{RSA Key Size} \\
    \hline
    128-bit security & 256 bits & 3072 bits \\
    \hline
    192-bit security & 384 bits & 7680 bits \\
    \hline
    256-bit security & 512 bits & 15360 bits \\
    \hline
\end{tabular}
\end{center}

As this table shows, ECC can achieve high levels of security with much smaller key sizes than RSA.

\subsection{Common Applications of ECC}
ECC has become widely adopted in a variety of real-world applications due to its efficiency and strong security properties.

\begin{itemize}
    \item \textbf{Mobile Device Encryption:} ECC is used in mobile devices to secure communications and data\cite{Dar2021LightweightAS}. Due to the limited computational power and battery life of mobile devices, ECC's efficiency makes it a preferred choice for securing messaging apps, email, and VPNs.
    \item \textbf{Cryptocurrency (e.g., Bitcoin):} ECC plays a crucial role in securing cryptocurrencies like Bitcoin\cite{Mehta2015ASO}. The Bitcoin protocol uses ECC to generate cryptographic keys and sign transactions, ensuring that only the owner of the private key can authorize spending.
    \item \textbf{SSL/TLS for Secure Web Communication:} ECC is used in SSL/TLS protocols to secure web communications\cite{Sebastian2015AdvantageOU}. When you visit websites with HTTPS, ECC may be part of the underlying cryptography that secures your connection.
\end{itemize}

\textbf{Example: ECC in Bitcoin}
In Bitcoin, elliptic curves are used to generate a pair of keys for every user. The public key is shared, while the private key remains secret. The keys are based on the secp256k1 curve, and ECC ensures that only the owner of the private key can sign transactions.

Here is a Python code example that demonstrates how ECC can be used to generate a Bitcoin address using the secp256k1 curve:
\begin{lstlisting}[style=python]
from ecdsa import SigningKey, SECP256k1
import hashlib

# Generate private key
private_key = SigningKey.generate(curve=SECP256k1)

# Get public key
public_key = private_key.get_verifying_key()

# Hash the public key to generate a Bitcoin address
public_key_bytes = public_key.to_string()
sha256_digest = hashlib.sha256(public_key_bytes).digest()
ripemd160 = hashlib.new('ripemd160')
ripemd160.update(sha256_digest)
bitcoin_address = ripemd160.hexdigest()

print("Bitcoin Address:", bitcoin_address)
\end{lstlisting}

In this example, we generate a private key, compute the corresponding public key, and then create a Bitcoin address using a combination of SHA-256 and RIPEMD-160 hashing algorithms.

ECC continues to be a cornerstone of modern cryptography, offering both security and efficiency, making it a critical technology for securing communication and transactions in the digital age.

\section{Diffie-Hellman Key Exchange}
    \subsection{Working Mechanism of Diffie-Hellman}
    The Diffie-Hellman key exchange\cite{Kumar2020DiffieHS} is a fundamental cryptographic protocol that allows two parties to securely generate a shared secret over an insecure communication channel. This shared secret can then be used for symmetric encryption of further communication. The beauty of this protocol is that it does not require the two parties to have previously exchanged any secret keys. Instead, they exchange some public values, from which the shared secret is derived.

    Here’s how the Diffie-Hellman key exchange works step-by-step:

    \begin{enumerate}
        \item \textbf{Public Parameters:} Both parties agree on two public parameters: a large prime number $p$ and a base $g$, which is a primitive root modulo $p$. These values do not need to be kept secret, and can be known by anyone.
        
        \item \textbf{Key Generation:}
        \begin{itemize}
            \item Party A selects a private key $a$ (a randomly chosen integer) and calculates their public key as $A = g^a \mod p$.
            \item Party B selects a private key $b$ and calculates their public key as $B = g^b \mod p$.
        \end{itemize}
        
        \item \textbf{Exchange of Public Keys:} Party A sends their public key $A$ to Party B, and Party B sends their public key $B$ to Party A. These public keys can be freely shared over an insecure channel.

        \item \textbf{Shared Secret Generation:}
        \begin{itemize}
            \item Party A computes the shared secret as $S_A = B^a \mod p$.
            \item Party B computes the shared secret as $S_B = A^b \mod p$.
        \end{itemize}
        \item \textbf{Result:} Both $S_A$ and $S_B$ are equal, as $B^a \mod p = (g^b)^a \mod p = (g^a)^b \mod p = A^b \mod p$. Thus, both parties now have the same shared secret $S$, which can be used for secure symmetric encryption.

    \end{enumerate}
    
    This process ensures that even if an attacker intercepts the public keys $A$ and $B$, they cannot easily determine the shared secret $S$ without knowing the private keys $a$ or $b$. This is because computing the private key from the public key requires solving the discrete logarithm problem, which is computationally infeasible for large values of $p$.

    \subsection{Security Analysis of the Key Exchange Process}
    The security of the Diffie-Hellman key exchange protocol relies heavily on the difficulty of solving the discrete logarithm problem\cite{GanjewarDieHK}. Even if an eavesdropper, often referred to as Eve, can observe the public values exchanged (the prime $p$, the base $g$, and the public keys $A$ and $B$), calculating the shared secret requires solving $g^a \mod p$ or $g^b \mod p$, which is practically impossible for large $p$.

    \textbf{Resistance to Eavesdropping:} The protocol is resistant to passive eavesdropping because the attacker cannot compute the shared secret without knowing the private keys. This protection stems from the fact that while the public keys are exchanged openly, the private keys are never transmitted, and recovering the private keys from the public keys is computationally difficult.

    \textbf{Vulnerability to Man-in-the-Middle (MITM) Attacks:} However, Diffie-Hellman is vulnerable to a man-in-the-middle attack if the public keys are not authenticated\cite{Naher2018AuthenticationOD}. In this attack, an adversary, often called "Mallory," intercepts the public key exchange between Party A and Party B. Mallory can then generate her own private key and public key, replacing Party A's public key with her own in the message to Party B and vice versa. This allows Mallory to establish two separate shared secrets with each party, enabling her to decrypt and re-encrypt all communications between them without their knowledge.

    To prevent man-in-the-middle attacks, it is essential to authenticate the public keys, typically through the use of certificates or digital signatures. Modern implementations of Diffie-Hellman, such as those in Transport Layer Security (TLS)\cite{Eronen2005PreSharedKC}, often use authenticated variants to address this issue.

    \subsection{Modern Applications of Diffie-Hellman}
    Diffie-Hellman is widely used in modern secure communication protocols. Its primary application is in establishing a shared secret for symmetric encryption in the following contexts:
    
    \begin{itemize}
        \item \textbf{Virtual Private Networks (VPNs):} VPNs often use Diffie-Hellman to establish a secure connection between the client and the server\cite{Carts2001ARO}. The shared secret derived from Diffie-Hellman is used to encrypt the data transmitted between the two endpoints.
        \item \textbf{TLS/SSL (Transport Layer Security):} Secure web communications, such as HTTPS, use a variant of Diffie-Hellman, known as Elliptic Curve Diffie-Hellman (ECDH)\cite{BlakeWilson2006EllipticCC}, to negotiate encryption keys during the handshake process. This ensures that even if the communication channel is intercepted, the eavesdropper cannot decrypt the data.
        \item \textbf{Secure Messaging:} Protocols like the Signal Protocol use Diffie-Hellman to provide forward secrecy in encrypted messaging applications\cite{CohnGordon2018OnEE}. This ensures that even if a user’s encryption keys are compromised in the future, previous communications remain secure.
    \end{itemize}

\section{Comparison of Symmetric and Asymmetric Encryption}
    \subsection{Comparison of Encryption Efficiency}
    Symmetric encryption and asymmetric encryption differ significantly in terms of their efficiency. Symmetric encryption algorithms, such as AES (Advanced Encryption Standard), are much faster than asymmetric algorithms like RSA (Rivest-Shamir-Adleman)\cite{Raigoza2016EvaluatingPO}. This is because symmetric encryption uses the same key for both encryption and decryption, allowing for efficient operations, typically involving block ciphers or stream ciphers.

    Asymmetric encryption, on the other hand, requires more complex mathematical operations, such as modular exponentiation, which are computationally intensive. This results in asymmetric encryption being slower and less efficient, especially for large volumes of data.

    \textbf{Why Symmetric Encryption is Faster:}
    \begin{itemize}
        \item Symmetric encryption typically operates on blocks of data (e.g., AES uses a 128-bit block size), allowing the algorithm to process data in parallel.
        \item It requires fewer computational steps, relying on substitution-permutation networks that can be implemented efficiently in hardware.
    \end{itemize}

    \textbf{Why Asymmetric Encryption is Slower:}
    \begin{itemize}
        \item Asymmetric encryption algorithms rely on number-theoretic problems like factoring large integers or solving the discrete logarithm problem, both of which are computationally intensive.
        \item The use of large key sizes (e.g., 2048-bit or 4096-bit RSA keys) further increases the computational overhead.
    \end{itemize}

    \subsection{Security Analysis}
    Both symmetric and asymmetric encryption methods have their strengths and weaknesses. Symmetric encryption offers high efficiency and strong encryption, but it suffers from the key distribution problem. Since the same key is used for encryption and decryption, securely sharing the key between parties is a challenge.

    Asymmetric encryption solves the key distribution problem by using two different keys: a public key for encryption and a private key for decryption. This makes key management simpler, but the increased computational complexity can be a disadvantage in performance-sensitive applications.

    \subsection{Use Cases of Asymmetric Encryption in Blockchain}
    Asymmetric encryption plays a crucial role in blockchain technology, particularly in the following areas:
    
    \begin{itemize}
        \item \textbf{Digital Signatures:} In blockchain systems like Bitcoin and Ethereum, transactions are signed using the sender’s private key\cite{Mayer2016ECDSASI}. This ensures that only the holder of the private key can initiate a transaction. The public key can be used by others to verify the authenticity of the transaction.
        
        \item \textbf{Key Management:} Asymmetric encryption allows users to securely manage their private and public keys. In the context of cryptocurrencies, users need to keep their private keys secure to control their assets, while their public keys can be shared freely for receiving payments.
        
        \item \textbf{Securing Transactions:} Cryptographic techniques based on asymmetric encryption, such as elliptic curve cryptography (ECC), are used to secure transactions in modern blockchain platforms\cite{Chandel2019AMA}. These techniques ensure that transactions are tamper-proof and can be validated by all participants in the network.
    \end{itemize}

\chapter{Hash Functions}

\section{Basic Definition and Properties of Hash Functions}

Hash functions play a fundamental role in many cryptographic systems, particularly in blockchain technology. At their core, a hash function is a mathematical algorithm that takes an input (or "message") and returns a fixed-size string of bytes, typically in the form of a hash value\cite{Smart2016HashFM}. This output is unique to the specific input, but the same input will always result in the same hash value. For example, in blockchain systems, hash functions are used to maintain data integrity, verify transactions, and protect the system from manipulation. There are several essential properties of hash functions, two of which are particularly important: one-wayness and collision resistance.

\subsection{One-Wayness and Irreversibility}

One of the most critical properties of a cryptographic hash function is one-wayness\cite{Boldyreva2009FoundationsON}. This means that, given the hash value, it is computationally infeasible to reverse-engineer the input that produced it. Essentially, hash functions are designed to be irreversible. Even if someone has the hash value, they cannot retrieve the original input data easily.

To understand this better, imagine a hash function as a meat grinder. Once you put a piece of meat into the grinder, it transforms into ground meat. While it's very easy to grind the meat, trying to take the ground meat and reassemble it back into the original piece of meat is practically impossible. Similarly, with hash functions, generating the hash value is straightforward, but reversing the process to get the original data is computationally infeasible.

This irreversibility is what protects data in blockchain systems\cite{Wang2024BlockchainSA}. For example, even if someone obtains the hash of a block or transaction, they cannot use it to figure out the details of the original transaction.

\subsection{Collision Resistance and Security}

Another critical property of cryptographic hash functions is collision resistance\cite{Stevens2021ASO}. A collision occurs when two different inputs produce the same hash output. A secure hash function is designed to make it computationally infeasible to find any two distinct inputs that result in the same hash value.

Collision resistance is crucial for security because if collisions were easy to find, attackers could exploit them to create fake data. For instance, in a blockchain, an attacker might try to modify a transaction but produce the same hash as the original. If such a collision were possible, the integrity of the entire blockchain could be compromised.

For example, imagine two different transactions—one transferring \$10 and another transferring \$1,000. If these two transactions could result in the same hash, an attacker might try to replace one transaction with the other. However, because of collision resistance, it's practically impossible for two different transactions to have the same hash.

In summary, one-wayness and collision resistance ensure the security and reliability of hash functions in cryptographic systems. Without these properties, it would be possible to reverse or manipulate the data stored in the blockchain, making the entire system vulnerable to attack.

\section{Simple Hash Algorithms}

While cryptographic hash functions are designed with complex mathematics to ensure security, there are simpler forms of hash functions that are used in various applications, especially when high security is not the main concern. These simpler algorithms are often used for tasks\cite{Wieder2017HashingLB} such as quick data indexing, load balancing, or data distribution across different storage locations.

\subsection{Modulo Operation and Hashing}

One of the most straightforward ways to generate a hash value is by using the modulo operation\cite{Girault1987HashFunctionsUM}. In this approach, the hash value is obtained by dividing the input value by a certain number (called the modulus) and taking the remainder.

For example, consider a simple hash function that calculates the hash value of an integer input by taking the input modulo 10 (i.e., input \% 10). This means that any input will be divided by 10, and the remainder will be the hash value. Here's a Python example to illustrate:

\begin{lstlisting}[style=python]
# Simple hash function using modulo operation
def simple_hash(value):
    return value % 10

# Example usage
data = [15, 27, 34, 89, 45]
hashed_values = [simple_hash(x) for x in data]
print(hashed_values)
\end{lstlisting}

In this example, the input numbers (15, 27, 34, 89, and 45) are divided by 10, and the remainders (5, 7, 4, 9, and 5) become the hash values. Notice that the number 15 and 45 both have the same hash value (5). This is an example of a collision, but for simple tasks like load balancing or small-scale data indexing, such collisions are often manageable.

\subsection{Real-World Application of the \%10 Hash Function}

Now that we've introduced the basic idea of a modulo-based hash function, let's explore a real-world scenario where this kind of simple hashing might be useful.

Imagine a small web application that needs to distribute requests evenly across 10 servers. One way to achieve load balancing is to assign each incoming request to a server based on the hash value of the user ID or session ID. For instance, the server could be selected by hashing the user ID with a \%10 function and assigning the request to the server corresponding to the hash value.

Here's a Python example simulating this scenario:

\begin{lstlisting}[style=python]
# Function to simulate load balancing across 10 servers
def assign_server(user_id):
    return user_id % 10

# Simulate assigning user requests to servers
user_ids = [123, 456, 789, 101, 212, 345, 678]
server_assignments = [assign_server(user_id) for user_id in user_ids]

# Output which server each user is assigned to
for user_id, server in zip(user_ids, server_assignments):
    print(f"User {user_id} is assigned to server {server}")
\end{lstlisting}

In this example, each user ID is hashed using the \%10 function, and the result determines which server the user will be assigned to. For instance, user 123 may be assigned to server 3 (since 123 \% 10 = 3), and user 456 may be assigned to server 6 (since 456 \% 10 = 6). This method provides a simple and effective way to distribute the load evenly across servers.

In summary, while modulo-based hashing is not suitable for cryptographic purposes due to its simplicity and vulnerability to collisions, it is a powerful tool for scenarios like load balancing or small-scale data indexing where security is not the primary concern. By using the \%10 hash function, we can efficiently map inputs to a manageable range of values, making it ideal for basic real-world applications.

\section{Cryptographic Hash Functions}

Cryptographic hash functions are mathematical algorithms that transform an arbitrary amount of input data into a fixed-size output, known as a hash or digest. These functions are essential in ensuring data integrity and are widely used in various applications like digital signatures, message authentication codes, and blockchain technologies\cite{Dang2009RecommendationFA}. In this section, we will discuss two popular cryptographic hash functions\cite{Pethe2016AnOO}: MD5 and the SHA family of hash functions.

\subsection{MD5 Algorithm}
MD5 (Message Digest Algorithm 5)\cite{Wang2005HowTB} is one of the earliest widely-used cryptographic hash functions. It was designed by Ronald Rivest\cite{Wang2004CollisionsFH} in 1991 to produce a 128-bit hash value from an arbitrary-length input.

\subsubsection{Working Principle of MD5}
The MD5 algorithm\cite{Wartik1992HashingA} processes data in 512-bit chunks, breaking it down into a series of operations over multiple rounds. The final output of MD5 is a 128-bit hash, often represented as a 32-character hexadecimal string. Here’s a high-level breakdown of how MD5 works:

\begin{itemize}
    \item \textbf{Step 1: Padding} - The original message is padded to ensure its length (in bits) is congruent to 448 modulo 512. The padding ensures that the message length becomes a multiple of 512 bits after appending the length of the message (in bits) to the end.
    
    \item \textbf{Step 2: Initialize MD5 buffer} - MD5 uses four buffers of 32 bits, initialized to specific constants. These buffers are referred to as \( A \), \( B \), \( C \), and \( D \).
    
    \item \textbf{Step 3: Process the message in 512-bit blocks} - The message is split into blocks of 512 bits. Each block is further divided into 16 words of 32 bits each. The algorithm then performs a series of operations across four rounds on each block.
    
    \item \textbf{Step 4: Update the buffers} - The buffers \( A \), \( B \), \( C \), and \( D \) are updated in each round using non-linear functions, modular addition, and bitwise operations. This process ensures a complex transformation of the data.
    
    \item \textbf{Step 5: Concatenate the buffers} - After processing all blocks, the final values of \( A \), \( B \), \( C \), and \( D \) are concatenated to form the final 128-bit hash.
\end{itemize}

Here is a Python implementation of MD5 using the built-in hashlib library to illustrate how it hashes an input string:

\begin{lstlisting}[style=python]
import hashlib

def md5_hash(input_string):
    # Create an MD5 hash object
    md5_hasher = hashlib.md5()
    
    # Update the hash object with the input string (encoded in bytes)
    md5_hasher.update(input_string.encode('utf-8'))
    
    # Return the hexadecimal digest
    return md5_hasher.hexdigest()

# Example usage
input_data = "Blockchain is transformative!"
print(f"MD5 Hash: {md5_hash(input_data)}")
\end{lstlisting}

\subsubsection{Limitations and Security Issues of MD5}
MD5 was widely used in the past, but its security has been severely compromised over time due to its vulnerability to collision attacks\cite{Stevens2012AttacksOH}. A collision occurs when two different inputs produce the same hash. In cryptographic applications, this can be catastrophic, leading to attacks such as creating fraudulent digital certificates.

\textbf{Key limitations of MD5:}
\begin{itemize}
    \item \textbf{Collision vulnerability:} In 2004, researchers demonstrated a practical method for finding MD5 collisions\cite{Mikle2004PracticalAO}, making it unsuitable for applications requiring high levels of security, such as digital signatures or SSL certificates.
    \item \textbf{Weakness in pre-image resistance:} Although finding the original input from a given MD5 hash is challenging, advances in cryptographic attacks have reduced the time needed for such attacks.
    \item \textbf{Not suitable for modern security needs:} Due to its flaws, MD5 is no longer recommended for use in secure systems, and it has been replaced by more robust hash functions, such as those from the SHA family.
\end{itemize}

\subsection{SHA Family of Hash Functions}
The SHA (Secure Hash Algorithm) family includes several cryptographic hash functions designed by the National Security Agency (NSA)\cite{Shah2014SECUREDHA,Khan2022EVOLUTIONAA}. The most common variants are SHA-1, SHA-2, and SHA-3. SHA-1 has been deprecated, while the SHA-2 family (SHA-224, SHA-256, SHA-512) is widely used today due to its improved security properties\cite{Eastlake2006USSH}.

\subsubsection{Principles and Security Issues of SHA-1}
SHA-1, developed in 1993, produces a 160-bit hash from input data. Its design is similar to MD5, but it processes data in 512-bit chunks, iterating over 80 rounds. SHA-1 was once widely used in digital signatures, SSL certificates, and other security protocols. However, it was officially deprecated by NIST in 2011\cite{MAlNawashi2024AnalysisAE} due to its vulnerability to collision attacks.

\textbf{Security Issues of SHA-1:}
\begin{itemize}
    \item \textbf{Collision attacks:} In 2017, researchers at Google and CWI Amsterdam successfully demonstrated a collision attack on SHA-1\cite{Stevens2017TheFC}, making it insecure for most cryptographic applications.
    \item \textbf{Limited security:} Although SHA-1 provides better security than MD5, advances in computing power and cryptographic techniques have made it vulnerable. It is now recommended to use stronger hash functions from the SHA-2 or SHA-3 families.
\end{itemize}

\subsubsection{SHA-2 Family: SHA-224, SHA-256, SHA-512}
The SHA-2 family improves upon SHA-1 by offering various hash sizes, each with stronger security guarantees. The most common variants include SHA-224, SHA-256, SHA-384, and SHA-512, where the numbers refer to the size of the hash output in bits. These algorithms are more secure and resistant to collision and pre-image attacks.

\textbf{Key differences between SHA-224, SHA-256, and SHA-512:}

\begin{itemize}
    \item \textbf{SHA-224:} This algorithm produces a 224-bit hash, offering a balance between security and efficiency for applications requiring shorter digests.
    \item \textbf{SHA-256:} The most commonly used variant, SHA-256, generates a 256-bit hash and is widely used in blockchain technology, including Bitcoin, to secure transactions and blocks.
    \item \textbf{SHA-512:} SHA-512 generates a 512-bit hash and is considered more secure for applications needing stronger protection against brute-force attacks. It is also more computationally intensive than SHA-256.
\end{itemize}

Here is an example Python code that demonstrates the use of SHA-256 hashing:

\begin{lstlisting}[style=python]
import hashlib

def sha256_hash(input_string):
    # Create a SHA-256 hash object
    sha256_hasher = hashlib.sha256()
    
    # Update the hash object with the input string (encoded in bytes)
    sha256_hasher.update(input_string.encode('utf-8'))
    
    # Return the hexadecimal digest
    return sha256_hasher.hexdigest()

# Example usage
input_data = "Blockchain is secure with SHA-256!"
print(f"SHA-256 Hash: {sha256_hash(input_data)}")
\end{lstlisting}

In conclusion, the SHA-2 family offers significant improvements in security over MD5 and SHA-1, making it the preferred choice for most modern cryptographic applications. The use of larger hash sizes like SHA-256 and SHA-512 helps to ensure a higher level of security and integrity, especially in applications such as blockchain, digital signatures, and data integrity verification.

\subsection{Detailed Analysis of SHA-256 Algorithm}

\subsubsection{Encryption Process of SHA-256}
SHA-256 (Secure Hash Algorithm 256) is a member of the SHA-2 family of cryptographic hash functions\cite{2019ASO}. It produces a fixed-size 256-bit (32-byte) hash value, often referred to as a message digest. SHA-256 is widely used in cryptography and blockchain technology, especially for ensuring the integrity of data.

The process of generating a hash using SHA-256 can be broken down into the following steps:

\paragraph{1. Padding the Message}
The input message is first padded to ensure that its length is congruent to 448 modulo 512. Padding is done by appending a single '1' bit to the message, followed by a number of '0' bits, and finally appending a 64-bit representation of the original message length.

\paragraph{2. Parsing the Message into Blocks}
After padding, the message is divided into blocks of 512 bits. Each block will be processed separately.

\paragraph{3. Message Schedule}
For each 512-bit block, SHA-256 expands the block into 64 32-bit words using the following rules:
\begin{itemize}
    \item The first 16 words are directly taken from the 512-bit block.
    \item The remaining 48 words are generated by applying bitwise operations (right shifts, rotations) and XOR operations to the previous words.
\end{itemize}

\paragraph{4. Compression Function}
SHA-256 uses eight fixed 32-bit words as the initial hash values. For each 512-bit block, these hash values are updated by processing the 64 words generated in the message schedule. The compression function consists of modular additions and bitwise logical operations (such as XOR, AND, OR, and NOT). The final result after processing all the blocks is the 256-bit hash value.

The following Python code demonstrates the hashing process using the \texttt{hashlib} library:
\begin{lstlisting}[style=python]
import hashlib

# Example message to hash
message = "Hello, Blockchain!"

# Convert message to bytes and apply SHA-256
hash_object = hashlib.sha256(message.encode())
hex_dig = hash_object.hexdigest()

print("SHA-256 Hash:", hex_dig)
\end{lstlisting}

\subsubsection{Application of SHA-256 in Merkle Trees}
A Merkle Tree is a binary tree structure used to efficiently verify the integrity of large datasets\cite{Jing2021ReviewAI}. In blockchain technology, Merkle Trees are used to ensure the integrity of transactions in blocks. SHA-256 plays a crucial role in constructing Merkle Trees by generating hash values for the data at each level of the tree.

Here’s how SHA-256 is applied in a Merkle Tree:
\begin{itemize}
    \item Each transaction in a block is hashed using SHA-256.
    \item The hashes of pairs of transactions are combined and hashed again to form the parent nodes.
    \item This process continues until only a single hash remains, called the Merkle Root.
\end{itemize}

The Merkle Root serves as a cryptographic fingerprint of all the transactions in the block. If any transaction changes, the Merkle Root will also change, thus ensuring data integrity.

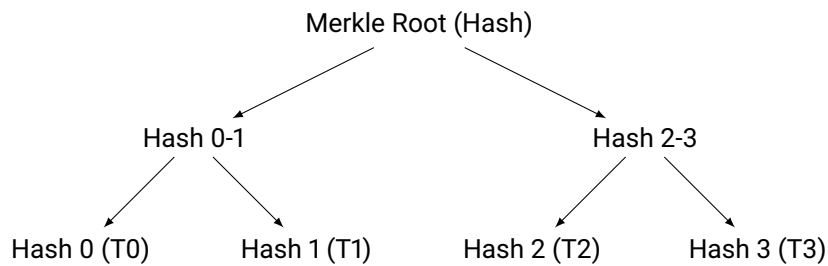
\begin{center}
\begin{tikzpicture}
  [level distance=1.5cm,
  level 1/.style={sibling distance=6cm},
  level 2/.style={sibling distance=3cm},
  edge from parent/.style={draw,-latex}]
  \node {Merkle Root (Hash)}
    child {node {Hash 0-1}
      child {node {Hash 0 (T0)}}
      child {node {Hash 1 (T1)}}
    }
    child {node {Hash 2-3}
      child {node {Hash 2 (T2)}}
      child {node {Hash 3 (T3)}}
    };
\end{tikzpicture}
\end{center}

In this diagram, each leaf node represents the SHA-256 hash of a transaction (T0, T1, etc.), and the internal nodes represent the hash of the concatenated hashes of its children. The Merkle Root at the top ensures that any modification to the transactions will result in a completely different Merkle Root, ensuring the integrity of the block.

\subsubsection{Use Cases of SHA-256}
SHA-256 has various real-world applications due to its security and efficiency. Some common use cases include:

\paragraph{1. Bitcoin Mining}
In Bitcoin, miners compete to find a value (called a nonce) such that the SHA-256 hash of the block header starts with a certain number of leading zeros\cite{Chow2017TheBM}. This process, known as Proof of Work (PoW), requires extensive computational effort but ensures the security of the network.

\paragraph{2. Digital Signatures}
SHA-256 is often used to create a digest of a message, which is then signed with a private key to create a digital signature\cite{Radack2009TheCH}. This process ensures both the integrity and authenticity of the message.

\paragraph{3. Secure Communications}
SHA-256 is used in various protocols such as TLS (Transport Layer Security) to ensure that data transmitted over the internet is secure and has not been tampered with\cite{Turner2014TransportLS}. It is also used in file integrity verification and software distribution to ensure that files have not been altered.

\subsection{RIPEMD160 and Bitcoin Address Generation}
RIPEMD160 (RACE Integrity Primitives Evaluation Message Digest)\cite{Preneel1TC} is a 160-bit cryptographic hash function that plays an important role in Bitcoin address generation. Bitcoin addresses are shorter than the 256-bit public keys used in Bitcoin transactions, and this is achieved by combining the SHA-256 and RIPEMD160 hash functions.

Here’s the process for generating a Bitcoin address:
\begin{itemize}
    \item First, the public key is generated from the private key using elliptic curve cryptography.
    \item The public key is then hashed using SHA-256.
    \item The result of the SHA-256 hash is then hashed again using RIPEMD160 to produce a shorter, 160-bit hash.
    \item This RIPEMD160 hash is encoded with a version byte and a checksum (calculated using another SHA-256 hash) to form the final Bitcoin address.
\end{itemize}

The following Python code demonstrates this process:

\begin{lstlisting}[style=python]
import hashlib
import base58

# Example public key (in hex format)
public_key = "04b0bd634234abbb1ba1e986e884185c5f7fd11ebf..."

# Step 1: Perform SHA-256 on the public key
sha256_hash = hashlib.sha256(bytes.fromhex(public_key)).digest()

# Step 2: Perform RIPEMD-160 on the result of SHA-256
ripemd160 = hashlib.new('ripemd160')
ripemd160.update(sha256_hash)
ripemd160_hash = ripemd160.digest()

# Step 3: Add version byte (0x00 for Main Network)
versioned_payload = b'\x00' + ripemd160_hash

# Step 4: Compute the checksum by performing SHA-256 twice
checksum = hashlib.sha256(hashlib.sha256(versioned_payload).digest()).digest()[:4]

# Step 5: Add the checksum to the versioned payload and encode it using Base58
bitcoin_address = base58.b58encode(versioned_payload + checksum)
print("Bitcoin Address:", bitcoin_address.decode())
\end{lstlisting}

This process ensures that Bitcoin addresses are both compact and secure. By combining the outputs of SHA-256 and RIPEMD160, Bitcoin takes advantage of the security features of both algorithms while keeping the addresses manageable in length\cite{Li2023NewRI}.

\section{The Role of Hash Functions in Blockchain}

\subsection{Ensuring Blockchain Data Integrity}
Hash functions are fundamental to the security and integrity of blockchain systems. A hash function is a mathematical algorithm that transforms an input (or "message") into a fixed-size string of bytes. The output is commonly referred to as the hash value, digest, or simply the hash. In the context of blockchain, a specific hash function like SHA-256 (Secure Hash Algorithm)\cite{Sharma2023DifferentCH} is often used.

When data is added to a blockchain, the hash function computes a unique hash for that specific data. Even a tiny change in the input data will result in a completely different hash. This property ensures the integrity of the data. If someone attempts to modify any part of the data, the hash will change, immediately alerting the system to potential tampering.

For example, consider a blockchain ledger where each block contains:
- Transaction data
- A timestamp
- The hash of the previous block
- A cryptographic hash of its own data

The structure of blockchain links each block to the previous one by including the previous block's hash in its own data. This creates a "chain" of blocks where altering one block would require changing all subsequent blocks, as the hash values would no longer match.

\begin{lstlisting}[style=python]
import hashlib

# Example: Hash function using SHA-256
def hash_function(data):
    return hashlib.sha256(data.encode('utf-8')).hexdigest()

# Data for Block 1
data_block_1 = "Alice sends 5 BTC to Bob"
hash_block_1 = hash_function(data_block_1)
print("Block 1 Hash:", hash_block_1)

# Tampering with the data (even a small change)
data_block_1_modified = "Alice sends 5.1 BTC to Bob"
hash_block_1_modified = hash_function(data_block_1_modified)
print("Tampered Block 1 Hash:", hash_block_1_modified)
\end{lstlisting}

In this code, the original data is hashed, and any slight modification to the input (e.g., changing "5 BTC" to "5.1 BTC") results in a completely different hash output, demonstrating how hash functions maintain data integrity.

\subsection{Application of Hash Functions in Proof-of-Work}
In a blockchain network, the Proof-of-Work (PoW) consensus mechanism is crucial for verifying transactions and adding new blocks to the chain\cite{Patil2023UnderstandingCM}. Hash functions, particularly SHA-256 in Bitcoin's case, play a central role in this process. PoW involves solving a complex mathematical puzzle, where miners compete to find a valid hash that is below a specific target value.

Miners take the block's data, including the hash of the previous block, transaction details, and a variable called a "nonce." They repeatedly change the nonce and re-hash the block's data until they find a hash value that meets the difficulty target (a hash starting with a certain number of leading zeros). This process is computationally expensive but ensures the immutability and security of the blockchain.

Example of Proof-of-Work:

\begin{lstlisting}[style=python]
import hashlib

# Simple Proof-of-Work example using SHA-256
def proof_of_work(block_data, difficulty):
    nonce = 0
    prefix = '0' * difficulty
    while True:
        text = block_data + str(nonce)
        new_hash = hashlib.sha256(text.encode('utf-8')).hexdigest()
        if new_hash.startswith(prefix):
            return nonce, new_hash
        nonce += 1

block_data = "Block data: Alice sends 3 BTC to Bob"
difficulty = 4  # Number of leading zeros required in the hash

nonce, valid_hash = proof_of_work(block_data, difficulty)
print(f"Nonce: {nonce}")
print(f"Valid hash: {valid_hash}")
\end{lstlisting}

In this example, the `proof\_of\_work` function continues to hash the block data by changing the nonce until it finds a hash that starts with a certain number of zeros (indicating the difficulty). The higher the difficulty, the more computational power is required, which forms the core of the Proof-of-Work mechanism.

\subsection{Combining Digital Signatures with Hash Functions}
In blockchain systems, digital signatures are often used alongside hash functions to guarantee the authenticity and integrity of transactions\cite{Zhang2019SecurityAP}. When a user initiates a transaction, the data of the transaction is first hashed. This hash is then signed using the user's private key, creating a unique digital signature. 

The recipient can verify the transaction by checking the digital signature against the user's public key and the hash of the original transaction data. If the transaction has been tampered with, the hash will not match, and the digital signature will be invalid.

Let's walk through the steps of how a transaction might work:

1. Alice wants to send 3 BTC to Bob. She creates a transaction containing this information.
2. The transaction data is hashed using SHA-256.
3. Alice signs this hash with her private key, generating a digital signature.
4. Bob (or any verifier) can verify the signature by hashing the original data again and using Alice's public key to confirm that the signature is valid.

Here is a simple example using Python to show how hashing and digital signatures might be combined:

\begin{lstlisting}[style=python]
from cryptography.hazmat.primitives.asymmetric import rsa, padding
from cryptography.hazmat.primitives import hashes
from cryptography.hazmat.primitives.asymmetric import utils

# Generate Alice's RSA key pair (private and public keys)
private_key = rsa.generate_private_key(
    public_exponent=65537,
    key_size=2048
)
public_key = private_key.public_key()

# Data (transaction) Alice wants to sign
transaction_data = "Alice sends 3 BTC to Bob"

# Step 1: Hash the transaction data
transaction_hash = hashlib.sha256(transaction_data.encode('utf-8')).digest()

# Step 2: Alice signs the hash using her private key
signature = private_key.sign(
    transaction_hash,
    padding.PSS(
        mgf=padding.MGF1(hashes.SHA256()),
        salt_length=padding.PSS.MAX_LENGTH
    ),
    utils.Prehashed(hashes.SHA256())
)

# Step 3: Bob verifies the signature using Alice's public key
try:
    public_key.verify(
        signature,
        transaction_hash,
        padding.PSS(
            mgf=padding.MGF1(hashes.SHA256()),
            salt_length=padding.PSS.MAX_LENGTH
        ),
        utils.Prehashed(hashes.SHA256())
    )
    print("The signature is valid.")
except:
    print("The signature is invalid.")
\end{lstlisting}

In this example:
- The transaction data is hashed using SHA-256.
- The hash is signed by Alice's private key.
- The recipient (or verifier) uses Alice's public key to check the signature and ensure that the transaction is authentic and hasn't been altered.

By combining digital signatures with hash functions, blockchain ensures that not only is the data tamper-evident (through hashing), but also that the authenticity of the data (via the digital signature) is verifiable.

\part{Bitcoin and Blockchain Technology}
\chapter{The Birth of Bitcoin}

\section{Historical Background of Digital Currency}

\subsection{Early Electronic Money and Its Limitations}

In the early days of digital communication, there were various attempts to create electronic forms of money. These early systems aimed to digitize money and make online transactions more efficient. Two prominent examples of early digital currency include eCash and DigiCash, which were proposed by cryptographer David Chaum\cite{Moore2013ThePA} in the 1980s and 1990s.

eCash was an anonymous electronic cash system that allowed users to make online payments without revealing their identity, making it a precursor to modern privacy-focused digital currencies\cite{Gouget2008RecentAI}. DigiCash, the company that developed eCash, aimed to ensure privacy in digital transactions by using cryptographic protocols that would blind the issuer (i.e., the bank) to the details of the transaction\cite{chaum2015beginnings}. However, eCash and DigiCash faced several challenges that ultimately prevented their widespread adoption.

The key limitations of early digital currencies like eCash included:

\begin{itemize}
    \item \textbf{Centralization:} Both eCash and DigiCash relied on a central authority, such as a bank or a financial institution, to verify and approve transactions. This centralization introduced a single point of failure, making the system vulnerable to issues like fraud, hacking, and mismanagement.
    \item \textbf{Double-Spending Problem:} One of the biggest challenges with digital currency is the risk of double spending, where the same digital token is spent more than once. In traditional banking systems, this issue is handled by central authorities who maintain a ledger of all transactions. However, in the absence of such authorities, it became difficult to prevent users from replicating digital tokens and spending them multiple times\cite{El2020BitcoinCI}.
    \item \textbf{Security:} Since these systems were centralized, they became attractive targets for hackers. If the central server was compromised, it could lead to significant financial loss and a breakdown of trust in the system.
    \item \textbf{Lack of Adoption:} Despite the innovative concepts behind eCash and DigiCash, they failed to gain widespread traction. Factors such as the complexity of the technology, lack of understanding by the public, and resistance from banks and governments all contributed to their limited success.
\end{itemize}

Ultimately, the inability of early electronic money systems to solve these issues, particularly the double-spending problem, paved the way for the development of more advanced systems like Bitcoin.

\subsection{The Emergence of Bitcoin: Solving the Double-Spending Problem}

Bitcoin was introduced in 2008 by an individual or group of individuals using the pseudonym Satoshi Nakamoto\cite{Meiklejohn2018TheLO}. What made Bitcoin revolutionary was its ability to solve the double-spending problem without relying on a central authority. Prior to Bitcoin, the double-spending problem was a fundamental obstacle that digital currencies faced, as digital tokens can be easily copied, unlike physical currencies.

Bitcoin addressed this issue using a decentralized peer-to-peer network and an innovative technology called the \textbf{blockchain}\cite{Lee2016NewKO}. The blockchain is a distributed ledger that records all transactions in a transparent and immutable manner. This means that once a transaction is added to the blockchain, it cannot be altered or deleted, ensuring that the same Bitcoin cannot be spent more than once.

At its core, Bitcoin's solution to the double-spending problem can be summarized as follows:

\begin{itemize}
    \item \textbf{Blockchain as a Public Ledger:} Every Bitcoin transaction is recorded on a public ledger, the blockchain. This ledger is maintained by a network of participants known as \textbf{miners}, who validate and confirm transactions.
    \item \textbf{Proof of Work:} In order to add a new block of transactions to the blockchain, miners must solve a complex mathematical puzzle, a process known as \textbf{Proof of Work}. This ensures that altering any transaction would require an immense amount of computational power, making it nearly impossible for a single entity to tamper with the blockchain.
    \item \textbf{Decentralization:} Unlike previous digital currencies, Bitcoin is completely decentralized. There is no single entity or organization controlling the network. Instead, the responsibility of maintaining the ledger and verifying transactions is distributed across thousands of nodes worldwide.
    \item \textbf{Immutability:} Once a transaction is confirmed and included in a block, it becomes part of the permanent ledger. This immutability guarantees that transactions are final and cannot be reversed, thus preventing double spending.
\end{itemize}

Bitcoin's unique design, combining decentralization, blockchain technology, and cryptographic security, offered a robust and scalable solution to the double-spending problem, which had plagued earlier digital currency systems\cite{Hunt2008BitcoinAP}.

\section{Interpretation of the Bitcoin Whitepaper}

\subsection{The Core Idea of Bitcoin: Decentralization}

The core idea behind Bitcoin, as outlined in Satoshi Nakamoto's whitepaper titled "\textit{Bitcoin: A Peer-to-Peer Electronic Cash System}," \cite{Hunt2008BitcoinAP} is the concept of decentralization. Unlike traditional financial systems that rely on intermediaries such as banks, Bitcoin removes the need for a trusted third party to validate transactions.

In the traditional banking model, when you make a transaction, the bank verifies and confirms the movement of funds from one account to another. This process ensures trust between parties, but it also requires trust in the bank itself. If the bank experiences issues—such as technical failures, mismanagement, or corruption—the entire system can break down.

Bitcoin operates on a peer-to-peer network, meaning that transactions occur directly between users, without the need for intermediaries. This peer-to-peer model is enabled by the blockchain, which allows every participant in the network to have a copy of the transaction history. In this way, Bitcoin achieves decentralization, where no single party holds control over the system.

To illustrate how Bitcoin's decentralization works, let's consider an example:

\begin{itemize}
    \item \textbf{Traditional Banking System:} Alice wants to send Bob \$100. She initiates the transaction through her bank, which verifies the transaction and deducts \$100 from Alice's account, then credits Bob's account with \$100. The bank is the central authority in this transaction.
    \item \textbf{Bitcoin System:} Alice wants to send Bob 1 Bitcoin (BTC). She broadcasts this transaction to the Bitcoin network, where miners work to validate the transaction. Once validated, the transaction is recorded on the blockchain, and Bob receives the 1 BTC. There is no central authority in this transaction—only the peer-to-peer network and the consensus of miners.
\end{itemize}

This decentralized model offers several benefits:

\begin{itemize}
    \item \textbf{Elimination of Middlemen:} Bitcoin removes the need for banks, payment processors, and other intermediaries, reducing transaction fees and increasing efficiency.
    \item \textbf{Increased Transparency:} All transactions are recorded on the public blockchain, making it easy to trace the flow of funds without exposing personal identities.
    \item \textbf{Resilience and Security:} Since the network is decentralized, it is much harder to attack or disrupt. Even if some nodes go offline, the network continues to function.
\end{itemize}

\subsection{Satoshi Nakamoto's Design Goals}

Satoshi Nakamoto's design for Bitcoin was shaped by several key goals, which are reflected in the architecture of the Bitcoin network\cite{Hunt2008BitcoinAP}:

\begin{itemize}
    \item \textbf{Security:} One of Nakamoto's main objectives was to ensure that the Bitcoin network is secure from malicious actors. This is achieved through the Proof of Work mechanism, cryptographic signatures, and the decentralized nature of the network. These features make it exceedingly difficult to alter or tamper with the blockchain.
    \item \textbf{Transparency:} Bitcoin transactions are transparent and can be verified by anyone. Although the identities of the participants are pseudonymous, the transaction details—such as the amount and timestamp—are recorded on the blockchain for all to see. This transparency helps build trust in the system.
    \item \textbf{Reducing Reliance on Centralized Institutions:} Bitcoin was designed to function without the need for banks or governments. By decentralizing the process of transaction validation and currency issuance, Bitcoin empowers individuals to control their own money.
    \item \textbf{Global Accessibility:} Bitcoin was designed to be a global currency that could be accessed by anyone with an internet connection. Traditional banking systems can be restrictive, especially for people in underbanked regions. Bitcoin aims to provide a financial system that is inclusive and open to all.
\end{itemize}

The architecture of Bitcoin reflects these design goals. By relying on a decentralized, secure, and transparent ledger, Bitcoin provides an alternative to the traditional financial system, offering a new form of money that is independent of central authorities.

\section{Innovative Features of Bitcoin}

\subsection{Proof-of-Work Mechanism}
The Proof-of-Work (PoW) mechanism is fundamental to ensuring the security and integrity of the Bitcoin network\cite{Garay2017ProofsOW}. PoW works by requiring participants, known as miners, to solve complex cryptographic puzzles in order to validate transactions and add new blocks to the blockchain. This process is computationally intensive, meaning that it requires significant computational power and energy.

The PoW system makes it incredibly difficult for malicious actors to take control of the network because they would need to control more than 50

For example, when a miner attempts to validate a block, they must find a hash (a long string of numbers and letters) that starts with a specific number of leading zeros. The difficulty of this task is adjusted periodically by the network, ensuring that blocks are added approximately every 10 minutes, no matter how much total computing power is present on the network. This proof is then easily verified by other nodes, ensuring that the network only accepts valid blocks, keeping the blockchain secure and resistant to tampering.

\subsection{Chain-Based Block Structure}
Bitcoin’s blockchain structure is an innovative way to ensure data integrity and immutability. Each block in the Bitcoin blockchain contains a list of transactions, and every block is cryptographically linked to the previous one by including the hash of the prior block within its header\cite{Senthilkumar2021DataCI}. This creates a continuous chain of blocks (hence the term blockchain), forming an immutable ledger of all transactions since the network's inception.

This structure ensures that altering any information in a block would invalidate that block and all subsequent blocks because the hash of each block depends on the hash of the previous one. For example, if someone tried to change the amount of a particular transaction from 1 BTC to 100 BTC, the hash of that block would change, breaking the chain and alerting the network to the tampering attempt.

As blocks are linked in a chronological order, it becomes computationally infeasible to go back and change the contents of previous blocks once they have been added to the blockchain. This feature provides strong security guarantees, preventing tampering and maintaining the integrity of the data.

\subsection{Peer-to-Peer Network and Distributed Consensus}
Bitcoin operates on a decentralized, peer-to-peer (P2P) network\cite{Donet2014TheBP}. This means that no single entity, such as a central bank or government, controls the system. Instead, every participant (node) in the network has equal access and authority. Nodes work together to validate and propagate transactions across the network, creating a trustless environment.

Decentralization is achieved through the process of distributed consensus, where nodes independently validate new blocks and transactions by adhering to the rules set by the Bitcoin protocol\cite{Jayabalan2021ASO}. Each node holds a copy of the entire blockchain, which they use to verify incoming transactions.

For example, when a user broadcasts a transaction to the network, it gets propagated to all nodes, which verify the transaction against their local copy of the blockchain. If the transaction is valid (i.e., the user has enough Bitcoin and the transaction structure is correct), the network reaches consensus, and the transaction is eventually confirmed by being included in a block. This decentralized consensus mechanism ensures that the system operates without requiring trust in any single intermediary.

\section{How to Create a Bitcoin Wallet?}

\subsection{Creating a Bitcoin Wallet Using Tools}
A Bitcoin wallet is a software application or hardware device that stores the private keys you need to access and spend your Bitcoin\cite{Vishawjyoti2021EWallet}. There are several types of wallets, each offering different levels of security, usability, and features. Below, we will guide you through selecting a wallet and the steps required to create one.

\subsubsection{Choosing Bitcoin Wallet Software}
The first step in creating a Bitcoin wallet is to select the appropriate type of wallet for your needs. Here are the most common types:

\begin{itemize}
    \item \textbf{Desktop Wallets}: These are software applications that you download and install on your computer. Desktop wallets offer a good balance between security and convenience but are vulnerable to malware if your computer is compromised. Examples include Electrum\cite{electrumElectrumBitcoin} and Bitcoin Core\cite{bitcoinBitcoinCore}.
    \item \textbf{Mobile Wallets}: These wallets are designed for smartphones and provide a convenient way to manage your Bitcoin on the go. However, they are less secure than hardware or desktop wallets. Popular mobile wallets include Mycelium\cite{myceliumMyceliumBitcoin} and Breadwallet\cite{breadwallet}.
    \item \textbf{Hardware Wallets}: These are physical devices designed to store your private keys offline, providing maximum security against hacking. Examples include Ledger Nano S\cite{ledgerHardwareWallet} and Trezor\cite{trezorTrezorHardware}.
    \item \textbf{Paper Wallets}: A paper wallet is a piece of paper that contains your Bitcoin private keys and public addresses. This is an offline, highly secure option if stored properly, but if lost or destroyed, access to the Bitcoin is permanently lost.
\end{itemize}

\textbf{Example Scenario}: If you are a new user and value both convenience and security, you might opt for a mobile wallet like Mycelium, which offers ease of use and secure storage. For someone with a larger amount of Bitcoin who is more security-conscious, a hardware wallet like Ledger Nano S would be a better option.

\subsubsection{Installing and Setting Up Wallet Software}
Once you have chosen your preferred wallet type, the next step is to install and configure the wallet software. Below is a step-by-step guide for setting up a desktop wallet:

\begin{enumerate}
    \item \textbf{Download the Software}: Visit the official website of the wallet you’ve chosen, for example, the Bitcoin Core wallet (\url{https://bitcoincore.org}). Download the latest version compatible with your operating system.
    \item \textbf{Install the Wallet}: After downloading, open the installer file and follow the on-screen instructions to install the wallet on your computer. Once installed, open the wallet application.
    \item \textbf{Set Up a New Wallet}: When you first open the wallet software, you will be prompted to create a new wallet. The software will generate a set of private keys, which are necessary to manage your Bitcoin.
    \item \textbf{Back Up Your Wallet}: After setting up the wallet, it's critical to create a backup of your wallet’s private keys. The software will prompt you to write down a seed phrase (typically 12 or 24 random words) that can be used to restore your wallet if your computer is lost or damaged.
    \item \textbf{Set Up Security Features}: Enable additional security features such as password protection and two-factor authentication (2FA). These steps add extra layers of security to prevent unauthorized access to your wallet.
    \item \textbf{Receive Bitcoin}: To receive Bitcoin, go to the “Receive” tab in the wallet software. Here, you will find your public Bitcoin address, which you can share with others to receive payments. The format will look like this: \texttt{1A1zP1eP5QGefi2DMPTfTL5SLmv7DivfNa}.
    \item \textbf{Send Bitcoin}: To send Bitcoin, go to the “Send” tab, enter the recipient’s Bitcoin address, the amount of Bitcoin you wish to send, and confirm the transaction.
\end{enumerate}

With this step-by-step process, you can easily create, configure, and start using your Bitcoin wallet securely.

\subsubsection{Backing Up Mnemonics and Private Keys}
One of the most important steps in securing a Bitcoin wallet is backing up both the mnemonic phrase and the private keys. Mnemonics and private keys are essential to accessing your wallet, as they give full control over the funds stored within the wallet. Losing these keys means losing access to your funds permanently, as there is no central authority in Bitcoin that can help recover your wallet.

\paragraph{Mnemonic Backup:} When creating a Bitcoin wallet, the wallet software generates a mnemonic phrase (sometimes called a seed phrase), which is a set of 12, 18, or 24 randomly generated words\cite{Sans2022ADM}. This mnemonic phrase is a human-readable representation of the private key, and it can be used to recover the wallet if the device where the wallet was created is lost or destroyed.

\paragraph{Private Key Backup:} The private key is a cryptographic key that grants access to your Bitcoin funds. It is crucial to keep this key secure and backed up in multiple safe locations, preferably offline. Exposing your private key to the internet can result in unauthorized access to your wallet and the loss of your funds.

\paragraph{Best Practices for Backup Storage:} 
\begin{itemize}
    \item Write down the mnemonic phrase and private keys on paper and store them in a safe location. Avoid digital backups such as cloud storage or photos on your phone.
    \item Consider using multiple backup locations, such as a safe or bank vault, to mitigate the risk of theft or damage to a single location.
    \item Never share your private keys or mnemonic phrase with anyone. The person holding the private key has full control of the funds in the wallet.
\end{itemize}

\subsubsection{Generating and Verifying a Wallet Address}
Creating a Bitcoin wallet involves generating a wallet address that others can use to send you Bitcoin. This address is derived from your public key, which in turn is derived from your private key.

\paragraph{Steps to Generate a Wallet Address:}
When you generate a new wallet, the wallet software creates a pair of cryptographic keys: a private key and a public key. These keys are mathematically related. The public key is used to create your Bitcoin wallet address through a series of hashing algorithms.

\begin{enumerate}
    \item A private key is randomly generated.
    \item The public key is derived from the private key using elliptic curve cryptography (ECC).
    \item The public key is then hashed using the SHA-256 algorithm followed by the RIPEMD-160 algorithm to create the wallet address.
\end{enumerate}

\paragraph{Verifying the Address:} 
To verify that your wallet address is correctly set up, you can perform a test transaction by sending a small amount of Bitcoin to the address and confirming that it is received. Another method is to check the address using a blockchain explorer, which can show whether the address is valid and recognized on the Bitcoin network.

\subsection{How Bitcoin Wallets Work}

\subsubsection{Elliptic Curve Cryptography Algorithm}
Bitcoin wallets rely on Elliptic Curve Cryptography (ECC) for generating public and private key pairs\cite{Bolfing2020Bitcoin}. ECC is a powerful cryptographic method that provides a high level of security with relatively small key sizes. In the case of Bitcoin, ECC operates over a specific curve called secp256k1. The curve allows for secure key generation and ensures that it is computationally infeasible to derive the private key from the public key.

\paragraph{Overview of ECC:} The elliptic curve is a mathematical structure that enables the generation of keys\cite{Yan2022TheOO}. Each point on the curve corresponds to a valid key pair. The private key is simply a randomly chosen large number, while the public key is a point on the elliptic curve derived from the private key using a multiplication operation. This one-way function ensures that while it is easy to derive the public key from the private key, it is nearly impossible to reverse the process.

\subsubsection{Public and Private Key Generation}
The process of generating a private key and the corresponding public key in Bitcoin wallets is entirely based on mathematics. Here’s how it works:

\paragraph{Private Key Generation:} 
A private key is generated by selecting a random 256-bit number. This number is kept secret and is the basis for all security in the wallet.

\begin{lstlisting}[style=python]
import os
import hashlib

# Generating a private key (256 bits)
private_key = os.urandom(32)
print("Private Key:", private_key.hex())
\end{lstlisting}

\paragraph{Public Key Derivation:} 
The public key is derived from the private key using an elliptic curve multiplication operation. In Bitcoin, this multiplication is done over the secp256k1 curve\cite{Balasubramanian2024SecurityOT}.

\begin{lstlisting}[style=python]
from ecdsa import SigningKey, SECP256k1

# Generating public key from private key
private_key = SigningKey.generate(curve=SECP256k1)
public_key = private_key.get_verifying_key()
print("Public Key:", public_key.to_string().hex())
\end{lstlisting}

\subsubsection{Generating a Wallet Address}
The Bitcoin wallet address is generated from the public key through a series of cryptographic hash functions. First, the public key is hashed with SHA-256, and then the result is hashed with RIPEMD-160 to produce the final Bitcoin address.

\paragraph{Steps to Generate a Wallet Address:}

\begin{enumerate}
    \item Apply SHA-256 to the public key.
    \item Apply RIPEMD-160 to the result of SHA-256 to create the public key hash.
    \item Add a network byte (0x00 for Bitcoin mainnet) to the beginning of the public key hash.
    \item Compute the checksum by hashing the result twice with SHA-256 and taking the first 4 bytes.
    \item Append the checksum to the public key hash and network byte.
    \item Encode the final result in Base58 to obtain the wallet address.
\end{enumerate}

\begin{lstlisting}[style=python]
import hashlib
import base58

# Example public key (compressed format)
public_key = b"\x02\xca\x3a\x52\x5f\xdb\xe4\xbd\x3e\xb1..."

# Step 1: SHA-256 hashing of the public key
sha256_result = hashlib.sha256(public_key).digest()

# Step 2: RIPEMD-160 hashing of the SHA-256 result
ripemd160 = hashlib.new('ripemd160')
ripemd160.update(sha256_result)
public_key_hash = ripemd160.digest()

# Step 3: Add network byte (0x00 for Bitcoin mainnet)
network_byte = b'\x00' + public_key_hash

# Step 4: Double SHA-256 hash for checksum
checksum = hashlib.sha256(hashlib.sha256(network_byte).digest()).digest()[:4]

# Step 5: Append checksum to public key hash + network byte
final_address_bytes = network_byte + checksum

# Step 6: Encode in Base58
wallet_address = base58.b58encode(final_address_bytes)
print("Bitcoin Wallet Address:", wallet_address.decode())
\end{lstlisting}

\subsubsection{Role of Private Keys and Signatures}
In the Bitcoin network, private keys are used to sign transactions. A signature verifies that the transaction was created by the owner of the Bitcoin without revealing the private key. When you send Bitcoin, your wallet software uses the private key to generate a digital signature for the transaction, ensuring it is both valid and authenticated.

\paragraph{Process of Signing a Transaction:}
\begin{enumerate}
    \item The transaction data is hashed using SHA-256.
    \item The private key is used to sign the hashed transaction, producing a digital signature.
    \item The signature is sent along with the transaction to the network for verification.
\end{enumerate}

\paragraph{Verification:} Other nodes on the Bitcoin network use the public key to verify that the signature corresponds to the private key without ever needing to know the private key itself.

\subsubsection{Principles and Uses of Mnemonics}
Mnemonic phrases are human-readable representations of private keys that make it easier for users to back up and recover their wallets. These phrases are usually a series of 12, 18, or 24 words chosen from a standardized word list.

\paragraph{Role in Wallet Recovery:}
If a user loses access to their wallet (e.g., due to device failure), the mnemonic phrase can be used to restore the wallet on a new device. The phrase is input into the wallet software, which then regenerates the private key and corresponding public key, giving the user access to their Bitcoin again.

\paragraph{Importance of Security:} It is vital to store mnemonic phrases in a secure, offline location, as anyone with access to the phrase can recover the wallet and steal the funds.

\chapter{Basic Structure of Blockchain}

\section{Definition of Blockchain}

\subsection{What is Blockchain?}
Blockchain is a decentralized, distributed ledger technology that allows data (typically transactions) to be recorded and maintained across many computers, known as \textit{nodes}\cite{Gutlapalli2016CommercialAO}. Unlike traditional centralized systems where a central authority (such as a bank) manages records, blockchain operates without any single entity in control. Instead, it is managed by a network of participants, each of whom holds a copy of the entire ledger.

The key characteristics of blockchain include:

\begin{itemize}
    \item \textbf{Decentralization}: No central authority controls the blockchain; all participants have equal power.
    \item \textbf{Distributed Ledger}: The ledger is replicated across all nodes, ensuring that each participant has a complete and identical copy.
    \item \textbf{Consensus Mechanism}: Nodes must agree (achieve consensus) on the validity of transactions before they can be added to the blockchain.
    \item \textbf{Immutability}: Once data is added to the blockchain, it cannot be altered or deleted, ensuring the integrity of the records.
\end{itemize}

For example, if a person wants to send cryptocurrency to another, a transaction will be created, broadcast to the network, and validated through the consensus mechanism\cite{Peng2023ARO}. Once validated, the transaction is grouped with others into a \textit{block}, and this block is added to the existing chain of blocks.

\subsection{How Blockchain Works}
To understand how blockchain works, consider the following process:

\begin{enumerate}
    \item \textbf{Transaction Creation}: A user initiates a transaction, such as transferring cryptocurrency. This transaction contains details such as the sender's address, the recipient's address, and the amount being transferred.
    \item \textbf{Broadcasting the Transaction}: The transaction is broadcast to the network, where it is visible to all nodes. Nodes are computers that participate in maintaining the blockchain.
    \item \textbf{Transaction Validation}: Nodes validate the transaction using cryptographic methods and consensus mechanisms. One common consensus mechanism is \textit{Proof of Work}, where nodes (miners) solve complex mathematical puzzles to prove that the transaction is valid.
    \item \textbf{Grouping Transactions into Blocks}: Once validated, transactions are grouped together into a block. The block contains a list of transactions as well as a reference to the previous block (in the form of a \textit{previous block hash}).
    \item \textbf{Adding the Block to the Chain}: The block is added to the blockchain, forming a chain of blocks that cannot be altered. Each new block is cryptographically linked to the one before it, ensuring the integrity of the entire chain.
\end{enumerate}

The following is a Python code snippet that simulates the creation of a simple block in a blockchain:

\begin{lstlisting}[style=python]
import hashlib

class Block:
    def __init__(self, previous_hash, transaction_data):
        self.previous_hash = previous_hash
        self.transaction_data = transaction_data
        self.block_hash = self.calculate_hash()

    def calculate_hash(self):
        data_to_hash = self.previous_hash + self.transaction_data
        return hashlib.sha256(data_to_hash.encode()).hexdigest()

# Create a block and print its details
previous_block_hash = "abc123def456"
transaction_data = "Alice pays Bob 10 BTC"
new_block = Block(previous_block_hash, transaction_data)

print("Previous Hash:", new_block.previous_hash)
print("Transaction Data:", new_block.transaction_data)
print("Block Hash:", new_block.block_hash)
\end{lstlisting}

\section{Components of a Block}

\subsection{Block Header and Block Body}
Each block in a blockchain consists of two main parts:

\begin{itemize}
    \item \textbf{Block Header}: The block header contains metadata about the block\cite{Gupta2021BlockchainTP}. It includes:
    \begin{itemize}
        \item The \textbf{previous block hash}, which links the current block to the previous one.
        \item A \textbf{timestamp}, indicating when the block was created.
        \item A \textbf{nonce}, a random number used in Proof of Work to vary the hash output.
        \item The \textbf{Merkle root}, which is a hash that represents the root of the Merkle tree and summarizes all transactions in the block.
    \end{itemize}
    
    \item \textbf{Block Body}: The block body contains the actual data, typically a list of validated transactions\cite{Gupta2021BlockchainTP}. Each transaction records details like the sender, receiver, and the amount transferred.
\end{itemize}

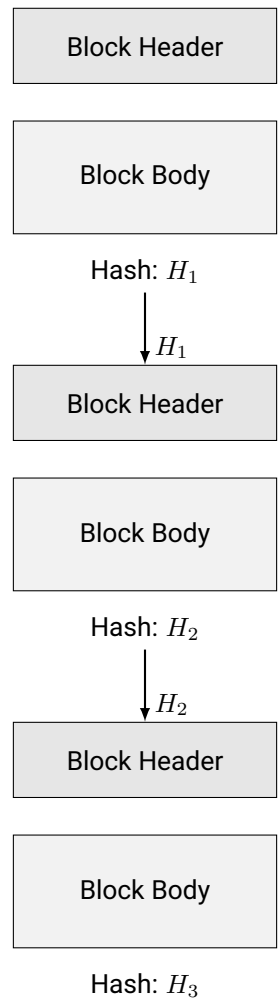
\begin{center}
\begin{tikzpicture}[
    block/.style = {draw, rectangle, minimum height=1.5cm, minimum width=3.5cm, align=center},
    header/.style = {draw, rectangle, fill=gray!20, minimum height=1cm, minimum width=3.5cm},
    body/.style = {draw, rectangle, fill=gray!10, minimum height=1.5cm, minimum width=3.5cm},
    arrow/.style = {draw, -latex, thick},
    every label/.append style = {font=\small}
]

\node[header] (block1header) at (0,0) {Block Header};
\node[body] (block1body) at (0,-1.75) {Block Body};
\node at (0,-3) (block1hash) {Hash: $H_1$};

\node[header] (block2header) at (0,-4.75) {Block Header};
\node[body] (block2body) at (0,-6.5) {Block Body};
\node at (0,-7.75) (block2hash) {Hash: $H_2$};

\node[header] (block3header) at (0,-9.5) {Block Header};
\node[body] (block3body) at (0,-11.25) {Block Body};
\node at (0,-12.5) (block3hash) {Hash: $H_3$};

\draw[arrow] (block1hash.south) -- ++(0,-0.5) -- (block2header.north) node[midway,right] {$H_1$};
\draw[arrow] (block2hash.south) -- ++(0,-0.5) -- (block3header.north) node[midway,right] {$H_2$};

\end{tikzpicture}
\end{center}

\bigskip
\noindent
\textbf{Blockchain Diagram Description:}

In the above diagram, each block in the blockchain is composed of two main parts: the \textbf{Block Header} and the \textbf{Block Body}. The Block Header contains metadata, such as the hash of the previous block, the timestamp, and other relevant information. The Block Body contains the actual transaction data or any other payload associated with that block.

In this diagram:
\begin{itemize}
    \item The hash of each block is displayed below it (e.g., $H_1$, $H_2$, and $H_3$).
    \item The hash of a block is calculated based on the content of the block, including the hash of the previous block.
    \item The arrows indicate the relationship between blocks: each block's header contains the hash of the previous block, forming a chain of blocks.
\end{itemize}

\bigskip
\noindent
\textbf{Why Blockchain is Immutable:}

The immutability of the blockchain arises from the fact that each block is cryptographically linked to the previous block through its hash\cite{Kizza2020BlockchainsCA}. If any data within a block is tampered with, the hash of that block will change, breaking the link with the subsequent block. This cascading effect requires recalculating the hashes of all subsequent blocks, which is computationally infeasible in most blockchain networks, particularly those using proof-of-work or similar consensus mechanisms. Therefore, the structure of the blockchain ensures that once data is added to the chain, it cannot be altered without redoing the entire chain, making it effectively tamper-resistant.

\subsection{Block Hash and Previous Block Hash}
Every block has a unique identifier, known as the \textbf{block hash}\cite{C2018AnOO}. This hash is created by applying a cryptographic function (such as SHA-256) to the block's contents, including its header and body. 

\textbf{The previous block hash} is a critical component in the block header because it links the current block to the one before it, forming a chain of blocks. This linkage ensures that if any block is altered, the hash of the altered block would no longer match, and all subsequent blocks would become invalid. This property guarantees the immutability and security of the blockchain.

Here’s an example of how blocks are cryptographically linked:

\begin{lstlisting}[style=python]
class Blockchain:
    def __init__(self):
        self.chain = []

    def add_block(self, transaction_data):
        previous_hash = self.chain[-1].block_hash if self.chain else '0'
        new_block = Block(previous_hash, transaction_data)
        self.chain.append(new_block)

    def display_chain(self):
        for block in self.chain:
            print(f"Block Hash: {block.block_hash}, Previous Hash: {block.previous_hash}")

# Initialize blockchain and add blocks
blockchain = Blockchain()
blockchain.add_block("Alice pays Bob 10 BTC")
blockchain.add_block("Bob pays Charlie 5 BTC")

blockchain.display_chain()
\end{lstlisting}

\section{Merkle Tree}

\subsection{Structure of the Merkle Tree}
A \textbf{Merkle tree} (also known as a binary hash tree) is a data structure used in blockchain to efficiently and securely verify the integrity of transactions\cite{Liu2021MerkleTA}. Each \textbf{leaf node} of the Merkle tree contains a hash of a single transaction, while each \textbf{non-leaf node} contains a hash of its two child nodes.

For example, consider four transactions: T1, T2, T3, and T4. The Merkle tree is constructed as follows:

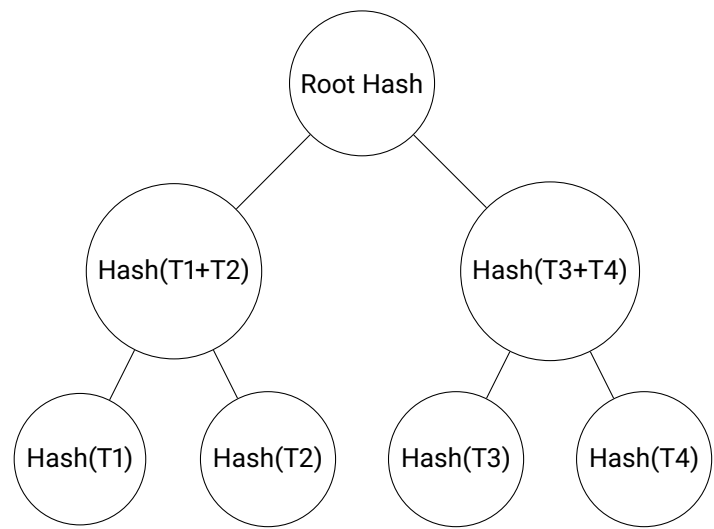
\begin{center}
\begin{tikzpicture}
  [level distance=25mm, 
   level/.style={sibling distance=50mm/#1}] 
  \node [circle,draw] {Root Hash}
    child {node [circle,draw] {Hash(T1+T2)}
      child {node [circle,draw] {Hash(T1)}}
      child {node [circle,draw] {Hash(T2)}}
    }
    child {node [circle,draw] {Hash(T3+T4)}
      child {node [circle,draw] {Hash(T3)}}
      child {node [circle,draw] {Hash(T4)}}
    };
\end{tikzpicture}
\end{center}

Here’s how the Merkle tree ensures the integrity of transactions:

\begin{itemize}
    \item Each leaf node stores the hash of a single transaction.
    \item Parent nodes store the hash of their two child nodes.
    \item The root node (also known as the \textbf{Merkle root}) contains a hash that summarizes all transactions in the block.
\end{itemize}

If even a single transaction changes, the hash at the leaf node changes, which in turn changes all parent hashes up to the root. Thus, the Merkle root serves as a compact proof of the integrity of all transactions in the block.

\subsection{Process of Generating a Merkle Root}
A Merkle root is generated through the use of a Merkle tree, a binary tree structure that helps in summarizing and verifying the integrity of large sets of data, such as transactions in a blockchain. The process of generating a Merkle root begins by taking all the transaction hashes from a block, which represent individual transactions. These hashes are then paired and concatenated before being hashed again. This process continues iteratively by pairing the resulting hashes and hashing them once more until a single hash, known as the Merkle root, is obtained.

\begin{itemize}
    \item Step 1: Start with the list of all transaction hashes from the block.
    \item Step 2: Pair the hashes (e.g., Hash 1 is paired with Hash 2, Hash 3 with Hash 4, and so on).
    \item Step 3: Concatenate the paired hashes and hash them again using a cryptographic hash function like SHA-256.
    \item Step 4: Continue this process, halving the number of resulting hashes with each step, until a single hash is left. This final hash is the Merkle root.
\end{itemize}

For example, assume there are four transactions in a block, with hashes `A`, `B`, `C`, and `D`. The process looks like this:

\begin{align*}
    \text{Hash of (A, B)} &= H(H(A) || H(B)) \\
    \text{Hash of (C, D)} &= H(H(C) || H(D)) \\
    \text{Merkle Root} &= H(H(H(A) || H(B)) || H(H(C) || H(D)))
\end{align*}

This final Merkle root is stored in the block header and acts as a concise summary of all transactions in the block, allowing efficient verification.

\subsection{Application of Merkle Tree in Data Verification}
Merkle trees enable efficient data verification without requiring the entire dataset\cite{Wang2023ATD}. In the context of blockchain, a Merkle tree is used to verify if a specific transaction is part of a block without needing to download the entire block. This is particularly useful for light nodes, which only store the block headers rather than the full blockchain data.

To verify a transaction, a node can request a Merkle proof, which consists of the hashes necessary to reconstruct the Merkle root from the transaction in question. This allows the node to check if the computed root matches the one in the block header.

For example, if we want to verify transaction A in the previous tree:

\begin{itemize}
    \item The node receives `H(B)` and `H(H(C) || H(D))` from the full node.
    \item The node then calculates `H(H(A) || H(B))` and compares it with the given Merkle root.
    \item If the calculated root matches the root in the block header, transaction A is verified as part of the block.
\end{itemize}

This mechanism allows nodes to operate efficiently and reduces bandwidth and storage requirements, as they don't need to download or store entire blocks.

\section{Nonce and Proof of Work (PoW)}
\subsection{Definition and Purpose of PoW}
Proof of Work (PoW) is a consensus algorithm used in blockchain networks, such as Bitcoin, to ensure network security and verify transactions\cite{ferdous2020blockchainconsensusalgorithmssurvey}. It requires miners (nodes that validate transactions) to solve a computational puzzle, where the goal is to find a value that, when hashed, meets a difficulty target set by the network.

The primary purpose of PoW is twofold:
\begin{itemize}
    \item \textbf{Prevent double-spending:} By making it computationally difficult to alter transaction history, PoW ensures that once a block is added, altering it would require redoing the work for that block and all subsequent blocks, making it economically infeasible to double-spend.
    \item \textbf{Ensure decentralized security:} Since PoW requires significant computational resources, it prevents any single entity from easily taking control of the network without incurring significant costs.
\end{itemize}

\subsection{Nonce Selection and Hash Calculation}
The key element of PoW is finding a nonce—a random number that miners adjust to solve the computational puzzle\cite{Raza2021EnergyEM}. The task is to find a nonce such that when the block header (including the nonce) is hashed, the resulting hash is below a certain threshold, known as the difficulty target.

Miners try different nonce values, incrementing them until they find a hash that satisfies the difficulty condition. The process looks like this:

\begin{lstlisting}[style=python]
import hashlib

def calculate_hash(block_header):
    return hashlib.sha256(block_header.encode()).hexdigest()

block_header = "previous_block_hash + merkle_root + timestamp + nonce"
difficulty_target = "0000ffffffffffffffffffffffffffffffffffffffffffffffffffffffffffff"

nonce = 0
while True:
    new_header = block_header + str(nonce)
    block_hash = calculate_hash(new_header)
    if block_hash < difficulty_target:
        print(f"Block solved with nonce: {nonce}, Hash: {block_hash}")
        break
    nonce += 1
\end{lstlisting}

This example shows how miners modify the nonce and rehash the block header until the hash value meets the difficulty target.

\section{Concept of Chain-Linked Blocks}
\subsection{Timestamp in Blockchain}
In a blockchain, each block contains a timestamp that indicates when the block was created. The timestamp plays a crucial role in ordering blocks and transactions chronologically. This is essential because the timestamp ensures that the chain of blocks maintains a proper sequence and prevents issues like replay attacks, where older transactions could be reused maliciously\cite{Hunt2008BitcoinAP}.

The timestamp is included in the block header, along with other data such as the previous block hash, the Merkle root, and the nonce\cite{Regnath2020BlockchainWT}. Here's an example of a block header structure:

\begin{lstlisting}[style=cmd]
Block Header:
    - Previous Block Hash: 00000000000000000007878ec04bb2b2e12317804df077010e5c3ad07e9b8b41
    - Merkle Root: 4d5ef7611a84f8c9bfc1c45de2274a1a42d71694a9f8b3548397b634f1e4c154
    - Timestamp: 1631293747
    - Nonce: 945395
\end{lstlisting}

The timestamp is crucial for tracking the order of transactions, ensuring the continuity of the chain, and allowing nodes to validate when each block was mined.

\subsection{Immutability of Blockchain}
One of the fundamental principles of blockchain is immutability, meaning that once data is written to a block and added to the blockchain, it cannot be altered or deleted without invalidating the entire chain. This is achieved through the use of cryptographic hashes and the PoW consensus mechanism.\cite{Rahman2020DesignPF}

Each block in the blockchain contains the hash of the previous block, linking the blocks together in a chain. If someone tries to alter a block’s data, the hash of that block changes, and consequently, all subsequent block hashes also change. Since PoW requires a computational puzzle to be solved for each block, redoing the work for the entire chain would be practically impossible without controlling the majority of the network's computational power.

In summary, immutability is achieved by:
\begin{itemize}
    \item \textbf{Cryptographic hashes:} Each block contains the hash of the previous block, creating a link that ensures tampering is detectable.
    \item \textbf{Consensus mechanism:} PoW ensures that altering any block would require enormous computational effort, making changes economically and technically infeasible.
\end{itemize}

\chapter{Bitcoin Network and Nodes}
    \section{Bitcoin Transaction Process}
        \subsection{Definition of Inputs and Outputs}
        A Bitcoin transaction is a transfer of value between Bitcoin addresses. It consists of two main components: \textbf{inputs} and \textbf{outputs}. The \textbf{input} defines the source of the funds, and the \textbf{output} defines where the funds are being sent. More precisely, inputs refer to references to previous unspent outputs (UTXOs - Unspent Transaction Outputs), and outputs specify the destination addresses and the amount of Bitcoin being transferred\cite{bitcoinDoesBitcoin}.

        \textbf{Inputs:} Each input in a Bitcoin transaction points to an unspent output from a previous transaction\cite{Kang2023BitcoinDA}. This input consumes the value of that output to be spent in the current transaction. An input also includes a cryptographic signature, which proves ownership of the funds associated with the output.

        \textbf{Outputs:} An output in a Bitcoin transaction specifies the amount of Bitcoin to be sent and the address of the recipient\cite{9122595}. The total amount of all inputs must either equal or exceed the total amount of all outputs. If there is a difference, that difference is returned as "change" to the sender in the form of an additional output. Outputs in Bitcoin transactions become new UTXOs for future transactions.

        \textbf{Example:} Imagine Alice has 1 BTC in her wallet (from a previous transaction output). She wants to send 0.5 BTC to Bob. Alice's transaction would look like this:
        \begin{itemize}
            \item \textbf{Input:} Alice's 1 BTC UTXO from a previous transaction.
            \item \textbf{Outputs:} 
            \begin{itemize}
                \item 0.5 BTC to Bob's address.
                \item 0.49 BTC as change back to Alice's address (assuming a small transaction fee of 0.01 BTC).
            \end{itemize}
        \end{itemize}
        
        After this transaction, Bob has 0.5 BTC in his wallet (in the form of a new UTXO), and Alice has 0.49 BTC (also as a new UTXO).

        \subsection{Explanation of the UTXO Model}
        The Bitcoin network relies on the \textbf{UTXO (Unspent Transaction Output)} model to track the ownership of Bitcoin\cite{DelgadoSegura2018AnalysisOT}. Unlike traditional accounts where balances are directly managed, Bitcoin transactions use UTXOs to ensure accuracy, security, and the prevention of double-spending.

        \textbf{How the UTXO Model Works:} 
        \begin{itemize}
            \item Every time a transaction is made, the outputs of that transaction become the inputs for future transactions.
            \item UTXOs are the pieces of Bitcoin that a user owns and can spend in new transactions.
            \item A Bitcoin wallet doesn't hold a balance but instead tracks the UTXOs that belong to the user. To compute a balance, the wallet sums up the values of all UTXOs associated with the user's addresses.
        \end{itemize}
        
        When constructing a new transaction, the user selects one or more UTXOs as inputs, and any leftover funds are returned as a new output (called "change"). The UTXO model simplifies the verification process because only the unspent outputs need to be checked, and no central ledger tracks balances.

        \textbf{Double-Spending Protection:} 
        UTXOs are used to prevent double-spending because once an output is spent, it cannot be used again. Nodes in the Bitcoin network keep a record of all UTXOs and can quickly determine if an output has already been spent by checking their UTXO database.

        \subsection{Constructing and Signing a Bitcoin Transaction}
        To create a Bitcoin transaction, the following steps are performed:
        
        \begin{enumerate}
            \item \textbf{Selecting Inputs:} The sender selects one or more UTXOs from their wallet to fund the transaction. These UTXOs must total at least the amount being sent, plus any transaction fees.
            
            \item \textbf{Determining Outputs:} The transaction outputs specify where the funds are being sent. Typically, this will include:
            \begin{itemize}
                \item The recipient’s address and the amount to send.
                \item A change address (usually the sender’s own address) if the total value of the inputs exceeds the amount being sent.
            \end{itemize}
            
            \item \textbf{Signing the Transaction:} The sender uses their private key to sign the transaction. This signature proves ownership of the inputs being spent and ensures that only the legitimate owner can transfer the funds. The private key is never shared, and the signature is verified by Bitcoin nodes before the transaction is accepted into the network.

        \end{enumerate}
        
        The following Python code snippet demonstrates a simplified structure of constructing and signing a Bitcoin transaction using a Python library:
        \begin{lstlisting}[style=python]
        from bitcoin import SelectParams
        from bitcoin.wallet import CBitcoinSecret, P2PKHBitcoinAddress
        from bitcoin.core import b2lx, COIN, CMutableTransaction, CMutableTxOut, CMutableTxIn

        # Select the network (mainnet/testnet)
        SelectParams('testnet')

        # Define private key, recipient address, and UTXO to spend
        my_private_key = CBitcoinSecret('your_private_key_here')
        recipient_address = P2PKHBitcoinAddress('recipient_address_here')
        utxo_txid = 'utxo_txid_here'  # Transaction ID of UTXO to spend
        utxo_vout = 0  # Output index of UTXO

        # Construct transaction input
        txin = CMutableTxIn(COutPoint(b2lx(utxo_txid), utxo_vout))

        # Construct transaction output (send 0.5 BTC to recipient)
        amount_to_send = 0.5 * COIN  # 0.5 BTC
        txout = CMutableTxOut(amount_to_send, recipient_address.to_scriptPubKey())

        # Create the transaction and sign it
        tx = CMutableTransaction([txin], [txout])
        txin.scriptSig = my_private_key.sign(tx)

        # Display the transaction
        print(tx)
        \end{lstlisting}
        
    \section{Types of Nodes in the Bitcoin Network}
        \subsection{Role of Full Nodes}
        A \textbf{full node} is a computer that fully validates transactions and blocks in the Bitcoin network\cite{Abe2018MitigatingBN}. Full nodes download and store the entire blockchain and participate in transaction validation, ensuring that the rules of the Bitcoin protocol are strictly followed.

        Full nodes are the backbone of the Bitcoin network because they:
        \begin{itemize}
            \item Verify transactions and blocks independently without relying on any third party.
            \item Relay validated transactions and blocks to other nodes in the network.
            \item Reject any transaction or block that violates Bitcoin's consensus rules.
        \end{itemize}
        
        By running a full node, users contribute to the network's decentralization and security, ensuring that no one can change the Bitcoin protocol rules or create invalid transactions. Full nodes prevent tampering with the blockchain and protect against fraud.

        \subsection{Light Nodes and Simplified Payment Verification (SPV)}
        \textbf{Light nodes}, also known as SPV (Simplified Payment Verification) nodes\cite{AlBassam2018FraudAD}, are a type of node that does not store the full blockchain but instead relies on full nodes for transaction verification.

        Light nodes only download the block headers (which are much smaller in size) instead of the entire blockchain\cite{Zamyatin2020TxChainEC}. This allows them to verify transactions without needing to download and store all the data. When a light node wants to verify a transaction, it can request proof from full nodes using Merkle trees, a cryptographic structure that ensures the integrity of the data.

        \textbf{Trade-offs:}
        \begin{itemize}
            \item \textbf{Advantages:} Light nodes are easier to run on devices with limited storage and bandwidth, such as smartphones.
            \item \textbf{Disadvantages:} Light nodes rely on full nodes for verification, meaning they trust the information provided by full nodes. This introduces a slight risk of relying on malicious or dishonest full nodes.
        \end{itemize}

    \section{Transaction Propagation and Confirmation}
        \subsection{Transaction Propagation in the P2P Network}
        Bitcoin uses a peer-to-peer (P2P) network to propagate transactions and blocks\cite{Donet2014TheBP}. When a user broadcasts a transaction, it is first sent to the nearest nodes in the network. Each node then verifies the transaction and relays it to its peers. This process continues until the transaction reaches miners, who will include it in the next block.

        Transaction propagation ensures that all nodes in the network have a copy of the transaction and that the transaction eventually reaches the miners for inclusion in the blockchain.

        \subsection{Transaction Confirmation and Block Packaging}
        Once a transaction reaches miners, it is added to a \textbf{block}, which is essentially a group of transactions. The miner competes to solve a cryptographic puzzle, and once a solution is found, the block is added to the blockchain.

        When a transaction is included in a block, it receives its first \textbf{confirmation}. Each additional block added after this provides another confirmation, increasing the security of the transaction. More confirmations make it increasingly difficult for an attacker to reverse the transaction.

        \subsection{Six Confirmations Rule in Blockchain}
        In the Bitcoin network, there is a commonly accepted practice known as the \textbf{six confirmations rule}\cite{Wan2019EvaluatingTI}. This rule states that after a transaction has been confirmed by six blocks, it is considered extremely secure and unlikely to be reversed. The reasoning behind this rule is based on the probability of an attacker successfully rewriting the blockchain. With each added block, the chances of an attacker succeeding diminish, making six confirmations a highly secure threshold.

\section{Bitcoin Miners and Mining}
    \subsection{Mechanism of Mining}
    Bitcoin mining is the process by which new transactions are added to the Bitcoin blockchain. It involves miners competing to solve a complex cryptographic puzzle based on the Proof of Work (PoW) consensus mechanism\cite{Ankalkoti2017ARS}. The puzzle requires miners to find a hash that, when applied to the block’s header, meets certain conditions (i.e., it must be lower than a certain target). This process is computationally intensive and ensures that adding new blocks to the blockchain requires substantial effort.

    The PoW system works as follows:
    \begin{itemize}
        \item Miners collect and verify transactions, bundling them into a block.
        \item The block header, which contains metadata about the block, is hashed.
        \item Miners must adjust a variable called the \textit{nonce}, and rehash the block header, until they find a hash that satisfies the network's difficulty target.
    \end{itemize}

    The \textbf{difficulty adjustment mechanism} is crucial to Bitcoin’s stability\cite{Zhang2018AnAO}. It ensures that blocks are added to the blockchain approximately every 10 minutes, regardless of the total computational power in the network. Every 2016 blocks (approximately every two weeks), the difficulty is adjusted based on how quickly or slowly the previous blocks were mined. If the blocks were mined too quickly, the difficulty increases; if too slowly, the difficulty decreases. This self-regulating process ensures a stable and predictable issuance of new bitcoins.

    \subsection{Miner Rewards and Block Rewards}
    Miners are incentivized to participate in the mining process through a system of rewards. When a miner successfully adds a new block to the blockchain, they receive a reward known as the \textbf{block reward}. This reward consists of two components:
    \begin{itemize}
        \item \textbf{Block Subsidy}: The block subsidy is a fixed amount of newly minted bitcoins. Initially set at 50 bitcoins per block, the subsidy is halved approximately every four years during an event called the \textit{halving}\cite{Meynkhard2019FairMV}. After the third halving in 2020, the block subsidy is now 6.25 bitcoins per block. The subsidy will continue to halve every 210,000 blocks until it eventually reaches zero, around the year 2140.
        \item \textbf{Transaction Fees}: In addition to the block subsidy, miners also receive transaction fees. Users include fees with their transactions as an incentive for miners to prioritize their transactions in the block. Over time, as the block subsidy diminishes, transaction fees will become the primary source of income for miners.
    \end{itemize}

    This combination of newly created bitcoins and transaction fees forms the total reward miners receive, making it economically viable for miners to dedicate their computing resources to the Bitcoin network.

    \subsection{Bitcoin's Total Supply and Scarcity}
    Bitcoin’s economic model is built on the principle of scarcity\cite{Sedliaik2023FactorsSB}. There will only ever be 21 million bitcoins in existence. This limit is hardcoded into the Bitcoin protocol and cannot be changed without a complete overhaul of the system, which would require consensus from the entire network.

    The scarcity is maintained through the block subsidy halving mechanism. As mentioned earlier, every 210,000 blocks, the reward for mining a new block is halved. This creates a deflationary pressure, meaning that over time, fewer bitcoins are introduced into the economy. The fixed supply of 21 million bitcoins is expected to create long-term value as demand increases while supply remains static. This scarcity is a key factor in Bitcoin’s value proposition as a form of “digital gold.”

\section{Bitcoin Wallets}
    \subsection{Generation of Private Keys, Public Keys, and Addresses}
    A Bitcoin wallet is essentially a pair of cryptographic keys: a private key and a public key\cite{Gutoski2015HierarchicalDB}. These keys are generated through the use of cryptographic algorithms, primarily elliptic curve cryptography (ECC).

    \begin{itemize}
        \item \textbf{Private Key}: A private key is a randomly generated 256-bit number, represented as a string of alphanumeric characters. It is crucial for authorizing transactions and must be kept secret.
        \item \textbf{Public Key}: The public key is derived from the private key using a mathematical operation called elliptic curve multiplication. While the private key is secret, the public key can be shared with others.
        \item \textbf{Bitcoin Address}: A Bitcoin address is a shorter, hashed version of the public key. It is generated by applying a series of cryptographic hash functions (SHA-256 and RIPEMD-160) to the public key, making it more user-friendly for transactions. This address is what users share to receive bitcoins.
    \end{itemize}

    The process of generating a Bitcoin address can be illustrated with the following Python code:

    \begin{lstlisting}[style=python]
    import hashlib
    
    def private_key_to_public_key(private_key):
        # Convert private key to public key (simplified)
        return 'PublicKey'  # Placeholder for actual ECC operations
    
    def public_key_to_address(public_key):
        # Apply SHA-256 and then RIPEMD-160 hash functions
        sha256_hash = hashlib.sha256(public_key.encode()).hexdigest()
        ripemd160_hash = hashlib.new('ripemd160', sha256_hash.encode()).hexdigest()
        return ripemd160_hash

    private_key = 'YourPrivateKey'
    public_key = private_key_to_public_key(private_key)
    address = public_key_to_address(public_key)
    print("Bitcoin Address:", address)
    \end{lstlisting}

    \subsection{Cold Wallets vs. Hot Wallets}
    Bitcoin wallets can be classified into two categories based on their level of connectivity:
    \begin{itemize}
        \item \textbf{Cold Wallets}: Cold wallets are offline storage solutions, such as hardware wallets or paper wallets. Because they are not connected to the internet, they are immune to hacking attacks and are generally considered more secure\cite{Barakat2022COMPARISONOH}. However, they are less convenient for frequent transactions. An example of a cold wallet is a hardware wallet like Trezor\cite{trezorTrezorHardware} or Ledger\cite{ledgerHardwareWallet}.
        \item \textbf{Hot Wallets}: Hot wallets are online wallets connected to the internet, such as software wallets or exchange-based wallets. While they provide easy access for frequent transactions, they are more vulnerable to security risks like hacking. Popular examples include mobile apps like Trust Wallet or web-based wallets provided by exchanges like Binance\cite{binance} or Coinbase\cite{coinbase}.
    \end{itemize}

    The trade-off between cold and hot wallets is one of security versus convenience. Best practices recommend using cold wallets for long-term storage of large amounts of Bitcoin, while hot wallets can be used for smaller, everyday transactions.

    \subsection{Multi-Signature Wallets}
    Multi-signature (multisig) wallets provide an additional layer of security by requiring multiple private keys to authorize a transaction\cite{Goel2023MultisignatureCW}. In a traditional Bitcoin wallet, a single private key is used to sign transactions. In contrast, a multisig wallet requires approval from multiple parties\cite{Tran2023DeploymentOT}.

    For example, in a 2-of-3 multisig wallet, three private keys are generated, but any two of them are required to authorize a transaction. This setup is useful for:
    \begin{itemize}
        \item \textbf{Security Enhancement}: By distributing keys across multiple devices or individuals, multisig wallets reduce the risk of theft.
        \item \textbf{Shared Control}: In business environments, multisig wallets allow for shared control over funds, ensuring that no single individual has unilateral access.
    \end{itemize}

    The structure of a multisig wallet can be visualized in the following diagram:

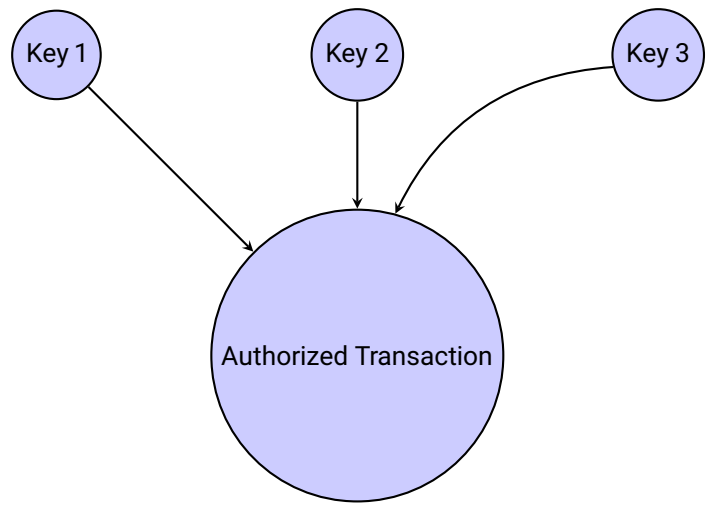
\begin{center} 
\begin{tikzpicture}
    [->, >=stealth, auto, node distance=4.0cm, thick, 
    node/.style={circle, draw, fill=blue!20, text centered}]
        \node[node] (A) {Key 1};
        \node[node] (B) [right of=A] {Key 2};
        \node[node] (C) [right of=B] {Key 3};
        \node[node] (D) [below of=B] {Authorized Transaction};
        \path (A) edge (D)
              (B) edge (D)
              (C) edge[bend right] (D);
    \end{tikzpicture}
\end{center}

    Multi-signature wallets are a powerful tool for improving the security and flexibility of Bitcoin transactions, especially in collaborative environments.

\chapter{History and Evolution of Bitcoin}
\section{The Genesis Block}

\subsection{Satoshi Nakamoto's Genesis Message}
One of the most remarkable features of Bitcoin's origin is the message embedded in the first block, known as the Genesis Block or Block 0\cite{C2018AnOO}, by the pseudonymous creator Satoshi Nakamoto. The Genesis Block, mined on January 3, 2009, contains a hidden message in its coinbase parameter:

\begin{lstlisting}[style=cmd]
"The Times 03/Jan/2009 Chancellor on brink of second bailout for banks"
\end{lstlisting}

This message is a reference to the headline of an article published in the UK newspaper *The Times* on the same day\cite{raskin2018digital}. It reflects the financial instability during the 2008 global financial crisis, where many banks were bailed out by governments in response to their near-collapse.

The message carries symbolic significance, representing Bitcoin's foundational goal: to create an alternative financial system that is decentralized, transparent, and free from the control of central banks and governments\cite{Maurer2015perhapsTR}. By embedding this message, Satoshi Nakamoto highlighted Bitcoin's intent to be a solution to the failings of the traditional financial system, which had shown vulnerabilities during the crisis. 

Bitcoin was designed to allow peer-to-peer transactions without the need for intermediaries, like banks, which are often subject to economic mismanagement and excessive control over individuals' financial freedom. The embedding of this message in the Genesis Block underscores this mission, illustrating Bitcoin’s role as a response to centralized financial institutions.

\subsection{Symbolic Meaning of the Bitcoin Genesis Block}
The Genesis Block is not only important because of the embedded message, but also because it marks the official beginning of the Bitcoin blockchain. In blockchain technology, every new block contains a reference to the block before it, forming a chain. The Genesis Block is unique because it is the very first block and thus does not reference any previous block.

This block is hardcoded into the Bitcoin protocol, and it contains a reward of 50 bitcoins for the miner (which was Satoshi Nakamoto). However, these bitcoins are unspendable, a symbolic gesture that marks the beginning of the Bitcoin system while setting a precedent for all future blocks.

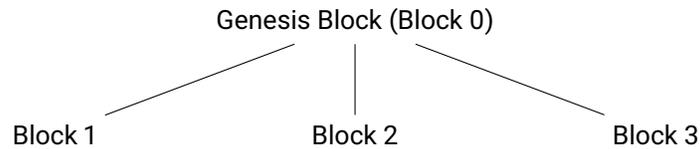
\begin{center}
\begin{tikzpicture}
[level 1/.style={sibling distance=40mm},
level 2/.style={sibling distance=20mm}]
\node {Genesis Block (Block 0)}
child {node {Block 1}}
child {node {Block 2}}
child {node {Block 3}};
\end{tikzpicture}
\end{center}

As shown in the above illustration, Block 1 and all subsequent blocks reference the Genesis Block, making it the foundation of the entire Bitcoin network. The security and immutability of the blockchain depend on this chain structure. If any block in the chain were to be modified, it would invalidate all the subsequent blocks, making it impossible to alter past transactions.

The Genesis Block symbolizes the start of a new era in financial systems—one that is decentralized, transparent, and resistant to censorship. It sets the precedent for all future blocks by providing a permanent, unchangeable record of the beginning of the Bitcoin network. 

\section{Bitcoin's Scalability Issues}

\subsection{Block Size Limit and Transaction Speed}
Bitcoin's original design includes a 1MB limit on the size of each block\cite{GroupBlockSI}, meaning that only a certain number of transactions can fit into a block. On average, Bitcoin processes around 7 transactions per second (TPS)\cite{Gbel2017IncreasedBS}. During times of low network activity, this transaction rate is adequate. However, during periods of high demand, the limited block size causes a bottleneck, leading to slower transaction speeds and higher fees.

To better understand this, consider the following analogy: imagine a highway with a limited number of lanes (the 1MB block size). As more cars (transactions) try to get through, traffic jams occur, causing delays. Drivers (users) might be willing to pay a toll (transaction fee) to use a faster lane, but the number of lanes remains the same, limiting overall throughput. 

Bitcoin's block size constraint directly impacts its ability to scale\cite{Croman2016OnSD}. When more transactions are submitted than the network can process, users experience longer waiting times for their transactions to be confirmed. This issue became particularly noticeable during the 2017 bull market, when transaction fees skyrocketed due to network congestion.

\subsection{Technical Challenges of Bitcoin Scalability}
Scaling Bitcoin presents several technical challenges. One of the core principles of Bitcoin is decentralization, meaning that no single entity controls the network. However, increasing the transaction throughput often requires modifications that could compromise decentralization.

The following are some of the major challenges in scaling Bitcoin:

\begin{itemize}
    \item \textbf{Trade-off between decentralization and scalability:} As Bitcoin's block size is increased, more data is required to be processed and stored by full nodes. This can lead to centralization, as only those with significant computing power and storage capabilities can afford to run a full node. In contrast, a smaller block size allows more users to participate in the network, maintaining decentralization but at the cost of slower transaction speedse\cite{Chohan2019TheLT}.
    
    \item \textbf{Security risks:} A key aspect of Bitcoin’s security is its Proof of Work (PoW) consensus mechanism. Increasing the block size or changing how transactions are processed could introduce new vulnerabilities. For example, larger blocks might take longer to propagate across the network, leading to temporary forks and security riskse\cite{Lewenberg2015InclusiveBC}.
    
    \item \textbf{Proposed solutions:} Several solutions have been proposed to address Bitcoin's scalability issues:
    \begin{itemize}
        \item \textbf{Segregated Witness (SegWit):} SegWite\cite{Kedziora2022AnalysisOS} was introduced to increase the effective block size without increasing the actual block size limit. It achieves this by removing certain parts of the transaction data, making more space for transactions in each block.
        
        \item \textbf{The Lightning Network:} The Lightning Networke\cite{Bartolucci2019APM} is a second-layer solution that allows for off-chain transactions. By creating payment channels between users, the Lightning Network can handle a large number of transactions outside the blockchain, only settling on-chain when necessary. This significantly reduces the burden on the Bitcoin network, allowing for faster and cheaper transactions.
        
        \item \textbf{Hard forks (e.g., Bitcoin Cash):} In 2017, a group of developers initiated a hard fork of Bitcoin, resulting in the creation of Bitcoin Cash (BCH)e\cite{Aggarwal2019ASA}. Bitcoin Cash increased the block size to 8MB and later to 32MB, allowing for more transactions per block. However, this came with its own set of challenges, including reduced decentralization and potential security risks.
    \end{itemize}
\end{itemize}

While solutions like SegWit and the Lightning Network have alleviated some of the scalability issues, Bitcoin's core design still faces inherent trade-offs between maintaining decentralization and improving transaction throughput. The ongoing challenge is to scale the network in a way that preserves its security and decentralization, which are the foundational principles behind Bitcoin’s success.

\section{Bitcoin Soft Forks and Hard Forks}
    \subsection{Definition and Examples of Soft Forks}
    Soft forks are backward-compatible upgrades to a blockchain protocole\cite{Zamyatin2018AWV}. This means that non-upgraded nodes (those that have not adopted the changes) can still interact with upgraded nodes without breaking the network's overall functionality. Soft forks typically introduce new rules that are more restrictive than before, allowing older nodes to continue functioning as long as they follow the new rules.

    For example, the BIP66 soft fork was implemented in Bitcoin in 2015\cite{bitcoinSoftforkBitcoin}. It introduced strict DER (Distinguished Encoding Rules) signatures, which enforced stricter rules for the way digital signatures were encoded on the Bitcoin network. Before BIP66, there was a degree of flexibility in how these signatures were generated. This flexibility sometimes caused issues, such as minor differences in how signatures were processed, leading to bugs and potential vulnerabilities. By enforcing a strict standard, BIP66 helped prevent these issues while remaining compatible with non-upgraded nodes.

    Another notable soft fork was the SegWit (Segregated Witness)\cite{bitcoin0141Bitcoin} upgrade, which will be explained in more detail in a later subsection.

    \subsection{Definition and Examples of Hard Forks}
    Hard forks, on the other hand, introduce non-backward-compatible changes to a blockchain protocol. In a hard fork, the upgraded nodes adopt a new set of rules that non-upgraded nodes cannot follow. As a result, the two groups (upgraded and non-upgraded) form two separate networks, which cannot communicate with each other. This leads to a permanent split in the blockchain, where each network continues independently with its own chain of transactions.

    A well-known example of a hard fork is the creation of Bitcoin Cash (BCH) in 2017\cite{Aggarwal2019ASA}. This fork was driven by disagreements within the Bitcoin community about how to scale the network, particularly regarding the block size limit. Bitcoin Cash increased the block size to allow more transactions per block, whereas Bitcoin (BTC) retained its smaller block size. The two networks became entirely separate, with their own coins and ecosystems.

    Another example of a hard fork is the Ethereum DAO fork in 2016\cite{Mehar2021UnderstandingAR}. After a vulnerability in the DAO (Decentralized Autonomous Organization) contract was exploited, a large portion of Ethereum’s funds was stolen. The community decided to implement a hard fork to reverse the effects of the hack, creating two separate blockchains: Ethereum (ETH)\cite{ethereumCompleteGuide} and Ethereum Classic (ETC)\cite{ethereumclassicEthereumClassic}.

    \subsection{SegWit Upgrade and Segregated Witness}
    Segregated Witness, commonly known as SegWit\cite{bitcoin0141Bitcoin}, was a major Bitcoin upgrade implemented through a soft fork in 2017. It aimed to solve a critical problem known as transaction malleability, which made it possible for attackers to change small parts of a transaction's identifier without altering its content. This issue made it difficult for users to track and verify transactions accurately.

    SegWit introduced a solution by separating (or segregating) the witness data — the part of the transaction that contains the digital signature — from the main transaction data. By moving the signature data outside the base block, SegWit reduced the transaction size, effectively increasing the block's capacity. This allowed more transactions to fit into each block, improving Bitcoin’s scalability without needing to increase the block size itself.

    Another significant impact of SegWit was its role in paving the way for second-layer solutions like the Lightning Network, which rely on the improvements in transaction malleability and efficiency brought by the upgrade.

\section{The Hard Fork of Bitcoin Cash}
    \subsection{The Birth of Bitcoin Cash and Its Causes}
    Bitcoin Cash (BCH) was created in August 2017\cite{Aggarwal2019ASA} as a result of a hard fork from Bitcoin. The primary reason for the fork was a long-standing debate within the Bitcoin community over how to best scale the network to accommodate increasing transaction demand. Bitcoin’s block size, capped at 1MB, limited the number of transactions that could be processed in each block, leading to higher fees and slower transaction times during periods of high demand.

    One side of the debate argued that increasing the block size was the simplest and most effective solution to improve Bitcoin's scalability\cite{GroupBlockSI}. However, another faction, including Bitcoin Core developers, opposed increasing the block size, citing concerns about centralization (larger blocks require more storage and processing power) and the potential risks of altering Bitcoin’s fundamental design\cite{deFilippi2016TheIP}.

    These disagreements ultimately led to the creation of Bitcoin Cash, which increased the block size to 8MB to handle more transactions per block.

    \subsection{Technical Changes in Bitcoin Cash: 8MB Block Size}
    The primary technical change introduced by Bitcoin Cash was the increase of the block size limit from 1MB to 8MB\cite{Johnson2021DoesBC}. By allowing blocks to be up to 8 times larger, Bitcoin Cash aimed to process more transactions in each block, thereby reducing transaction confirmation times and lowering transaction fees during periods of high network activity.

    This increase in block size allowed Bitcoin Cash to handle a higher volume of transactions per second (TPS) compared to Bitcoin. While Bitcoin typically supports around 3-7 TPS, Bitcoin Cash can handle a significantly higher number of transactions, which theoretically could reach up to 24-32 TPS, depending on network conditions.

    \subsection{Market and Community Split Post-Fork}
    The Bitcoin Cash hard fork led to a significant split in both the community and the market. Users, miners, and developers had to choose which version of the blockchain they supported: Bitcoin (BTC) with its smaller block size and conservative scaling approach, or Bitcoin Cash (BCH) with its larger block size and aggressive scaling solution.

    In terms of market value, Bitcoin remained the dominant chain, retaining its symbol (BTC) and most of its market capitalization. Bitcoin Cash (BCH), although less valuable than Bitcoin, still gained substantial traction, especially among those who believed in its vision of scaling through larger blocks. Over time, Bitcoin Cash developed its own ecosystem of users, merchants, and exchanges, establishing itself as a separate entity within the cryptocurrency space.

\section{The Proposal of the Lightning Network}
    \subsection{Layer 2 Solutions}
    Layer 2 solutions are protocols built on top of the base blockchain (Layer 1) to improve scalability and reduce fees\cite{Mandal2023InvestigatingLS}. Instead of changing the core blockchain protocol, these solutions create an additional layer where transactions can occur off-chain, reducing the load on the base layer.

    The Lightning Network is a prime example of a Layer 2 solution designed to address Bitcoin's scalability problems. By moving the majority of small transactions off-chain, the Lightning Network aims to reduce transaction fees and improve transaction speed, while still allowing users to settle the final transaction on the Bitcoin blockchain.

    \subsection{Lightning Network Channels and Fast Payments}
    The Lightning Network operates by creating payment channels between users. A payment channel is a private, off-chain ledger that allows two parties to conduct multiple transactions without having to record each one on the Bitcoin blockchain. Instead, only the opening and closing of the channel are recorded on-chain.

    Once a payment channel is opened, users can send payments back and forth instantly and with minimal fees. These transactions are not broadcast to the entire Bitcoin network, reducing congestion on the main blockchain. The Lightning Network can also facilitate payments across multiple channels, enabling a web of interconnected payment pathways that increase flexibility.

    For example, if Alice wants to send money to Bob but doesn’t have a direct payment channel with him, she can route the payment through a series of intermediaries who do. The Lightning Network will automatically find the shortest and cheapest route for the transaction, ensuring that payments are fast and efficient.

    The main advantage of the Lightning Network is that it enables almost instant, low-cost payments, making it suitable for everyday transactions such as buying coffee or paying for services, without waiting for confirmations on the Bitcoin blockchain.

\chapter{Security of Bitcoin}
    \section{Bitcoin's Security Mechanisms}
        \subsection{Public Key Cryptography and Digital Signature Assurance}
        Bitcoin relies heavily on public key cryptography to secure its transactions and ensure the integrity and ownership of the funds\cite{Conti2017ASO}. Each Bitcoin user has a pair of cryptographic keys: a public key and a private key. The public key can be shared with anyone, and it serves as an address to receive Bitcoin. However, the private key must remain confidential, as it is used to sign transactions and authorize the transfer of Bitcoin.

        When a user wants to send Bitcoin, they create a transaction message that includes details such as the amount of Bitcoin and the recipient's public address\cite{Puranam2019AnatomyAL}. This message is then signed using the sender's private key. This digital signature is crucial because it proves that the transaction was indeed initiated by the rightful owner of the Bitcoin without revealing the private key itself\cite{Vesel2021BitcoinAQ}. The signature is verified using the corresponding public key, ensuring that the transaction is authentic and not tampered with.

        A real-world analogy would be like signing a contract. The signature proves that you agree to the terms without sharing any sensitive information. In Bitcoin, the cryptographic digital signature assures that only the owner of the private key could have signed the transaction.

        To illustrate this process in Python, we can simulate public key cryptography using the `ecdsa` (Elliptic Curve Digital Signature Algorithm) library\cite{Johnson1998EllipticCD}. This simple example demonstrates generating a private-public key pair, signing a message, and verifying the signature.

        \begin{lstlisting}[style=python]
        from ecdsa import SigningKey, VerifyingKey, NIST384p

        # Generate private and public keys
        private_key = SigningKey.generate(curve=NIST384p)
        public_key = private_key.get_verifying_key()

        # Sign a message
        message = b"This is a Bitcoin transaction."
        signature = private_key.sign(message)

        # Verify the signature
        try:
            public_key.verify(signature, message)
            print("The signature is valid.")
        except:
            print("The signature is invalid.")
        \end{lstlisting}

        In Bitcoin, this process occurs at a much larger scale, where every node in the network can verify the digital signature associated with a transaction to ensure its authenticity before including it in the blockchain.

        \subsection{Hash Locking and Time Locking}
        Hash locking and time locking are advanced mechanisms to provide additional security and flexibility to Bitcoin transactions. These mechanisms are part of Bitcoin’s scripting capabilities, allowing users to create conditional transactions. Let’s break down each of these mechanisms in detail.

        \subsubsection{Hash Locking}
        Hash locking ensures that certain conditions are met before funds can be transferred\cite{Churchill2015WhySW}. It works by locking the Bitcoin transaction with a specific cryptographic condition, typically a hash pre-image. A hash lock requires that the correct input (pre-image) be provided to unlock the funds. This mechanism is often used in "Atomic Swaps"\cite{Miraz2019AtomicCS} or "Hashed Time-Locked Contracts (HTLCs)"\cite{Bribing2021TimelockedB} in the context of the Lightning Network, where two parties exchange Bitcoin without the need for a third-party intermediary.

        The hash lock works as follows:

\begin{itemize}
    \item A hash of a secret value is created.
    \item The transaction is locked using this hash.
    \item The recipient can only unlock the funds by providing the original secret that, when hashed, matches the pre-locked hash.
\end{itemize}

        \subsubsection{Time Locking}
        Time locking is another mechanism that delays the spending of a transaction until a specified time or block height\cite{PrasannaSrinivasan2021NetworkVI}. Bitcoin has two types of time locks:

\begin{itemize}
    \item \textbf{CheckLockTimeVerify (CLTV):} Prevents a transaction from being spent until a specified block height or time\cite{lightningTimelocksBuilders}.
    \item \textbf{CheckSequenceVerify (CSV):} Introduces relative time locking, where a transaction can only be spent after a certain number of blocks have passed\cite{lightningTimelocksBuilders}.
\end{itemize}

        An example use case for time locks is a payment that is locked until a future date, like an inheritance or savings account where Bitcoin cannot be withdrawn until a certain time has passed.

        The following pseudo-code demonstrates how a time-locked transaction can be created using Bitcoin Script:

        \begin{lstlisting}[style=cmd]
        OP_CHECKLOCKTIMEVERIFY
        OP_DROP
        OP_DUP
        OP_HASH160 <Public Key Hash>
        OP_EQUALVERIFY
        OP_CHECKSIG
        \end{lstlisting}

        In this example, the transaction will only be valid after the specified time condition has been met (defined by `OP\_CHECKLOCKTIMEVERIFY`).

    \section{Common Attack Methods}
        \subsection{Double-Spending Attack}
        A double-spending attack occurs when a malicious actor attempts to spend the same Bitcoin more than once\cite{Kumar2023ARO}. Bitcoin's decentralized nature and the delay between transaction broadcasts create a potential window for such an attack. If an attacker can convince one party to accept a transaction and then spends the same Bitcoin elsewhere before the network has confirmed the original transaction, they have effectively double-spent.

        Bitcoin mitigates this risk using its \textbf{Proof of Work (PoW)} consensus mechanism\cite{ferdous2020blockchainconsensusalgorithmssurvey}, where miners expend computational power to validate transactions and include them in blocks. Once a block is added to the blockchain, it becomes increasingly difficult to alter or reverse the transactions within that block. As more blocks are added on top, the security of those transactions increases.

        A typical example of double-spending prevention:

\begin{itemize}
    \item Alice sends Bitcoin to Bob and broadcasts the transaction.
    \item Miners compete to include this transaction in the next block.
    \item Once Bob sees the transaction in a confirmed block, he can be confident that the Bitcoin is truly his.
\end{itemize}

        The probability of a successful double-spend decreases significantly after multiple confirmations. Generally, six confirmations are considered safe, meaning the transaction is buried under six additional blocks\cite{Wan2019EvaluatingTI}.

        \subsection{Principle and Impact of a 51\% Attack}
        A 51\% attack refers to a situation where a single entity or group of miners gains control of more than 50\% of the network's total mining hash rate\cite{AponteNovoa2021The5A}. In such a scenario, the attacker could potentially:
        - Reverse recent transactions, allowing them to double-spend Bitcoin.
        - Prevent new transactions from gaining confirmations.
        - Halt the network temporarily by preventing new blocks from being mined.

        Although this type of attack is theoretically possible, it is extremely difficult and expensive to execute due to the immense computational power required to control more than half of the network. Moreover, executing such an attack would undermine the value of Bitcoin itself, thus disincentivizing miners from attempting it.

        For example, if an attacker manages to control 51\% of the Bitcoin network, they could send Bitcoin to a merchant, wait for confirmation, and then use their majority power to reorganize the blockchain, excluding the initial transaction and effectively double-spending those funds elsewhere.

        The impact of a 51\% attack would be catastrophic for trust in Bitcoin, but it is considered highly unlikely due to the decentralized and highly distributed nature of mining operations globally. To date, Bitcoin has successfully resisted such attacks, in part because of the economic and technical challenges involved.

        \textbf{Visualization of a 51\% attack}:

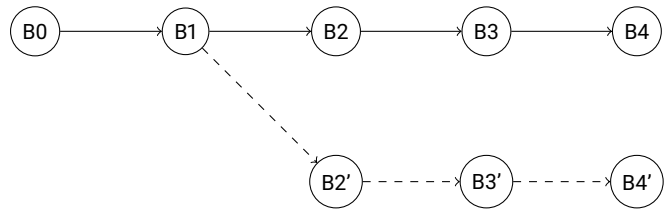
\begin{center}
        \begin{tikzpicture}
        [scale=1, every node/.style={scale=0.8}]
            \node (B0) at (0,0) [draw, circle] {B0};
            \node (B1) at (2,0) [draw, circle] {B1};
            \node (B2) at (4,0) [draw, circle] {B2};
            \node (B3) at (6,0) [draw, circle] {B3};
            \node (B4) at (8,0) [draw, circle] {B4};

            \node (B2a) at (4,-2) [draw, circle] {B2'};
            \node (B3a) at (6,-2) [draw, circle] {B3'};
            \node (B4a) at (8,-2) [draw, circle] {B4'};

            \draw[->] (B0) -- (B1);
            \draw[->] (B1) -- (B2);
            \draw[->] (B2) -- (B3);
            \draw[->] (B3) -- (B4);

            \draw[->, dashed] (B1) -- (B2a);
            \draw[->, dashed] (B2a) -- (B3a);
            \draw[->, dashed] (B3a) -- (B4a);
        \end{tikzpicture}
\end{center}

        In this diagram, an attacker creates an alternative chain (B2', B3', B4') and uses their control of the network to make it the longest chain, thus reversing previous transactions on the main chain.

\section{Bitcoin Privacy}
    \subsection{Pseudonymity and User Privacy}
    Bitcoin operates on a pseudonymous system, where users are not required to reveal their real identities. Instead, they are identified by unique addresses, which are derived from cryptographic keys\cite{Chang2020ImprovingBO}. Each Bitcoin address is a long string of alphanumeric characters, and it serves as a pseudonym for the user. 

    This pseudonymous nature provides a level of privacy, as no direct personal information is tied to the addresses. However, it is important to understand that this privacy is limited. Bitcoin’s blockchain is a public ledger, meaning all transactions are visible to anyone. Although the real identity behind an address is not immediately known, sophisticated analysis of the transaction patterns and interactions can potentially lead to the deanonymization of users.

    \textbf{Example:} Imagine Alice sends Bitcoin to Bob. Both Alice and Bob have pseudonymous Bitcoin addresses, let’s say Alice’s address is \texttt{1A1zP1...} and Bob’s address is \texttt{1BvBMSE...}. While their real identities are not directly tied to these addresses on the blockchain, anyone can see that an amount of Bitcoin was transferred from Alice’s address to Bob’s. With enough transaction data and analysis (for instance, connecting patterns from exchanges, services, or real-world interactions), a third party could eventually infer who Alice and Bob are.

    This is why Bitcoin provides only pseudonymity, not true anonymity. Users are therefore vulnerable to techniques like transaction analysis, which could reveal their identities.

    \subsection{Transaction Traceability and Coin Mixing Services}
    Every transaction made on the Bitcoin blockchain is traceable, thanks to the transparent and immutable nature of the public ledger. Each Bitcoin address and transaction is recorded and can be tracked back through the history of the blockchain, from the most recent transaction to the very first one. This transparency is one of the key features of Bitcoin, but it also poses privacy challenges for users who may want to keep their transaction history confidential.

    To address these concerns, several privacy-enhancing tools have emerged. One of the most popular methods is the use of coin mixing services, such as CoinJoin\cite{coinjoinCoinJoinBitcoin}. These services help obfuscate the trail of Bitcoin transactions by combining the transactions of multiple users into a single large transaction. This makes it difficult to trace the flow of funds from one user to another.

    \textbf{Example:} Alice, Bob, and Charlie all want to enhance the privacy of their Bitcoin transactions. Using a service like CoinJoin, they pool their transactions together into a single, larger transaction. On the blockchain, it will look like a single transaction occurred, with multiple inputs and outputs. This mixing of inputs and outputs makes it difficult for outside observers to determine which output belongs to which input, thereby enhancing privacy for all parties involved.

    However, it’s important to note that while coin mixing services can provide additional layers of privacy, they are not foolproof. In some cases, advanced analysis techniques can still identify patterns and connections, which may reveal the original sender or recipient. Furthermore, some services require trust in the service provider, introducing additional risks. For these reasons, users seeking absolute privacy may explore more advanced methods or consider alternative cryptocurrencies specifically designed with privacy features.

\section{Bitcoin Governance and Community}
    \subsection{Bitcoin Improvement Proposal (BIP) Process}
    Bitcoin’s development and improvement are guided by a structured process known as the Bitcoin Improvement Proposal (BIP)\cite{githubGitHubBitcoinbips}. The BIP process allows developers to propose new features, enhancements, or changes to the Bitcoin protocol in an organized and transparent manner. This system ensures that changes to Bitcoin are thoroughly vetted and reviewed by the community before they are implemented.

    The BIP process starts with a developer drafting a formal proposal. This document outlines the technical specifications and reasoning behind the proposed change. Once the draft is complete, it is submitted to the wider Bitcoin community for review. During this phase, other developers, miners, and users provide feedback, suggest improvements, or raise concerns.

    If a proposal gains enough support and passes the review process, it may be implemented into the Bitcoin protocol. However, it’s important to understand that the implementation of a BIP is not mandatory. The decentralized nature of Bitcoin means that each node on the network is free to choose whether or not to adopt the changes. If enough nodes adopt a new feature or update, it becomes part of the consensus protocol.

    \textbf{Example:} One of the most famous BIPs is \textbf{BIP 141}\cite{bitcoin0141Bitcoin}, which proposed the implementation of Segregated Witness (SegWit). SegWit introduced changes to the way transactions were structured, increasing the capacity of Bitcoin’s blocks and solving some issues related to transaction malleability. The proposal underwent extensive community discussion and testing before being adopted by the network.

    The BIP process is crucial for ensuring that changes to Bitcoin are well thought out and receive broad community support, maintaining the stability and security of the network.

    \subsection{Decentralized Governance of the Bitcoin Community}
    One of the defining characteristics of Bitcoin is its decentralized governance. Unlike traditional financial systems or organizations, there is no central authority that controls Bitcoin. Instead, the decision-making process is distributed across various stakeholders, including developers, miners, and users. Each group plays a significant role in shaping the future of Bitcoin.

    \textbf{Developers:} The developers maintain and improve the Bitcoin codebase. They propose changes through the BIP process, review code, and ensure the security of the network. However, developers alone cannot enforce changes; they rely on the wider community for support and adoption.

    \textbf{Miners:} Miners are responsible for securing the network by validating transactions and adding them to the blockchain. They also play a critical role in governance by signaling their support or opposition to proposed changes. For example, miners may indicate their support for a new protocol upgrade by including special data in the blocks they mine. If a significant majority of miners support an upgrade, it can be adopted by the network.

    \textbf{Users:} Bitcoin users, including individuals and businesses, also have a voice in the governance process. They can choose which version of the Bitcoin software to run on their nodes, effectively participating in the decision of which protocol rules to follow. This user choice is a powerful form of governance, as widespread user adoption is necessary for any protocol upgrade to succeed.

    Consensus in Bitcoin is achieved through collaboration and coordination among these different groups. For a protocol change to be implemented, it must gain broad support across the community. This decentralized governance model is designed to prevent any single entity from controlling the network, ensuring that Bitcoin remains open, secure, and resilient.

    \textbf{Example:} The 2017 block size debate highlighted how Bitcoin’s decentralized governance works\cite{Atzori2017BlockchainTA}. The community was divided on whether to increase the block size to allow for more transactions per block. Some developers and miners supported the change, while others opposed it, leading to a hard fork that resulted in the creation of Bitcoin Cash. This event showcased how the decentralized governance model allows for differences of opinion, with each faction free to pursue its vision of Bitcoin’s future.

\part{Ethereum and Smart Contracts}
\chapter{The Birth of Ethereum}
    \section{Historical Background of Ethereum}
        \subsection{Vitalik Buterin's Vision}

        Vitalik Buterin, a programmer and visionary in the blockchain space, created Ethereum in 2015\cite{Buterin2015ANG}. His journey towards developing Ethereum began in 2013 when he recognized the limitations of Bitcoin, the first and most popular cryptocurrency at the time. Bitcoin was primarily designed as a peer-to-peer electronic cash system, allowing for secure and decentralized money transfers. However, it was limited in its functionality, as its blockchain could only process simple transactions and had no ability to support complex applications.

        Buterin saw an opportunity to expand on Bitcoin’s concept by creating a new blockchain that could serve as a platform for not just cryptocurrency transactions, but also decentralized applications (dApps)\cite{Buterin2015ANG}. His vision was to build a system where developers could create applications that operate autonomously on a blockchain, with no reliance on intermediaries. These applications would use "smart contracts," which are self-executing contracts with the terms of the agreement directly written into code.

        In 2015, Ethereum was launched, providing developers with a more versatile and programmable blockchain platform. While Bitcoin’s blockchain was only capable of processing financial transactions, Ethereum allowed for a more extensive range of use cases, enabling the creation of decentralized applications, financial services, gaming, and even decentralized governance.

        \subsection{Turing Completeness and Ethereum's Design Goals}

        A key element of Ethereum's innovation is its Turing-complete programming environment\cite{Tikhomirov2017EthereumSO}. Turing completeness refers to a system's ability to perform any calculation or solve any problem that can be computed, provided enough time and resources are available. In simpler terms, a Turing-complete system can execute a full range of computational instructions\cite{Tikhomirov2017EthereumSO}.

        In contrast to Bitcoin’s limited scripting language, which could only handle basic transaction logic, Ethereum was designed to be a fully programmable platform. This Turing-complete environment allowed for complex logic to be coded into smart contracts and executed on the Ethereum blockchain. 

        The design goals of Ethereum centered around creating a decentralized, global computer that could:
        \begin{itemize}
            \item Execute smart contracts autonomously.
            \item Enable decentralized applications (dApps) that run without any central control.
            \item Provide developers with a flexible environment to build applications across various industries, such as finance, healthcare, and governance.
        \end{itemize}
        
        In this context, Ethereum became much more than just a cryptocurrency. It evolved into a platform that could support decentralized finance (DeFi), supply chain management, digital identity systems, and countless other applications, all without the need for intermediaries or centralized control.

    \section{Differences Between Ethereum and Bitcoin}
        \subsection{Smart Contracts and Decentralized Applications (dApps)}

        The primary difference between Ethereum and Bitcoin lies in Ethereum’s ability to support smart contracts and decentralized applications (dApps). 

        \textbf{Smart Contracts:} A smart contract is a self-executing piece of code with the terms of the agreement written directly into the code\cite{Ene2020SmartC}. When certain predefined conditions are met, the contract automatically executes without the need for intermediaries. For example, a smart contract could be written to automatically transfer ownership of a digital asset once payment is received. This automation eliminates the need for third parties like banks or escrow services.

        Here's a simple example of a smart contract written in Solidity, Ethereum’s programming language:
        \begin{lstlisting}[style=solidity]
        pragma solidity ^0.8.0;

        contract SimpleContract {
            address public owner;
            uint public balance;

            constructor() {
                owner = msg.sender;
                balance = 0;
            }

            function deposit() public payable {
                balance += msg.value;
            }

            function withdraw(uint amount) public {
                require(msg.sender == owner, "Only owner can withdraw");
                require(amount <= balance, "Insufficient funds");
                payable(owner).transfer(amount);
                balance -= amount;
            }
        }
        \end{lstlisting}
        
        In this example, the smart contract manages deposits and withdrawals. The contract’s logic ensures that only the owner can withdraw funds, and only if the balance is sufficient.

        \textbf{Decentralized Applications (dApps):} Developers can build decentralized applications (dApps) on Ethereum\cite{Ta2019BuildingAD}, which operate without centralized servers. These dApps are powered by smart contracts, enabling them to function automatically based on the code written into the blockchain. This allows for a wide range of applications, from decentralized finance (DeFi) platforms to games and social networks, all running without a central authority.

        \subsection{Role of the Ethereum Virtual Machine (EVM)}

        At the heart of Ethereum's functionality is the Ethereum Virtual Machine (EVM)\cite{Xu2018BuildingAE}. The EVM is a global, decentralized computer that enables developers to execute smart contracts on the Ethereum blockchain. It provides a runtime environment for smart contracts, ensuring that they are executed consistently across all nodes in the network.

        Every Ethereum node runs its own copy of the EVM, and when a smart contract is deployed or a transaction is executed, the EVM processes the code in the contract. The decentralized nature of the EVM guarantees that all smart contracts run in a consistent manner, regardless of which node is processing the transaction. This consistency ensures that the output of a smart contract on one node will be the same on every other node.

        Without the EVM, Ethereum would not be able to function as a decentralized platform for smart contracts and dApps. It is what allows Ethereum to serve as a "world computer," where anyone can deploy code that will run autonomously, without interference or downtime.

        \subsection{Ethereum's Gas Mechanism}

        One of Ethereum’s critical innovations is the concept of "Gas."\cite{Albert2018RunningOF} Gas is a unit that measures the computational work required to execute transactions and smart contracts on the Ethereum network. 

        When a user wants to perform a transaction, deploy a smart contract, or interact with a dApp, they must pay a certain amount of Gas. The amount of Gas required depends on the complexity of the operation. For example, a simple Ether transfer might cost a small amount of Gas, while executing a complex smart contract could require much more.

        Gas is paid in Ether, and the cost of Gas is calculated using the following formula:
        \begin{equation}
        \text{Total Gas Cost} = \text{Gas Limit} \times \text{Gas Price}
        \end{equation}

        \begin{itemize}
            \item \textbf{Gas Limit:} The maximum amount of Gas that the user is willing to pay for a transaction.
            \item \textbf{Gas Price:} The price the user is willing to pay per unit of Gas, usually denominated in gwei (1 gwei = $10^{-9}$ Ether).
        \end{itemize}

        Gas plays a crucial role in Ethereum’s ecosystem for several reasons:
        \begin{itemize}
            \item It prevents network abuse by requiring users to pay for computational resources. Without Gas, malicious users could run infinite loops or highly complex computations, causing the network to slow down or crash.
            \item It incentivizes miners (or validators, in Ethereum 2.0) to prioritize and process transactions. Users can set a higher Gas price to ensure their transaction is processed more quickly.
        \end{itemize}

        The Gas mechanism is essential for maintaining the stability, security, and efficiency of the Ethereum network. By attaching a cost to computational work, Ethereum ensures that resources are used judiciously and that the network remains operational even as it scales\cite{Yang2019EmpiricallyAE}.

\chapter{Principles and Implementation of Smart Contracts}

\section{Definition of Smart Contracts}
Smart contracts are digital agreements that are automatically enforced by code. Unlike traditional contracts, which depend on third parties (such as lawyers or courts) for enforcement, smart contracts are self-executing. This concept was first introduced by Nick Szabo in the 1990s\cite{Ene2020SmartC}.

\subsection{Nick Szabo's Concept of Smart Contracts}
Nick Szabo is a computer scientist, legal scholar, and cryptographer. In the 1990s, Szabo proposed the idea of smart contracts to improve the way agreements are made and enforced\cite{Laubscher2020SmartC}. He described smart contracts as a set of promises, specified in digital form, including protocols within which the parties perform on these promises. 

The core idea behind Szabo's concept was that contractual clauses can be embedded in hardware and software systems, making them self-enforcing\cite{Szabo1997FormalizingAS}. For example, in a vending machine, a customer inserts coins and selects a product, and the machine automatically delivers the product when the correct amount is inserted. In this analogy, the machine acts as a smart contract, enforcing the terms (payment and delivery) without the need for a human intermediary. This concept has been greatly expanded with the advent of blockchain technology.

\subsection{Automatic Execution and Self-Enforcement of Smart Contracts}
One of the key features of smart contracts is their ability to automatically execute and self-enforce when predefined conditions are met. For instance, in a real estate transaction, if the buyer deposits the required amount of cryptocurrency to the contract, the ownership of the property is automatically transferred to them. There is no need for a third party to verify or approve the transaction.

This automation is made possible through blockchain technology. Blockchain is a distributed ledger where each node (participant) has a copy of the entire ledger, ensuring that the contract's terms are visible and verifiable by everyone. When conditions are fulfilled, the smart contract automatically executes the transaction, thereby eliminating the need for intermediaries like banks, lawyers, or escrow services.

For example, consider the following Solidity code for a simple smart contract that releases payment after a condition is met:

\begin{lstlisting}[style=solidity]
pragma solidity ^0.8.0;

contract ConditionalPayment {
    address payable public recipient;
    uint public releaseTime;

    constructor(address payable _recipient, uint _releaseTime) {
        recipient = _recipient;
        releaseTime = _releaseTime;
    }

    function releasePayment() public payable {
        require(block.timestamp >= releaseTime, "Cannot release yet");
        recipient.transfer(address(this).balance);
    }
}
\end{lstlisting}

In this example, the smart contract automatically transfers funds to the recipient when the specified time is reached, without the need for manual intervention.

\section{Smart Contracts on Ethereum}

\subsection{Writing and Deploying Smart Contracts}
On Ethereum, smart contracts are written using a programming language called Solidity\cite{soliditylangHomeSolidity}. Developers create smart contracts by writing code that specifies the contract's terms and conditions. Once the code is written, it is compiled and deployed to the Ethereum blockchain.

Deploying a smart contract involves creating a special transaction that includes the contract's bytecode. This transaction is broadcast to the Ethereum network, and when it is mined, the contract is stored on the blockchain. Once deployed, the contract is assigned a unique address, which can be used by other users to interact with it.

Users interact with smart contracts by sending transactions to the contract's address. These transactions can trigger functions within the contract, allowing users to engage with the contract's logic. Each transaction is recorded on the blockchain, ensuring transparency and immutability.

Here's an example of a simple contract deployment:

\begin{lstlisting}[style=solidity]
pragma solidity ^0.8.0;

contract SimpleStorage {
    uint public storedData;

    function set(uint x) public {
        storedData = x;
    }

    function get() public view returns (uint) {
        return storedData;
    }
}
\end{lstlisting}

In this contract, users can set and retrieve a value. To deploy this contract, a developer would compile the Solidity code and send the bytecode as a transaction to the Ethereum network.

\subsection{Solidity Programming Language}
Solidity is the most widely-used language for writing smart contracts on Ethereum\cite{ethereumCompleteGuide}. It is a statically-typed language that is designed to work seamlessly with the Ethereum Virtual Machine (EVM). Solidity features a syntax similar to JavaScript, which makes it approachable for developers familiar with web development.

The language is tailored for writing contracts with specific features such as:

\begin{itemize}
    \item Data types (integers, strings, etc.)
    \item Functions (public, private, view, etc.)
    \item Modifiers (conditions placed on function execution)
    \item Event logging (to trigger external notifications)
    \item Error handling (using \texttt{require}, \texttt{assert}, and \texttt{revert} functions)
\end{itemize}

For example, the following Solidity contract demonstrates how you can implement a voting system:

\begin{lstlisting}[style=solidity]
pragma solidity ^0.8.0;

contract Voting {
    mapping(address => uint) public votesReceived;
    address[] public candidates;

    function addCandidate(address candidate) public {
        candidates.push(candidate);
    }

    function vote(address candidate) public {
        votesReceived[candidate] += 1;
    }

    function getVotes(address candidate) public view returns (uint) {
        return votesReceived[candidate];
    }
}
\end{lstlisting}

\subsection{Working Mechanism of the Ethereum Virtual Machine (EVM)}
The Ethereum Virtual Machine (EVM) is the decentralized computational environment that executes smart contracts on the Ethereum network\cite{ethereumEthereumVirtual}. Every node on the Ethereum blockchain runs its own copy of the EVM, ensuring that the execution of contracts is consistent and verifiable across all nodes.

When a transaction is sent to a smart contract, the EVM processes the contract's code to determine the appropriate actions based on the input parameters and the contract's state. The EVM ensures that all operations are deterministic, meaning that they will produce the same result regardless of where and when they are executed.

Each smart contract execution consumes "gas," which is a measure of computational effort. The gas cost ensures that developers write efficient code and prevents abuse of the network by limiting the number of operations that can be executed within a single transaction.

\section{Use Cases of Smart Contracts}

\subsection{Decentralized Finance (DeFi)}
Smart contracts have revolutionized the financial industry through the creation of Decentralized Finance (DeFi) applications\cite{Jensen2021AnIT}. DeFi allows users to access traditional financial services like lending, borrowing, and trading without intermediaries like banks or brokers. All transactions are governed by smart contracts, ensuring transparency and security.

For example, platforms like Aave\cite{aaveAave} and Compound\cite{compoundv2CompoundUnderstanding} allow users to lend and borrow cryptocurrency using smart contracts. Users deposit funds into smart contract-based pools, and borrowers can take loans by locking collateral into the same system. Interest rates are determined algorithmically based on supply and demand, and all transactions are executed automatically by the contracts.

\subsection{Decentralized Autonomous Organizations (DAOs)}
Decentralized Autonomous Organizations (DAOs) are organizations that are governed by smart contracts rather than centralized entities\cite{Ding2023ASO}. In a DAO, decision-making is automated and decentralized. Members can vote on proposals, and the rules for decision-making are enforced by smart contracts.

For instance, a DAO might be used to manage a shared investment fund. Members contribute funds to the DAO, and smart contracts control how the funds are invested or distributed based on the outcome of votes. One well-known example of a DAO is the MakerDAO\cite{makerdaoMakerDAOUnbiased}, which governs the DAI stablecoin.

\subsection{Gaming and Non-Fungible Tokens (NFTs)}
Smart contracts have also made a significant impact on the gaming industry through the creation of Non-Fungible Tokens (NFTs)\cite{Arora2021SmartCA}. NFTs represent unique digital assets, such as in-game items, artwork, or collectibles, that can be bought, sold, and traded on blockchain platforms.

Smart contracts power NFTs by managing their creation, transfer, and ownership. For example, in games like "Axie Infinity,"\cite{axieinfinityAxieInfinity} players can collect and battle unique creatures (represented as NFTs). Each creature is distinct, and its ownership is recorded on the blockchain, ensuring scarcity and authenticity.

The following is an example of a smart contract used to create NFTs using the ERC721 standard\cite{ethereumERC721NonFungible}:

\begin{lstlisting}[style=solidity]
pragma solidity ^0.8.0;

import "@openzeppelin/contracts/token/ERC721/ERC721.sol";

contract MyNFT is ERC721 {
    uint public nextTokenId;
    address public admin;

    constructor() ERC721('MyNFT', 'MNFT') {
        admin = msg.sender;
    }

    function mint(address to) external {
        require(msg.sender == admin, 'only admin can mint');
        _safeMint(to, nextTokenId);
        nextTokenId++;
    }
}
\end{lstlisting}
This contract allows the creation of unique NFTs, which can be traded or used within various applications.

\chapter{Development and Challenges of Ethereum}

\section{The DAO Incident}

    \subsection{The Intent and Design of The DAO}
    The DAO (Decentralized Autonomous Organization) was one of the most ambitious projects launched on the Ethereum platform in 2016\cite{Mehar2017UnderstandingAR}. Its original intent was to create a decentralized investment fund that would operate entirely through smart contracts, without any intermediaries. The vision was to allow investors to pool their funds into The DAO, and vote on which projects should receive funding. This organization was revolutionary because it relied entirely on code to manage the fund’s governance, removing the need for a centralized authority\cite{DuPont2017ExperimentsIA}.

    The design of The DAO was simple in concept but complex in implementation. Participants could buy DAO tokens, which represented voting power in the organization\cite{PeaCalvin2024ConcentrationOP}. Every decision was to be made through token-holder voting, and smart contracts would automatically execute the decisions based on the outcome of these votes. For example, if a project proposal was approved by a majority vote, funds would automatically be sent to the project team, all without human intervention.

    One of the key advantages of The DAO was transparency and trustlessness\cite{Kondova2019GovernanceOD} — since all of its operations were governed by code on the Ethereum blockchain, anyone could inspect the smart contract to verify how funds would be managed. In theory, this was a major step forward in decentralized governance, offering the possibility of creating a new way to fund and govern projects democratically.

    \subsection{Vulnerability Exploitation and Hacker Attack}
    Despite its ambitious design, The DAO’s code had a critical vulnerability. A flaw in the smart contract logic allowed an attacker to exploit a reentrancy bug\cite{Liu2018ReGuardFR}, which enabled them to repeatedly withdraw funds from The DAO without properly updating the balance. This exploit worked by making recursive calls to the smart contract, draining funds each time before the contract could update its records to reflect the withdrawal.

    The hacker was able to siphon off approximately 3.6 million Ether, worth about \$60 million at the time\cite{Dingman2019DefectsAV}. Here is a simplified explanation of the vulnerability:

    \begin{lstlisting}[style=solidity]
    // Vulnerable withdrawal function in The DAO
    function withdraw(uint amount) public {
        if(balance[msg.sender] >= amount) {
            msg.sender.call.value(amount)();
            balance[msg.sender] -= amount;
        }
    }
    \end{lstlisting}

    In the above code, the key issue is the order of operations. The contract sends the Ether to the user before updating their balance, allowing the attacker to recursively call the `withdraw` function, draining funds multiple times before the balance is updated.

    This incident had significant implications for the Ethereum community, raising concerns about the security of smart contracts and decentralized applications (dApps). The attack brought to light the risks associated with complex smart contract systems and the need for rigorous security audits.

    \subsection{Hard Fork: Ethereum and Ethereum Classic}
    In the aftermath of the hack, the Ethereum community faced a difficult decision. To address the situation and return the stolen funds, Ethereum developers proposed a hard fork\cite{Mehar2017UnderstandingAR}. A hard fork is a radical change to the blockchain protocol that results in the creation of two separate chains. In this case, the hard fork would effectively reverse the transactions made by the hacker, restoring the stolen Ether to its original owners.

    However, the proposal was highly controversial. While many supported the hard fork as a necessary step to undo the damage caused by the hack, others argued that it violated the principle of blockchain immutability—the idea that once a transaction is confirmed, it should never be changed. As a result, the community split into two camps:

    \begin{itemize}
        \item \textbf{Ethereum (ETH)}: This chain implemented the hard fork, effectively rolling back the DAO hack and returning the stolen funds to the original owners.
        \item \textbf{Ethereum Classic (ETC)}: This chain rejected the hard fork, continuing on the original blockchain that included the hacker’s transactions. The ETC community upholds the principle of immutability, believing that "code is law" and that the blockchain should remain unchanged.
    \end{itemize}

    The split led to the coexistence of two Ethereum-based blockchains, each with its own set of supporters and philosophies. Ethereum (ETH) remains the more popular and widely adopted blockchain, while Ethereum Classic (ETC) continues to exist as a reminder of the ideological divide within the Ethereum community.

\section{Ethereum 2.0 and Proof of Stake (PoS)}

    \subsection{Design Goals of Ethereum 2.0}
    Ethereum 2.0, also known as Eth2 or "The Merge," is a major upgrade aimed at improving the scalability, security, and sustainability of the Ethereum network\cite{Asif2023ShapingTF}. The current Ethereum blockchain, based on Proof of Work (PoW), has faced limitations such as high energy consumption, slow transaction speeds, and network congestion. Ethereum 2.0 seeks to address these issues through several key improvements:

    \begin{itemize}
        \item \textbf{Scalability}: By increasing the network’s capacity to process more transactions per second, Ethereum 2.0 aims to alleviate the bottlenecks that currently cause high transaction fees and slow processing times.
        \item \textbf{Security}: Ethereum 2.0 introduces new cryptographic techniques and mechanisms to make the network more resistant to attacks.
        \item \textbf{Sustainability}: By moving away from the energy-intensive Proof of Work (PoW) consensus mechanism to the more efficient Proof of Stake (PoS), Ethereum 2.0 dramatically reduces its carbon footprint.
    \end{itemize}

    These changes will ensure that Ethereum can continue to grow and support a wide range of decentralized applications (dApps) without being hampered by performance issues or environmental concerns.

    \subsection{Transition from PoW to PoS}
    The most significant change in Ethereum 2.0 is the transition from Proof of Work (PoW) to Proof of Stake (PoS) as the consensus mechanism\cite{Asif2023ShapingTF}. In PoW, miners use computational power to solve complex mathematical puzzles and validate transactions. However, this approach is highly energy-intensive, leading to criticism about its environmental impact.

    Proof of Stake (PoS) offers an alternative that requires far less energy. In PoS, instead of miners, there are validators who are chosen to create new blocks and validate transactions based on the amount of cryptocurrency they have staked as collateral\cite{Wendl2022TheEI}. The more tokens a validator stakes, the higher the chance they will be selected to validate the next block. This mechanism reduces the need for computational power and lowers energy consumption.

    \begin{lstlisting}[style=python]
    # Example of selecting a validator in Proof of Stake:
    import random

    def select_validator(validators):
        total_stake = sum(validator['stake'] for validator in validators)
        selection = random.uniform(0, total_stake)
        current = 0
        for validator in validators:
            current += validator['stake']
            if current >= selection:
                return validator['name']

    validators = [
        {'name': 'Validator1', 'stake': 100},
        {'name': 'Validator2', 'stake': 200},
        {'name': 'Validator3', 'stake': 300}
    ]

    print(select_validator(validators))  # Randomly selects a validator based on their stake
    \end{lstlisting}

    PoS also reduces the risk of a 51

    \subsection{Sharding and Scalability}
    A key innovation of Ethereum 2.0 is the introduction of sharding, a technique designed to improve scalability\cite{CortesGoicoechea2020ResourceAO}. Sharding divides the blockchain into smaller pieces, called shards, which can process transactions and smart contracts in parallel. This allows the network to handle a significantly larger volume of transactions, improving overall throughput.

    In a traditional blockchain, every node processes every transaction, which limits the network’s speed. With sharding, different nodes are assigned to different shards, so they only need to process a subset of transactions. This increases the capacity of the network without requiring each node to maintain the entire blockchain.

    Imagine Ethereum as a city with a single highway (traditional blockchain). Traffic (transactions) moves slowly because all vehicles (nodes) use the same road. Sharding adds multiple lanes (shards), allowing cars to move faster and reducing congestion.

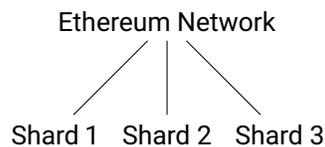
\begin{center}
    \begin{tikzpicture}
    \node {Ethereum Network}
        child {node {Shard 1}}
        child {node {Shard 2}}
        child {node {Shard 3}};
    \end{tikzpicture}
\end{center}

    Each shard operates independently, but they are interconnected, allowing transactions to be shared and validated across the entire network. This architecture enables Ethereum 2.0 to scale while maintaining security and decentralization.

\section{Scalability Challenges of Ethereum}
Ethereum, as one of the most popular blockchain platforms, faces significant scalability challenges\cite{Bez2019TheSC}. As the number of users and decentralized applications (dApps) increases, the network's limitations become more evident. These challenges can lead to high gas fees, slow transaction times, and network congestion. In this section, we will explore the key scalability issues Ethereum faces and examine some of the proposed solutions to improve its performance.

    \subsection{High Gas Fees}
    Gas fees are a critical aspect of the Ethereum network. They are necessary to compensate miners for the computational resources they use to validate transactions and execute smart contracts. However, gas fees can become prohibitively expensive, especially during periods of high demand. This section will explain the gas mechanism, the reasons for the surge in gas fees, and strategies to optimize gas usage.
    
        \subsubsection{How the Gas Mechanism Works}
        Gas in Ethereum measures the computational effort required to execute a transaction or smart contract\cite{Khan2021GasCA}. Every operation in the Ethereum Virtual Machine (EVM), from basic arithmetic to storing data on the blockchain, has a specific gas cost. The total gas cost for a transaction is determined by summing the costs of all the operations required to process it.
        
        When a user initiates a transaction, they specify the amount of gas they are willing to pay for its execution\cite{Fajge2021WaitOR}. Miners prioritize transactions that offer higher gas fees since it is more profitable for them. This market-driven mechanism ensures that network resources are allocated efficiently, with users paying more during periods of high demand to have their transactions processed faster.
        
        For example, if Alice wants to send 1 ETH to Bob, she needs to set both a gas limit (the maximum amount of gas she is willing to spend) and a gas price (the amount she is willing to pay per unit of gas). If the total gas required for the transaction is less than or equal to the gas limit Alice set, her transaction will succeed. Otherwise, it will fail, but Alice will still have to pay for the gas used up to the point of failure.
        
        \subsubsection{Reasons for the Surge in Gas Fees}
        Several factors contribute to the rise in gas fees on the Ethereum network. The primary reasons include:
        
        \begin{itemize}
            \item \textbf{Network Congestion:} During times of high activity, such as popular dApp launches or NFT drops, the Ethereum network can become congested. With more users competing to have their transactions processed, gas prices rise as people are willing to pay more to get their transactions included in a block\cite{Ko2022AnAO}.
            \item \textbf{High Demand for dApps:} Decentralized applications (dApps) like decentralized finance (DeFi) platforms and non-fungible token (NFT) marketplaces place significant demand on the network. Popular platforms like Uniswap\cite{uniswapUniswapInterface} or OpenSea\cite{opensea} can cause a spike in gas fees, as they require multiple complex smart contract interactions that consume a lot of gas.
            \item \textbf{Limited Block Space:} Each block in Ethereum has a maximum gas limit, meaning there is a finite amount of computational work that can be included in each block. When the demand exceeds this limit, users have to compete by paying higher fees to get their transactions included.
        \end{itemize}
        
        The surge in gas fees can have a negative impact on the user experience, making small transactions economically unfeasible. It also limits the scalability of the network, as higher transaction costs deter users from engaging with dApps and other blockchain activities.
        
        \subsubsection{Gas Optimization Strategies and Improvement Proposals}
        Several strategies can be employed to optimize gas usage and reduce fees. These strategies include:
        
        \begin{itemize}
            \item \textbf{Code Optimization:} Smart contract developers can optimize their code to minimize gas usage. For example, they can avoid redundant operations, use more efficient data structures, or batch multiple operations into a single transaction to reduce gas costs.
            \item \textbf{Batching Transactions:} Batching allows users to group multiple transactions into one, saving on the total gas costs\cite{Wang2021iBatchSE}. This is especially useful for frequent users or dApps that perform many small transactions.
            \item \textbf{EIP-1559:} One of the most significant proposals to address high gas fees is Ethereum Improvement Proposal (EIP) 1559\cite{ethereumEIP1559Market}. This proposal introduces a base fee that is burned, reducing the total supply of Ether over time, and a tip that incentivizes miners to include transactions. EIP-1559 also makes gas fees more predictable, reducing the chances of extreme price spikes during network congestion.
        \end{itemize}
        
        The ongoing development of Ethereum 2.0, which will introduce a proof-of-stake (PoS) consensus mechanism and shard chains, is expected to further alleviate gas fee issues by increasing the network's capacity and throughput\cite{Chen2023ComparisonOP}.

    \subsection{Exploration of Layer 2 Solutions}
    Layer 2 solutions are scaling methods that operate on top of the Ethereum main chain\cite{Neiheiser2023PracticalLO}. They help offload transaction processing from the main chain, reducing congestion and lowering gas fees. In this section, we will explore several Layer 2 technologies\cite{Neiheiser2023PracticalLO}, including Rollups, state channels, and sidechains, which are designed to improve Ethereum's scalability.

        \subsubsection{Rollup Technologies (Optimistic Rollups and ZK-Rollups)}
        Rollups are one of the most promising Layer 2 solutions for Ethereum scalability\cite{Thibault2022BlockchainSU}. They allow transactions to be processed off-chain and then submit only compressed data or proofs to the Ethereum main chain. This significantly reduces the amount of data stored on-chain and increases transaction throughput.
        
        There are two main types of Rollups:
        
        \begin{itemize}
            \item \textbf{Optimistic Rollups:} These Rollups assume that all transactions are valid by default\cite{9862815}. They rely on fraud proofs, where users can challenge suspicious transactions within a specific timeframe. If a fraud is detected, the fraudulent transaction is rolled back, and the user who submitted the fraud proof is rewarded.
            \item \textbf{ZK-Rollups (Zero-Knowledge Rollups):} ZK-Rollups use cryptographic proofs (called zero-knowledge proofs) to validate transactions\cite{Fernando2023PosterWA}. These proofs ensure that the off-chain transactions are valid without the need to trust any third party. ZK-Rollups are generally faster and more efficient than Optimistic Rollups but are more complex to implement.
        \end{itemize}
        
        Rollups have gained significant traction in the Ethereum ecosystem due to their ability to reduce gas fees and increase transaction throughput while maintaining security. For example, Arbitrum\cite{arbtirumArbitrumScalable} and Optimism\cite{optimismHome} are leading platforms that use Optimistic Rollups, while zkSync\cite{zksyncZKsync} and StarkWare\cite{starkwareHomepage} leverage ZK-Rollups.
        
        \subsubsection{Application of State Channels}
        State channels allow participants to conduct multiple off-chain transactions, with only the final state being committed to the blockchain\cite{McCorry2019YouSM}. This method is highly efficient for repeated interactions, such as payments or gaming applications, where the parties involved can settle their accounts off-chain and only submit the final balance to Ethereum.
        
        For example, imagine two users, Alice and Bob, engaging in a series of micro-transactions. Instead of submitting every transaction to the Ethereum network, they can open a state channel. They can interact freely off-chain, and once they decide to close the channel, only the final transaction reflecting the net balance will be posted on-chain. This drastically reduces gas costs and enhances scalability.
        
        \subsubsection{Plasma and Sidechain Solutions}
        Plasma is a framework that allows for scalable and secure off-chain transactions\cite{Poon2017PlasmaS}. Plasma chains are separate blockchains that run alongside the Ethereum main chain. They handle large volumes of transactions off-chain and interact with the main chain for security and settlement. Plasma chains can process thousands of transactions per second while ensuring that users can always retrieve their funds from the main chain if something goes wrong with the Plasma chain.
        
        Sidechains\cite{Singh2020SidechainTI}, like Plasma, operate independently of the Ethereum main chain but still interact with it for finality and security. Sidechains have their own consensus mechanisms and can offer faster transaction times and lower fees. A popular example of a sidechain is Polygon (formerly Matic), which has become one of the leading Layer 2 solutions by providing fast and low-cost transactions.
        
    \subsection{Network Load from Decentralized Applications (dApps)}
    Decentralized applications (dApps) are one of the key drivers of Ethereum's popularity but also a significant source of scalability challenges. As more users interact with dApps, especially during periods of high demand, the Ethereum network can struggle to handle the load efficiently.

        \subsubsection{Scalability Issues of dApps}
        dApps can suffer from the same scalability issues as the Ethereum network itself, including high gas fees and slow transaction times. Popular dApps, particularly those in the DeFi space, often experience congestion during peak periods, making it difficult for users to interact with them affordably. For instance, during the DeFi summer of 2020, gas fees skyrocketed as users rushed to participate in yield farming\cite{Cousaert2021SoKYA}, making smaller transactions prohibitively expensive.
        
        \subsubsection{Challenges Posed by High-Frequency Transactions}
        High-frequency activities, such as algorithmic trading on decentralized exchanges (DEXs) or frequent NFT minting, can place additional strain on the network\cite{Brolley2022OnDemandFT}. These types of activities require a high number of transactions in a short period, leading to network congestion and increased gas prices.
        
        \subsubsection{Integration of Layer 2 with dApps}
        To overcome scalability challenges, many dApps are beginning to integrate Layer 2 solutions such as Rollups, state channels, and Plasma. For example, Uniswap\cite{uniswapUniswapInterface}, a leading decentralized exchange, has integrated with Optimistic Rollups to offer faster and cheaper trades. Similarly, gaming dApps are using state channels to allow users to interact with the platform off-chain, reducing the cost of frequent in-game transactions.

\chapter{Proof of Stake (PoS)}
    \section{Basic Concepts of PoS}
        \subsection{Definition of Proof of Stake}
        Proof of Stake (PoS) is a consensus mechanism that is utilized in blockchain networks to determine how the next block in a blockchain is chosen and validated\cite{Wendl2022TheEI}. Unlike Proof of Work (PoW), where participants (miners) solve complex cryptographic puzzles to create new blocks, PoS selects validators based on the amount of cryptocurrency they have "staked" or locked up in the network\cite{Trivedi2023EnhancedSF}.

        In PoS, the idea is that those who have a larger stake in the network have more to lose if the system is compromised\cite{Trivedi2023EnhancedSF}. Hence, validators are selected to propose and validate new blocks in a way proportional to the amount of cryptocurrency they hold and are willing to stake. Validators are chosen through a pseudo-random selection process, with factors such as the size of the stake and the validator's age (the time they have been staking) playing a role.

        This method eliminates the need for intensive computational work, which is one of the main criticisms of PoW\cite{Miraz2021EvaluationOG}. Instead, PoS is considered energy-efficient, as it removes the requirement for high energy consumption seen in mining operations.

        \subsection{Ensuring Network Security through Staking}
        The PoS mechanism ensures network security by requiring validators to put their funds at risk. This act of staking is crucial because it aligns the interests of validators with the health and integrity of the blockchain.

        Validators are incentivized to act honestly because if they attempt to manipulate the network or validate fraudulent transactions, they risk losing part or all of their stake. This is known as a "slashing" mechanism, where a portion of the validator's staked funds is forfeited as a penalty for malicious behavior. Therefore, validators have a strong financial motivation to maintain the network's integrity.

        Additionally, validators are rewarded with transaction fees or newly minted cryptocurrency for their work in validating blocks. The combination of rewards for honest behavior and penalties for malicious actions helps maintain network security in PoS systems.

    \section{Comparison Between PoS and PoW}
        \subsection{Energy Consumption Comparison}
        One of the most significant differences between Proof of Stake (PoS) and Proof of Work (PoW) is their energy consumption. PoW relies on miners using substantial amounts of computational power to solve cryptographic puzzles, which requires a large energy expenditure. This has led to criticism, especially with popular PoW blockchains like Bitcoin consuming more energy than entire countries.

        In contrast, PoS is designed to be far more energy-efficient. Because validators in PoS are chosen based on the amount they have staked, there is no need for computationally intensive mining. As a result, PoS can significantly reduce the environmental impact of blockchain networks. For example, Ethereum's switch from PoW to PoS with Ethereum 2.0 is projected to reduce the energy consumption of the network by over 99\% \cite{Kapengut2022AnES} .

        \subsection{Network Security Comparison}
        The security models of PoS and PoW differ significantly. In PoW, the security of the network relies on the computational power of miners. An attack vector in PoW is the 51\% attack, where an entity that controls more than 51\% of the network's hash rate can theoretically manipulate the blockchain by double-spending or preventing transactions\cite{Sayeed2019AssessingBC}.

        PoS, on the other hand, secures the network by relying on economic incentives. The primary attack vector in PoS is the "Nothing at Stake" problem\cite{Li2017SecuringPB}, where validators may attempt to validate multiple chains simultaneously without any risk. However, PoS addresses this issue by introducing penalties such as slashing\cite{Roughgarden2024KeynotePS}, where dishonest validators can lose their staked funds if they try to attack the network.

        Although both PoS and PoW have different security risks, PoS provides a more energy-efficient solution with mechanisms in place to discourage malicious behavior.

        \subsection{Impact on Centralization and Decentralization}
        Both PoW and PoS face challenges related to centralization, but in different ways. PoW tends to favor those with access to more powerful hardware and cheaper electricity\cite{Romiti2019ADD}, leading to the rise of large mining pools that can dominate the network.

        In PoS, centralization can occur due to wealth concentration\cite{He2020StakingPC}. Since validators are chosen based on their stake, those with more cryptocurrency have a higher chance of being selected, which can lead to the rich becoming richer. However, some PoS systems implement measures to encourage decentralization, such as limiting the influence of any single validator and ensuring that smaller validators have a chance to participate\cite{Li2023RewardDI}.

    \section{Applications of PoS}
        \subsection{PoS Mechanism in Ethereum 2.0}
        Ethereum 2.0 marks one of the most prominent examples of a major blockchain transitioning from PoW to PoS\cite{Asif2023ShapingTF}. In Ethereum 2.0, participants must lock up at least 32 ETH to become a validator. Validators are responsible for proposing new blocks and validating transactions, and they are rewarded for doing so correctly.

        Ethereum 2.0 introduces sharding\cite{CortesGoicoechea2020ResourceAO}, which divides the blockchain into smaller parts called "shards" to improve scalability. By combining PoS with sharding, Ethereum aims to increase the number of transactions it can handle while also significantly reducing its energy consumption compared to PoW.

        Validators in Ethereum 2.0 are selected randomly to propose and attest to blocks, ensuring fairness and decentralization while securing the network through staking.

        \subsection{Other Blockchain Projects Using PoS}
        Many other blockchain projects also implement PoS as their consensus mechanism. Some examples include:

        \begin{itemize}
            \item \textbf{Cardano}: Cardano\cite{cardanoHomeCardanoorg} uses a PoS protocol called Ouroboros, where validators (or stake pool operators) are randomly chosen to validate blocks. Cardano emphasizes a research-driven approach to PoS and provides a unique staking delegation model, allowing users to delegate their stake to stake pools without losing control of their funds.
            \item \textbf{Polkadot}: Polkadot\cite{pokladotPolkadotInteroperable} employs a Nominated Proof of Stake (NPoS) system, where nominators back validators with their stake. This system promotes decentralization by distributing staking power among multiple validators.
            \item \textbf{Tezos}: Tezos\cite{tezosHomeTezos} uses a variation of PoS called Liquid Proof of Stake (LPoS), where validators, known as "bakers," are selected to validate transactions. Tezos also allows users to delegate their staking power without locking up their funds, providing flexibility to participants.
        \end{itemize}

        Each of these projects introduces unique features and governance models, but they all rely on PoS to secure their networks and reduce energy consumption compared to PoW-based blockchains.

    \section{Challenges and Improvements of PoS}
        \subsection{The “Nothing at Stake” Problem}
        One of the potential issues in PoS is the "Nothing at Stake" problem\cite{Li2017SecuringPB}. Since validators in PoS do not need to expend significant resources to validate blocks, they might be tempted to validate multiple chains, hoping that one will eventually be accepted by the network. This could undermine the security of the blockchain.

        To address this, many PoS systems implement "slashing" mechanisms. Validators who attempt to validate multiple chains or act maliciously can lose part or all of their stake, making this attack financially unappealing.

        \subsection{Long-Range Attacks and Their Prevention}
        Another challenge in PoS is long-range attacks\cite{Deirmentzoglou2019ASO}, where an attacker with control of old validator keys attempts to rewrite the blockchain history. These attacks are particularly concerning for PoS because validators are chosen based on stake, and old keys could be used to manipulate the chain.

        To prevent long-range attacks, blockchain systems employ strategies such as checkpointing\cite{Azouvi2022PikachuSP}. Checkpoints are specific points in the blockchain that are considered final, and the history before these checkpoints cannot be altered. This helps protect the blockchain from attackers attempting to rewrite older blocks.

        \subsection{Future Directions for PoS Improvement}
        Although PoS has proven to be a viable and energy-efficient consensus mechanism, there are areas where it can be further improved. These include:

        \begin{itemize}
            \item \textbf{Decentralization}: Finding ways to ensure that wealth concentration does not lead to centralization is a key challenge. Future PoS improvements may include more sophisticated validator selection methods that favor smaller stakers or introduce new mechanisms to distribute rewards more equitably.
            \item \textbf{Validator Incentives}: Enhancing the incentives for validators to participate in the network without compromising security is crucial. This could involve dynamic reward structures or more flexible staking models.
            \item \textbf{Governance}: Decentralized governance in PoS systems remains a topic of active research. Future systems may include more robust governance models that allow stakeholders to have a more direct say in protocol upgrades and decision-making processes.
        \end{itemize}

        These improvements will shape the future of PoS and blockchain technology, ensuring that it remains secure, decentralized, and scalable as more projects adopt PoS as their consensus mechanism.

\chapter{Delegated Proof of Stake (DPoS)}

\section{Basic Working Principles of DPoS}

Delegated Proof of Stake (DPoS) is a consensus mechanism that builds upon the traditional Proof of Stake (PoS) system\cite{Saad2021ComparativeAO}, designed to achieve a higher level of scalability and efficiency while maintaining security. DPoS introduces the concept of delegating power to a smaller group of trusted participants, referred to as delegate nodes, to validate transactions and maintain the blockchain. These delegates are chosen through a voting process, which is an integral part of the DPoS framework. This section will discuss the fundamental principles that define how DPoS works, focusing on the voting mechanism and the responsibilities of elected delegate nodes.

\subsection{Voting Mechanism for Electing Nodes}

In DPoS, the stakeholders of the network, often referred to as token holders, have the ability to vote for a specific number of delegates\cite{Misic2024TowardDI}. These delegates, also known as block producers or witnesses, are responsible for validating transactions, producing blocks, and securing the network.

The voting process in DPoS is straightforward but essential for the system's operation:

\begin{enumerate}
    \item \textbf{Stakeholder Participation:} All stakeholders in the network, meaning users who hold tokens, are eligible to participate in the voting process. Each stakeholder's voting power is proportional to the number of tokens they hold. For example, a user with 10\% of the total supply of tokens will have 10\% of the voting power. This gives more influence to users who have a greater stake in the network's success.
    
    \item \textbf{Voting for Delegates:} Stakeholders cast their votes for a predetermined number of delegates (e.g., 21 delegates in EOS). Depending on the system, voters can either vote directly for individual delegates or delegate their voting rights to a representative who votes on their behalf.
    
    \item \textbf{Weight of Votes:} The more tokens a stakeholder holds, the more influence their vote carries. For instance, if a stakeholder holds 1000 tokens, their vote has more weight than that of a user with 100 tokens. This system ensures that stakeholders with a larger financial interest in the network have a stronger say in who is responsible for maintaining the blockchain.
    
    \item \textbf{Dynamic Election:} Voting is continuous, meaning that delegate nodes are constantly under scrutiny. If a delegate node fails to perform its duties, stakeholders can vote to replace them with a more competent or trusted candidate. This ensures that delegates remain accountable to the community.
\end{enumerate}

\textit{Example:} In the EOS blockchain\cite{eosHomeEOSIO,Benhaim2021ScalingBC}, there are 21 active block producers at any given time. Every 126 seconds, new block producers are elected based on stakeholder votes. This means that if a block producer does not perform well, such as by failing to produce blocks on time or acting maliciously, they can be quickly replaced through a democratic voting process.

\subsection{Responsibilities of Delegate Nodes}

Once elected, delegate nodes (or block producers) in a DPoS system have several key responsibilities to ensure the network's smooth functioning. These roles are critical to maintaining the security, stability, and performance of the blockchain:

\begin{enumerate}
    \item \textbf{Transaction Validation:} Delegate nodes are responsible for verifying and validating transactions. Once validated, these transactions are grouped into blocks, which are then added to the blockchain. Since only a limited number of delegates are elected, this process is highly efficient, leading to faster transaction confirmation times compared to traditional PoS or Proof of Work (PoW) systems\cite{Benhaim2021ScalingBC}.
    
    \item \textbf{Block Production:} In DPoS, the elected delegates take turns producing blocks. Each delegate is assigned a specific time slot, and during this slot, they must produce a block\cite{Wang2020RevisitingTF}. If a delegate fails to produce a block during their slot, the network may skip that block, potentially impacting transaction throughput.
    
    \item \textbf{Securing the Network:} Delegate nodes are also tasked with securing the network by ensuring the integrity of the blockchain. They must ensure that transactions are not fraudulent and that the blocks they produce follow the network's rules and protocol.
    
    \item \textbf{Accountability:} One of the critical features of DPoS is the constant accountability of delegate nodes\cite{Li2023ResearchAI}. If a delegate node underperforms, such as by not producing blocks or engaging in malicious behavior, stakeholders can vote to remove them from their position. This feature promotes an environment of trust and responsibility, ensuring that only reliable delegates remain in power.
\end{enumerate}

\textit{Example:} In the Steem blockchain\cite{steemPoweringCommunities}, delegates (referred to as witnesses) are responsible for producing blocks every three seconds. If a witness fails to produce a block, they may be penalized, and stakeholders can vote them out of the witness position. This dynamic system helps ensure high network uptime and reliability.

\section{Comparison Between DPoS and PoS}

Delegated Proof of Stake (DPoS) and Proof of Stake (PoS) are both consensus mechanisms designed to secure blockchain networks, but they differ in how they approach decentralization, security, and efficiency\cite{SchaafAnalysisOP}. This section will explore the differences between DPoS and PoS, focusing on the issues of node centralization and the balance between security and efficiency.

\subsection{The Issue of Node Centralization}

One of the key distinctions between DPoS and PoS is the potential for centralization\cite{article}. In a PoS system, any stakeholder can participate in the consensus process, meaning that there could be a broad distribution of validators. The more decentralized the network is, the less likely it is to suffer from a concentration of power, making the network more resistant to corruption or manipulation.

In contrast, DPoS intentionally limits the number of nodes that participate in block production. Only the elected delegates are responsible for validating transactions and maintaining the network. While this improves efficiency, it can also lead to concerns about centralization, as a small number of nodes hold significant power.

\textit{Example:} In EOS, only 21 block producers are responsible for maintaining the network, which can lead to a concentration of power among a few delegates\cite{eosnetworkConsensusDeveloper}. While these delegates are held accountable by the voting system, the limited number of participants means that the network may be more centralized compared to a PoS system with hundreds or thousands of validators.

\subsection{Balancing Security and Efficiency}

DPoS is designed to balance security and efficiency, focusing on maximizing transaction throughput while maintaining a certain level of security. By reducing the number of nodes involved in the block validation process, DPoS can achieve much higher transaction speeds than PoS or PoW systems\cite{Bachani2022PreferentialDP}.

However, this increase in efficiency comes with potential trade-offs in security. Since the network relies on a smaller number of delegates, the system is more vulnerable if a few of these delegates act maliciously or collude to manipulate the blockchain.

\textit{Example:} In BitShares\cite{bitsharesBitSharesBlockchain}, the DPoS system allows the network to process thousands of transactions per second, making it highly efficient for financial applications. However, stakeholders must remain vigilant to ensure that block producers are acting in the network's best interest to avoid security risks.

\section{Applications of DPoS}

Several blockchain platforms use DPoS to enhance scalability and transaction throughput. This section will examine how DPoS is implemented in prominent platforms like EOS\cite{eosHomeEOSIO}, Steem\cite{steemPoweringCommunities}, and BitShares\cite{bitsharesBitSharesBlockchain}.

\subsection{DPoS Mechanism in EOS}

EOS is one of the most well-known platforms using the DPoS consensus mechanism\cite{eosHomeEOSIO}. It achieves high transaction speeds by allowing only 21 elected block producers to validate transactions. These block producers are voted in by EOS token holders and can be replaced if they do not perform their duties effectively.

\textit{Example:} The EOS blockchain can process thousands of transactions per second, making it suitable for large-scale applications such as decentralized apps (dApps) and enterprise solutions. The high efficiency of EOS comes at the cost of potential centralization, as only 21 block producers are responsible for maintaining the entire network.

\subsection{DPoS Structure in Steem and BitShares}

Steem\cite{steemPoweringCommunities} and BitShares\cite{bitsharesBitSharesBlockchain} also implement DPoS but with slightly different structures and approaches. In Steem, elected witnesses are responsible for producing blocks and managing the network's governance. Similarly, in BitShares, the DPoS mechanism ensures high transaction throughput while providing a decentralized voting system for governance decisions.

\textit{Example:} Steem focuses on social media and content sharing, using DPoS to ensure that content creators are rewarded efficiently. BitShares, on the other hand, focuses on decentralized finance (DeFi), using DPoS to maintain a highly efficient and scalable trading platform.

\section{Advantages and Challenges of DPoS}

DPoS offers several advantages, particularly in terms of efficiency and scalability, but it also faces challenges, particularly around centralization risks.

\subsection{Higher Transaction Processing Efficiency}

DPoS significantly improves transaction processing efficiency by reducing the number of nodes involved in block validation. This leads to faster transaction times compared to PoS or PoW systems. In a DPoS network, only the elected delegates are responsible for producing blocks, which streamlines the consensus process and enables higher throughput.

\subsection{Risks of Centralization}

While DPoS offers efficiency benefits, it also introduces the risk of centralization. Since a small number of delegates hold significant power, there is a risk that these delegates could collude or act in their own self-interest, potentially compromising the network's integrity. Furthermore, the concentration of power may reduce the decentralization that is central to blockchain technology's ethos.

\textit{Example:} In platforms like EOS, where only 21 block producers maintain the entire network, there is a risk that these few delegates could control or manipulate the network. While stakeholders can vote them out, the relatively small number of participants increases the risk of centralization.

\chapter{Other Consensus Mechanisms}

    \section{Proof of Authority (PoA)}
        Proof of Authority (PoA)\cite{ethereumProofofauthorityPoA} is a consensus mechanism that relies on a limited number of trusted validators, who are responsible for validating transactions and creating new blocks. This system is well-suited for private or consortium blockchains where decentralization is not the main concern, but trust, speed, and low cost are prioritized.

        \subsection{Working Principles of PoA}
        Proof of Authority operates differently from mechanisms like Proof of Work (PoW) and Proof of Stake (PoS). In PoA, a set of pre-approved validators with established identities is responsible for maintaining the integrity of the network\cite{ethereumProofofauthorityPoA}. Unlike PoW, where computational power is required to validate transactions, or PoS, where validators stake cryptocurrency, PoA leverages the reputation and authority of known validators. The validators are typically well-known and trusted entities, such as companies or individuals with a public reputation.

        The process of validating transactions and producing blocks in PoA networks is fast because the small number of validators reduces the complexity involved\cite{Islam2022ACA}. The validators are incentivized to maintain the network's trust, as any malicious behavior could harm their reputation, leading to their removal from the role of validator.

        Here is a simple analogy: Imagine a group of trusted officials in a small community who are responsible for verifying and approving important decisions. Since these officials are known and trusted by everyone, the approval process is fast and efficient, but it also requires careful selection of who these officials are.

        \subsection{Validator Selection and Network Governance}
        In a PoA network, the process of selecting validators is critical. Validators are typically selected based on their trustworthiness and reputation\cite{ZHANG2019181}. In many cases, validators are well-established companies or organizations that have an interest in maintaining the stability and security of the network.

        One of the key governance mechanisms in PoA systems is transparency\cite{fastercapitalProofAuthority}. Since the identities of the validators are known, the community can hold them accountable for their actions. Additionally, the governance system may include rules that allow for the replacement of validators if they act maliciously or lose the trust of the community.

        A practical example of validator selection is in enterprise blockchains, where several companies form a consortium. Each company might select one representative to act as a validator, and these validators are publicly known. The trust between the companies ensures that validators will act honestly, as any deviation could lead to legal or business consequences.

        \subsection{Applications of PoA}
        Proof of Authority is used in several real-world scenarios, particularly where speed, cost, and trust between participants are more important than decentralization. Some notable examples include:

        \begin{itemize}
            \item \textbf{Private Blockchains}: Many enterprises use PoA for their private blockchains\cite{Samuel2021ChoiceOE}. These blockchains are closed to the public, and only pre-approved participants (validators) are allowed to take part in the network. The fast transaction speed and low costs make PoA ideal for supply chain management and other enterprise applications.
            
            \item \textbf{Consortium Blockchains}: In consortium blockchains, multiple organizations come together to form a blockchain network. Validators are selected from the member organizations, and trust is maintained through pre-existing business relationships\cite{ZHANG2019181}. An example of this is VeChain, a blockchain designed for supply chain management that uses a PoA consensus mechanism.
        \end{itemize}

    \section{Hybrid Consensus Mechanisms}
        Hybrid consensus mechanisms combine different consensus algorithms to balance the trade-offs in security, energy efficiency, and decentralization\cite{KumarJha2023HybridCM}. The most common hybrid models combine Proof of Work (PoW) and Proof of Stake (PoS), where each system contributes its strengths to the overall network.

        \subsection{Combining PoW and PoS}
        In hybrid consensus systems that combine PoW and PoS, each mechanism has a distinct role. For example, PoW might be used for creating new blocks, while PoS is used for finalizing those blocks. This approach allows the network to leverage the security and decentralization of PoW, while also benefiting from the energy efficiency and economic incentives of PoS.

        \textbf{Example:} In a typical PoW/PoS hybrid system, miners use computational power (as in Bitcoin's PoW) to propose new blocks. Once a block is proposed, PoS validators—who have staked tokens—vote on the validity of the block. The voting process ensures that the block is final and cannot be reversed.

        \subsection{Operating Model of Hybrid Consensus}
        The practical implementation of a hybrid consensus model involves clear separation of roles between PoW miners and PoS validators. The PoW miners compete to solve complex cryptographic puzzles and propose new blocks, while the PoS validators, who have staked their tokens, validate these blocks before they are added to the blockchain.

        The combination of PoW and PoS offers several benefits:
        \begin{itemize}
            \item \textbf{Security}: PoW provides a high level of security, as it requires significant computational power to attack the network.
            \item \textbf{Energy Efficiency}: PoS reduces the energy consumption needed for finalizing transactions, as validation relies on economic incentives rather than computational power.
            \item \textbf{Reduced Centralization}: By incorporating both mechanisms, the system reduces the centralization risk inherent in PoS, where wealthy participants could potentially control the network.
        \end{itemize}

    \section{Emerging Consensus Mechanisms}
        As blockchain technology evolves, new consensus mechanisms are being developed to address limitations in scalability, security, and privacy. Two such mechanisms are Practical Byzantine Fault Tolerance (PBFT)\cite{Li2020ASM} and the integration of Zero-Knowledge Proofs (ZKPs)\cite{ZHOU2024103678} in consensus systems.

        \subsection{Practical Byzantine Fault Tolerance (PBFT)}
        PBFT is a consensus algorithm that is designed to handle Byzantine faults in a distributed system\cite{Sakho2020ResearchOA}. Byzantine faults occur when some participants in the system act maliciously or send conflicting information. PBFT achieves consensus through multiple rounds of voting between participants, ensuring that even if some participants are compromised, the network can still reach a consensus.

        PBFT is particularly suited for permissioned blockchains, where the participants are known and trusted to some extent. The algorithm works by having each node in the network communicate with every other node, confirming the validity of transactions through voting. This method is highly efficient for small-scale networks but can become more resource-intensive as the number of participants grows.

        \subsection{Zero-Knowledge Proofs and Consensus Mechanisms}
        Zero-Knowledge Proofs (ZKPs) \cite{Berentsen2023AnIT}are cryptographic techniques that allow one party to prove to another party that a statement is true without revealing any information beyond the fact that the statement is true. ZKPs are increasingly being explored as a way to enhance privacy and scalability in blockchain systems.

        By integrating ZKPs into consensus mechanisms, blockchain networks can improve both privacy and scalability. For example, ZKPs can be used to validate transactions without revealing any details about the transaction itself, such as the amount or the parties involved.

        \textbf{Potential Applications:}
        \begin{itemize}
            \item \textbf{Privacy-Enhanced Blockchains}: Some blockchain projects, like Zcash, are already using ZKPs to enhance privacy. In the context of consensus mechanisms, ZKPs could allow validators to confirm transactions without knowing their contents, thus maintaining privacy.
            \item \textbf{Scalability Improvements}: ZKPs can also reduce the amount of data that needs to be processed, allowing blockchains to handle more transactions without compromising on security or privacy.
        \end{itemize}

\part{Practical Applications of Blockchain}
\chapter{Blockchain Applications in Fintech}

\section{Decentralized Finance (DeFi)}
Decentralized Finance, commonly referred to as DeFi, is an emerging financial ecosystem built on blockchain technology. It seeks to replicate and enhance traditional financial services by removing intermediaries such as banks, brokers, and financial institutions\cite{Gupta2022DecentralizedFA}. Instead of relying on centralized entities, DeFi utilizes blockchain to create an open and permissionless financial network that anyone with an internet connection can access.

\subsection{Basic Concepts of DeFi}
At its core, DeFi is based on several foundational principles:
\begin{itemize}
    \item \textbf{Open Access:} DeFi protocols are typically open to everyone\cite{Salami2021ChallengesAA}. Unlike traditional financial systems where individuals may face geographical, regulatory, or identity barriers, anyone with internet access can interact with DeFi platforms.
    \item \textbf{Transparency:} All transactions on DeFi platforms are recorded on a public blockchain\cite{secSECgovStatement}. This transparency ensures that the protocols are auditable and verifiable by anyone.
    \item \textbf{Smart Contracts:} DeFi applications rely heavily on smart contracts—self-executing contracts with the terms of the agreement written directly into code\cite{John2023SmartCA}. These contracts automatically enforce agreements, reducing the need for human intervention and central authorities.
\end{itemize}

For instance, the Ethereum blockchain\cite{ethereumCompleteGuide} is one of the most popular platforms for DeFi due to its ability to support complex smart contracts. DeFi platforms like Compound\cite{compoundv2CompoundUnderstanding} and Aave\cite{aaveAave} allow users to lend and borrow cryptocurrencies, while protocols like Uniswap\cite{uniswapUniswapInterface} and SushiSwap\cite{sushisvapSushiSwapAdvanced} enable decentralized trading.

\subsection{Lending Platforms and Decentralized Exchanges (DEX)}
One of the key innovations in DeFi is the development of decentralized lending platforms and exchanges\cite{Carapella2022DecentralizedF}. These systems remove the need for banks or brokerage firms by allowing users to directly lend, borrow, and trade assets in a peer-to-peer fashion.

\textbf{DeFi Lending Platforms:} 
DeFi lending platforms enable users to lend their digital assets to others in exchange for interest. Conversely, users can borrow assets by using their own cryptocurrencies as collateral. For example:
\begin{itemize}
    \item \textbf{Compound:} Compound\cite{compoundv2CompoundUnderstanding} is a decentralized lending platform where users can supply or borrow assets like Ether (ETH) and stablecoins (such as DAI). The interest rates are determined algorithmically based on supply and demand.
    \item \textbf{Aave:} Another leading lending platform, Aave\cite{aaveAave}, offers similar functionality but includes features like "flash loans," which allow users to borrow without collateral if the loan is repaid in the same transaction.
\end{itemize}

\textbf{Decentralized Exchanges (DEX):}
Decentralized Exchanges (DEXs) like Uniswap\cite{uniswapUniswapInterface} and SushiSwap\cite{sushisvapSushiSwapAdvanced} are platforms that allow users to trade cryptocurrencies directly with each other. Unlike traditional exchanges that rely on an order book to match buyers and sellers, DEXs utilize liquidity pools—collections of funds provided by users that are used for trading. The benefits of DEXs include:
\begin{itemize}
    \item \textbf{No Central Control:} Users retain full control of their funds throughout the transaction process, avoiding risks associated with centralized exchanges, such as hacking or mismanagement.
    \item \textbf{Lower Costs:} DEXs typically have lower transaction fees because they don't require intermediaries.
    \item \textbf{Permissionless Trading:} Users can trade any supported token without the need for verification or authorization from a central authority.
\end{itemize}

\subsection{Role and Application of Stablecoins}
A significant challenge in DeFi is the volatility of most cryptocurrencies. To address this issue, stablecoins have been introduced. Stablecoins are digital currencies pegged to the value of stable assets, such as fiat currencies (like the US dollar), to reduce price fluctuations.

\textbf{Importance of Stablecoins in DeFi:}
\begin{itemize}
    \item \textbf{Liquidity:} Stablecoins provide liquidity for DeFi platforms. Since their value is stable, they can be used to lend, borrow, and trade without the risk of significant price swings.
    \item \textbf{Cross-Border Transactions:} Stablecoins facilitate faster and cheaper cross-border payments compared to traditional banking systems.
    \item \textbf{Store of Value:} Users in unstable economies may use stablecoins as a store of value, protecting their wealth from hyperinflation or devaluation.
\end{itemize}

Examples of stablecoins include:
\begin{itemize}
    \item \textbf{USDT (Tether)\cite{tether}:} Pegged to the US dollar, Tether is one of the most widely used stablecoins in DeFi.
    \item \textbf{DAI\cite{makerdaoMakerDAOUnbiased}:} DAI is a decentralized stablecoin issued by the MakerDAO protocol. It is backed by a pool of cryptocurrencies rather than fiat, making it unique in its operation.
\end{itemize}

\section{Cross-Border Payments and Remittances}
One of the most transformative applications of blockchain in fintech is its use in cross-border payments and remittances. Traditionally, international transfers involve high fees and delays due to the involvement of multiple intermediaries like banks and payment processors. Blockchain offers a faster, cheaper, and more efficient alternative\cite{forbesCouncilPost,Qiu2018RippleVS}.

\subsection{Application of Bitcoin in Cross-Border Payments}
Bitcoin, the first and most well-known cryptocurrency, is increasingly being used for cross-border payments\cite{bitcoinBitcoinOpen}. Due to its decentralized nature, Bitcoin allows individuals to send funds across borders without relying on banks or remittance services.

\textbf{Advantages of Bitcoin in Cross-Border Payments:}
\begin{itemize}
    \item \textbf{Lower Fees:} Bitcoin transactions often have lower fees compared to traditional remittance services like Western Union or PayPal, especially for large transfers.
    \item \textbf{Faster Transaction Times:} Traditional cross-border payments can take several days to clear. Bitcoin transactions, on the other hand, can be completed in minutes or hours, depending on network congestion.
    \item \textbf{Borderless Payments:} Bitcoin allows for direct peer-to-peer transfers, making it ideal for international transactions where intermediaries might not be available or are too expensive.
\end{itemize}

A notable example is the use of Bitcoin in countries with unstable local currencies\cite{he2024cryptocurrency} or where access to traditional banking is limited. Individuals can use Bitcoin to send or receive money across borders, bypassing expensive and slow conventional systems.

\subsection{Ripple and XRP as Cross-Border Payment Solutions}
Ripple\cite{rippleGlobalPayments} is a blockchain-based payment system designed specifically for fast, low-cost cross-border transactions. Its native cryptocurrency, XRP\cite{rippleDigitalAsset}, serves as a bridge currency for financial institutions, enabling liquidity without the need to hold pre-funded accounts in different currencies.

\textbf{Advantages of Ripple and XRP:}
\begin{itemize}
    \item \textbf{Speed:} Ripple's consensus mechanism allows for transactions to be confirmed in seconds, making it much faster than Bitcoin and traditional payment systems.
    \item \textbf{Low Costs:} Ripple's transaction fees are significantly lower than those of traditional remittance services.
    \item \textbf{Bank Partnerships:} Ripple has established partnerships with several major banks and financial institutions, including Santander\cite{rippleSantanderPartners} and American Express\cite{rippleAmericanExpress}, to facilitate cross-border settlements using XRP.
\end{itemize}

Ripple's approach to cross-border payments has the potential to reshape global finance by reducing the reliance on correspondent banking systems and enabling faster, cheaper international transfers.

\section{Digital Currency and Central Bank Digital Currencies (CBDC)}

\subsection{Concept of Digital Currency}
Digital currencies are currencies that exist purely in digital form, without any physical counterpart like cash or coins. These currencies can be decentralized, like Bitcoin, or centralized, like digital versions of fiat currencies issued by governments. Digital currencies are characterized by their ability to be transferred electronically without the need for intermediaries.

\textbf{Key Characteristics of Digital Currencies:}
\begin{itemize}
    \item \textbf{No Physical Form:} Digital currencies are intangible and exist only on computers or digital wallets.
    \item \textbf{Fast Transactions:} Digital currency transactions can be completed almost instantaneously, unlike traditional banking systems that may take days.
    \item \textbf{Reduced Costs:} Digital currencies eliminate many of the costs associated with physical currency production and distribution.
\end{itemize}

The rise of digital currencies has led to the development of Central Bank Digital Currencies (CBDCs)\cite{Griffoli2018CastingLO}, where central banks explore issuing state-backed digital money.

\subsection{Central Bank Digital Currency Experiments in Various Countries}
Many central banks are experimenting with CBDCs as a way to modernize their monetary systems and provide a digital alternative to cash. CBDCs are different from decentralized cryptocurrencies because they are issued and regulated by a central authority, typically the national government.

\textbf{Notable CBDC Initiatives:}
\begin{itemize}
    \item \textbf{China’s Digital Yuan (e-CNY)\cite{cnbcChinaGiven}:} China is one of the frontrunners in CBDC development. The digital yuan is being piloted in several cities across China and aims to complement physical currency while providing a more efficient and secure payment system.
    \item \textbf{European Central Bank's Digital Euro\cite{europaDigitalEuro}:} The European Central Bank (ECB) is actively researching the potential issuance of a digital euro, which would serve as a digital version of the euro for use in everyday transactions across Europe.
    \item \textbf{Bahamas' Sand Dollar\cite{sanddollar}:} The Sand Dollar, launched by the Central Bank of the Bahamas, is one of the world’s first fully operational CBDCs. It aims to improve financial inclusion and streamline payments in the country.
\end{itemize}

As countries continue to explore CBDCs, they will likely play a significant role in the future of global finance, offering a digital alternative to cash that is secure, efficient, and easy to use.

\chapter{Blockchain Applications in Supply Chain Management}

\section{Challenges in Supply Chain Management}

\subsection{Transparency and Traceability in Supply Chains}

In modern supply chains, which often span multiple countries and involve various stakeholders, transparency and traceability are major challenges\cite{Saxena2023BlockchainFS}. As products move from raw materials through different stages of production and logistics, it becomes increasingly difficult to maintain visibility into each stage of the process. Lack of transparency can result in inefficiencies, increased costs, and potential fraud\cite{Patterson2018UnderstandingAM}.

For example, in the food industry, it can be challenging to track the origin of raw materials. When a contamination issue arises, companies might struggle to identify the source of the problem, leading to costly recalls. The complexity of supply chains often means that by the time an issue is detected, the product may have passed through multiple hands, making it difficult to pinpoint the exact origin of the issue. 

In addition to inefficiencies, lack of traceability can lead to fraud\cite{DuHadway2022LightID}. Unscrupulous actors may replace genuine goods with counterfeit ones or misrepresent the origin of products. This is a significant problem in industries such as pharmaceuticals, where counterfeit drugs can pose serious health risks, or luxury goods, where counterfeit products can harm brand reputation.

\subsection{Trust Issues in Supply Chain Management}

Trust is another significant issue in supply chain management. In a globalized economy, supply chains are often comprised of multiple entities—manufacturers, suppliers, logistics providers, and retailers—many of whom do not fully trust each other. This lack of trust can lead to delays, errors, and disputes\cite{Imeri2019BlockchainAO}.

For instance, when a shipment is delayed, the supplier may blame the logistics provider, while the logistics provider might claim the issue is with customs clearance. Without a transparent and trusted system for verifying what actually occurred, disputes can escalate, leading to financial losses and damaged relationships.

Traditionally, intermediaries such as banks or third-party verification agencies are required to build trust between these parties\cite{Yaksick2019OvercomingSC}. These intermediaries validate transactions, confirm deliveries, and manage payments. However, this reliance on intermediaries adds cost and complexity to the process. Moreover, intermediaries themselves can be subject to fraud or corruption, introducing new vulnerabilities into the system.

\section{Blockchain Applications in Logistics}

\subsection{Role of Blockchain in Goods Tracking}

Blockchain technology can significantly enhance the tracking of goods in supply chains by providing a transparent, decentralized, and tamper-proof system. With blockchain, every transaction, movement, or change of ownership of goods can be recorded in a digital ledger that is accessible to all stakeholders\cite{Reddy2019EnhancingSC}.

For example, when goods are shipped from a manufacturer to a warehouse, the transaction can be recorded on the blockchain in real-time. The data can include the time of departure, the condition of the goods, and the expected delivery time. Once the goods arrive at the warehouse, another transaction is added to the blockchain to confirm their receipt. This chain of transactions provides an immutable history of the goods' journey, enhancing accountability and reducing the potential for disputes.

Real-time tracking on a blockchain can also provide significant efficiency improvements. Companies can track the exact location of goods, anticipate potential delays, and plan logistics accordingly. This improves the overall responsiveness of the supply chain and reduces costs associated with inefficiencies.

\subsection{Immutability of Supply Chain Data}

One of the most important features of blockchain is its immutability. Once data is recorded on a blockchain, it cannot be altered or deleted. This feature is particularly valuable in supply chain management, where the integrity of data is critical for ensuring transparency, trust, and accountability\cite{Wu2019DataMI}.

For example, if a shipment of goods is recorded on the blockchain as having been delivered in good condition, that record cannot be altered later to claim the goods were damaged upon arrival. This prevents fraudulent claims and ensures that all parties can trust the data stored on the blockchain.

Immutability also helps in regulatory compliance. For industries such as pharmaceuticals or food, where traceability is legally required, blockchain provides an unchangeable record of the movement of goods\cite{Scott2018EvaluatingFO}. This not only simplifies audits but also helps companies prove compliance with regulations, reducing the risk of fines or legal action.

\section{Product Authentication and Traceability}

\subsection{Using Blockchain to Ensure Product Authenticity}

Counterfeiting is a major issue in several industries, including luxury goods, electronics, and pharmaceuticals. Blockchain can be used to verify the authenticity of products by providing a permanent, verifiable record of their origin and history\cite{Thakkar2021ApplicationFC}.

For instance, a luxury watch manufacturer can record each step of the production process on the blockchain, from sourcing the raw materials to the final assembly of the watch. When the watch is sold to a retailer, the transaction is also recorded on the blockchain. Consumers can then use a blockchain-based application to verify the authenticity of the watch by scanning a unique QR code or serial number, which links to the product's blockchain history.

In the pharmaceutical industry, blockchain can be used to track the movement of drugs from the manufacturer to the pharmacy\cite{Haq2018BlockchainTI}. This helps prevent counterfeit drugs from entering the supply chain, as each step of the drug's journey is recorded on the blockchain and can be verified by healthcare providers and consumers.

\subsection{End-to-End Traceability from Raw Materials to Consumers}

Blockchain enables end-to-end traceability of products, from the sourcing of raw materials to the final delivery to the consumer\cite{Cruz2020BlockchainbasedTP}. This level of transparency helps build consumer trust and ensures product quality.

For example, in the food industry, blockchain can track the journey of a product from the farm where the raw ingredients were grown, through the processing facility, to the logistics provider, and finally to the supermarket shelf. Each stage of the process is recorded on the blockchain, creating a complete history of the product's journey.

Consumers can access this information through a mobile app or website by scanning a QR code on the product packaging. This allows them to verify the origin of the product and check for any certifications, such as organic or fair trade. In cases of foodborne illness outbreaks, blockchain can help trace the contaminated product back to its source more quickly, reducing the impact of recalls\cite{Oriekhoe2024ENSURINGGF}.

Blockchain-based traceability also enhances sustainability efforts. Companies can provide consumers with transparent information about the environmental and ethical practices involved in producing their products, fostering a deeper level of trust and engagement with their brand.

\chapter{Blockchain Applications in Identity Authentication}

\section{Decentralized Identity (DID)}

\subsection{Concept and Principles of DID}
Decentralized Identity (DID)\cite{NassarKyriakidou2023DecentralizedIW} is a groundbreaking concept that allows individuals to have full control over their digital identities without needing to rely on centralized authorities, such as governments or large corporations. Traditional identity systems often depend on central entities, which can lead to security risks, privacy issues, and a lack of control for individuals. DID, on the other hand, ensures that each person owns and manages their identity autonomously.

In a DID system, identities are created and managed using blockchain technology. A decentralized identifier (DID) is a unique, self-sovereign identifier that a user can generate. These identifiers are recorded on a blockchain, which serves as a tamper-proof ledger. Once a DID is created, users can link their personal data, credentials, and other relevant information to it.

For example, instead of logging into a service using a username and password controlled by a company (like Google or Facebook), a user could authenticate themselves using a DID, which is stored on a blockchain. The user controls this DID via cryptographic keys, ensuring that they are the only ones who can prove ownership of the identity.

To better illustrate, consider the following process:
\begin{enumerate}
    \item A user generates a DID using a blockchain-based identity service. 
    \item This DID is stored on a public blockchain, and the user holds a private cryptographic key that proves ownership.
    \item When logging into a new service, the user presents their DID, and the service can verify the identity using the blockchain.
    \item No central authority is required, and the user maintains control over their private key, ensuring their identity remains secure and tamper-resistant.
\end{enumerate}

Blockchain technology ensures that DID systems are decentralized, immutable, and secure, which addresses many of the limitations of traditional identity management systems.

\subsection{Role of DID in Data Privacy Protection}
One of the key benefits of DID systems is the enhanced protection of user privacy\cite{2019EIDASSS}. In traditional systems, users often have to share personal data (such as email, address, or full identification details) with multiple service providers, each of whom stores that information centrally. This increases the risk of data breaches, identity theft, and unauthorized access.

In a DID system, individuals can control exactly what information they share with service providers. For instance, if a service requires proof of age, a user could share a cryptographically signed proof that they are over a certain age, without needing to share their full birthdate or identity. This concept is known as "selective disclosure."\cite{Kim2021AnalysisOT}

Selective disclosure works by using cryptographic techniques that allow a user to share only the minimum necessary information. Here’s a step-by-step example:
\begin{enumerate}
    \item A user holds a DID that includes various verified credentials, such as a driver’s license or university degree.
    \item When applying for a job, the user can share only the credential that proves they have a university degree, without sharing other personal information.
    \item The employer verifies the credential using the blockchain, ensuring that it is genuine and issued by a legitimate authority.
    \item At no point does the employer receive unnecessary details, such as the user’s home address or birthdate.
\end{enumerate}

In this way, DID systems ensure that users can verify their identity or credentials without exposing sensitive data. The cryptographic methods used in DIDs, such as zero-knowledge proofs, ensure that privacy is maintained, even during verification processes.

\section{Digital Credentials and Certification}

\subsection{Generation and Verification of Digital Credentials}
Digital credentials, such as certificates, diplomas, or licenses, can be issued, stored, and verified on the blockchain. Traditional credentials are often in paper form or stored in centralized databases, which can be prone to forgery or unauthorized access. Blockchain provides a more secure and efficient way of managing these credentials.

When an organization issues a digital credential, it creates a tamper-proof record on the blockchain, which can be verified by any third party. These credentials are linked to a user's DID and are cryptographically signed to ensure authenticity.

The process of issuing and verifying digital credentials typically follows these steps:
\begin{enumerate}
    \item An organization, such as a university, creates a digital credential for a student, e.g., a diploma.
    \item The university signs the credential using its private key and stores it on a blockchain.
    \item The student can then store this credential in their digital wallet, which is linked to their DID.
    \item When the student applies for a job, they can share their credential with the employer.
    \item The employer verifies the credential by checking the blockchain, ensuring it was issued by the university and is not tampered with.
\end{enumerate}

This system offers several advantages:
\begin{itemize}
    \item Tamper-proof: Once stored on the blockchain, credentials cannot be altered or forged.
    \item Efficiency: Verification is automatic and can be done without needing to contact the issuing institution.
    \item Transparency: Anyone can verify the authenticity of a credential by checking the blockchain.
\end{itemize}

In blockchain-based systems, the process of issuing and verifying digital credentials can be automated using smart contracts. Below is a simple example of a smart contract written in Solidity that issues and verifies digital credentials.

\begin{lstlisting}[style=solidity]
// A simple contract to issue and verify digital credentials

pragma solidity ^0.8.0;

contract CredentialRegistry {
    struct Credential {
        string credentialHash;
        address issuer;
        uint256 issuedAt;
    }

    mapping(string => Credential) public credentials;

    // Function to issue a credential
    function issueCredential(string memory _credentialHash) public {
        credentials[_credentialHash] = Credential({
            credentialHash: _credentialHash,
            issuer: msg.sender,
            issuedAt: block.timestamp
        });
    }

    // Function to verify a credential
    function verifyCredential(string memory _credentialHash) public view returns (bool) {
        Credential memory cred = credentials[_credentialHash];
        return cred.issuer != address(0);
    }
}
\end{lstlisting}

This contract allows an issuer to store a credential's hash on the blockchain and allows third parties to verify its authenticity.

\subsection{Ensuring the Security of Digital Identity}
Ensuring the security of digital identities on the blockchain is critical, as identity theft or unauthorized access could have severe consequences. Several mechanisms are used to protect digital identities:

\begin{enumerate}
    \item \textbf{Encryption}: Cryptographic encryption techniques protect data associated with DIDs. Public-key cryptography is used to ensure that only the owner of a DID can control it\cite{Min2021ASO}. The private key is held by the user, and the public key is recorded on the blockchain.
    
    \item \textbf{Cryptographic Keys}: Each DID is associated with a pair of cryptographic keys (public and private keys). The private key remains with the user, and the public key is shared with others for verification\cite{Park2022ANA}. For example, when a user logs in to a service using their DID, the service can verify the user’s identity by checking their public key on the blockchain.
    
    \item \textbf{Decentralized Verification}: Unlike traditional systems where a single entity verifies identity, blockchain allows decentralized verification\cite{Thorve2022DecentralizedIM}. Multiple nodes in the blockchain network can validate the authenticity of an identity or credential, reducing the risk of fraud or unauthorized changes.
\end{enumerate}

These measures ensure that identity management using blockchain is highly secure, tamper-resistant, and less vulnerable to centralized attacks, thereby significantly reducing the risks of identity theft and unauthorized access to personal data.

\chapter{Non-Fungible Tokens (NFTs)}

\section{Basic Concepts of NFTs}

\subsection{Indivisibility and Uniqueness of NFTs}

Non-Fungible Tokens, or NFTs, are a type of digital asset that represent ownership of unique items on the blockchain. Two key characteristics of NFTs are their indivisibility and uniqueness\cite{Idelberger2022NonfungibleT}. Unlike cryptocurrencies such as Bitcoin or Ethereum, which can be divided into smaller units and are fungible (interchangeable), NFTs are unique and cannot be split into smaller parts\cite{Martin2021ATD}.

Each NFT has a unique identifier and metadata, which distinguishes it from other tokens. This uniqueness is guaranteed through the use of blockchain technology, ensuring that each NFT represents a distinct digital or physical asset. For example, in the case of digital art, each NFT corresponds to a specific piece of artwork that is distinguishable from any other piece, even if two artworks appear visually similar.

Consider a simple analogy: if you own a 1-dollar bill, it can easily be exchanged for another 1-dollar bill, and both hold the same value. However, if you own an original painting, it cannot be exchanged for another painting in the same way. Each painting is unique and irreplaceable. NFTs operate on this same principle of uniqueness.

\subsection{Differences Between NFTs and ERC-20 Tokens}

NFTs are often compared to ERC-20 tokens\cite{Parham2022NonfungibleTP}, another popular type of token used on the Ethereum blockchain. However, there are significant differences between the two:

\begin{itemize}
    \item \textbf{Fungibility:} ERC-20 tokens are fungible\cite{Zhang2023TheAO}, meaning one token is identical to another and can be exchanged without any difference. For instance, one DAI token is equivalent to another DAI token in every way\cite{Feng2024ModelingAA}. On the other hand, NFTs are non-fungible, meaning each token is unique and not interchangeable with another NFT.
    \item \textbf{Use Cases:} ERC-20 tokens are commonly used for digital currencies, governance tokens, and other assets where interchangeability is required\cite{Zhang2023TheAO}. NFTs, by contrast, are used to represent unique assets such as digital art, collectibles, virtual real estate, or music\cite{Gupta2023ReviewingTR}.
    \item \textbf{Market Dynamics:} The market for ERC-20 tokens behaves like traditional financial markets where tokens can be traded based on supply and demand with uniform pricing\cite{Somin2018NetworkAO}. NFTs, however, have markets more akin to the art world, where the value of each asset is determined individually based on rarity, creator reputation, and demand\cite{Mekacher2022HowRS}.
\end{itemize}

\section{Technical Architecture of NFTs}

\subsection{ERC-721 Standard}

The ERC-721 standard defines the technical foundation for NFTs on the Ethereum blockchain\cite{ethereumERC721NonFungible}. This standard outlines the methods that must be implemented for a contract to handle NFTs. The most critical feature of the ERC-721 standard is its ability to ensure the interoperability and uniqueness of NFTs across different platforms and applications.

An ERC-721 contract assigns a unique identifier, often referred to as `tokenId`, to each NFT. This ID, along with other metadata, ensures that no two tokens are exactly the same. The standard also provides functionalities for transferring ownership of tokens, querying ownership, and approving third-party transfers.

Here’s an example of a minimal ERC-721 contract in Solidity:

\begin{lstlisting}[style=solidity]
pragma solidity ^0.8.0;

import "@openzeppelin/contracts/token/ERC721/ERC721.sol";

contract SimpleNFT is ERC721 {
    uint256 public tokenCounter;

    constructor() ERC721("SimpleNFT", "SNFT") {
        tokenCounter = 0;
    }

    function createNFT() public returns (uint256) {
        uint256 newTokenId = tokenCounter;
        _safeMint(msg.sender, newTokenId);
        tokenCounter++;
        return newTokenId;
    }
}
\end{lstlisting}

In this example, each time the `createNFT` function is called, a new NFT is minted with a unique `tokenId`. The `safeMint` function ensures the safe creation of the token and assigns ownership to the caller.

\subsection{NFT Creation and Transaction Process}

The process of creating (or minting) an NFT involves deploying a smart contract that adheres to the ERC-721 standard. The metadata associated with the NFT, such as its name, description, and link to the digital file (e.g., an image or video), is also stored in the blockchain or in decentralized storage systems like IPFS\cite{ipfsOpenSystem}.

Once minted, the NFT can be listed on NFT marketplaces (such as OpenSea or Rarible) for trading. When a user purchases an NFT, a transaction is executed on the blockchain, transferring ownership of the token from the seller to the buyer. The transaction is secured by the blockchain, ensuring that ownership records cannot be tampered with.

\begin{lstlisting}[style=cmd]
# Example of minting and transferring an NFT (CLI commands)
mintNFT --contract 0x1234567890abcdef --from 0xabcdef123456 --metadata-uri ipfs://Qm123abc
transferNFT --contract 0x1234567890abcdef --from 0xabcdef123456 --to 0xfedcba654321 --token-id 1
\end{lstlisting}

\section{Use Cases of NFTs}

\subsection{Digital Art}

NFTs have gained significant attention in the digital art space. Artists can tokenize their artwork and sell it as an NFT, giving buyers ownership of a unique, verifiable piece of digital content. One of the key advantages of NFTs for artists is the ability to receive royalties from secondary sales. Each time the NFT is resold, the original artist can receive a percentage of the sale price through smart contract functionality.

For example, consider the digital artist Beeple\cite{beeplecrapHOMEBEEPLE}, who sold his NFT artwork "Everydays: The First 5000 Days" for over \$69 million\cite{christiesBeeple1981}. This sale exemplifies the growing interest in digital art NFTs and how they can provide new revenue streams for artists.

\subsection{In-Game Items and Virtual Real Estate}

NFTs have found valuable use cases in the gaming industry, where they represent unique in-game items, such as weapons, skins, or virtual land in metaverse platforms. Players can own, trade, or sell these items, enabling true ownership of digital assets. 

A popular example is the game \textit{Axie Infinity}\cite{axieinfinityAxieInfinity}, where players collect and battle digital creatures (called Axies) that are represented as NFTs. These creatures can be bought, sold, or bred, with some rare Axies fetching high prices.

Another example is \textit{Decentraland}\cite{decenrtalandDecentraland}, a virtual world where users can purchase plots of virtual land as NFTs, which they can develop, sell, or lease to other users.

\subsection{Digital Copyright for Music and Multimedia}

NFTs are revolutionizing the music and multimedia industry by allowing creators to tokenize their work and control how it is distributed. Musicians can release NFT versions of their albums, granting collectors ownership of limited editions or exclusive content.

For example, the musician 3LAU\cite{3lau3LAUOfficial} made headlines by selling a series of NFTs representing digital music albums, generating millions of dollars in sales. This model allows creators to bypass traditional music platforms and directly engage with fans, while also receiving royalties from subsequent sales.

By using NFTs, musicians and multimedia creators gain more control over their digital rights and open new avenues for monetization and licensing of their content.

\chapter{Decentralized Autonomous Organizations (DAO)}

\section{Basic Concepts of DAO}
Decentralized Autonomous Organizations (DAOs) are one of the most groundbreaking innovations enabled by blockchain technology\cite{Ding2023ASO}. They represent a new form of organization that operates without centralized control, relying on smart contracts and a decentralized governance structure. In this chapter, we will explore the fundamental concepts of DAOs, including their governance structure, voting mechanisms, and practical applications in both governance and decentralized finance (DeFi).

\subsection{Decentralized Governance Structure of DAOs}
At the heart of a DAO is its decentralized governance structure. Unlike traditional organizations, where decisions are typically made by a central authority or a board of directors, DAOs distribute decision-making power among the members\cite{Schneider2022DecentralizedAO}. This is achieved through the ownership of tokens, which act as voting shares in the organization.

The core of this decentralized governance is often implemented through smart contracts\cite{Balcerzak2022BlockchainTA}. A smart contract is a self-executing contract where the terms of the agreement are written directly into code. In a DAO, smart contracts define the rules of the organization and automatically enforce them without the need for intermediaries.

For example, a DAO may have a rule that requires a majority vote for any decision to be implemented. Token holders can propose changes or new initiatives, and voting is conducted transparently on the blockchain\cite{Han2023DAOG}. Once a proposal receives the required number of votes, the smart contract automatically enforces the decision, whether it’s allocating funds, changing a governance parameter, or implementing new features.

\begin{lstlisting}[style=solidity]
// Example of a simple voting smart contract in Solidity
pragma solidity ^0.8.0;

contract SimpleDAO {
    mapping(address => uint256) public shares;
    mapping(uint256 => Proposal) public proposals;
    uint256 public proposalCount;

    struct Proposal {
        string description;
        uint256 voteCount;
        bool executed;
    }

    constructor() {
        shares[msg.sender] = 100; // Assign initial shares to contract creator
    }

    function propose(string memory _description) public {
        proposals[proposalCount++] = Proposal(_description, 0, false);
    }

    function vote(uint256 _proposalId) public {
        require(shares[msg.sender] > 0, "Must have shares to vote");
        Proposal storage proposal = proposals[_proposalId];
        proposal.voteCount += shares[msg.sender];
    }

    function executeProposal(uint256 _proposalId) public {
        Proposal storage proposal = proposals[_proposalId];
        require(proposal.voteCount > 50, "Not enough votes to pass");
        require(!proposal.executed, "Proposal already executed");
        proposal.executed = true;
    }
}
\end{lstlisting}

This Solidity code demonstrates the basic structure of a DAO's proposal and voting system. Token holders can propose ideas, vote based on their token shares, and if enough votes are received, the proposal is executed automatically.

\subsection{Voting Mechanisms in DAOs}
Voting is one of the key features of DAOs, and there are several mechanisms that DAOs use to ensure fair participation and influence. The most common voting mechanisms include:

\subsubsection{Token-Weighted Voting:}
In many DAOs, voting power is proportional to the number of tokens held by a member. This is known as token-weighted voting\cite{Kurniawan2022VotingMS}. The more tokens you own, the greater your influence on the outcome of a decision. While this system is straightforward, it can lead to centralization, where large token holders have disproportionate control.

\subsubsection{Quadratic Voting:}
Quadratic voting is a more complex system that aims to balance the influence of participants\cite{Posner2014VotingSQ}. Instead of voting power being directly proportional to the number of tokens held, quadratic voting allows members to allocate votes in a way that reflects the intensity of their preferences. The cost of casting additional votes increases quadratically, making it expensive for large token holders to dominate the process.

For example, casting one vote costs 1 token, but casting two votes costs 4 tokens, and casting three votes costs 9 tokens. This mechanism allows smaller holders to express strong preferences without being overwhelmed by larger token holders.

\subsubsection{Delegated Voting:}
Some DAOs use a delegated voting mechanism, where token holders can delegate their voting power to a trusted representative. This representative can then vote on behalf of the delegator. This approach is often used in situations where the decisions are complex, and not all members have the time or expertise to make informed choices.

\section{Practical Applications of DAOs}
DAOs have a wide range of applications, particularly in governance and decentralized finance (DeFi). In this section, we will explore how DAOs are used in real-world projects and how they contribute to decentralized governance and financial systems.

\subsection{DAO Applications in Governance Projects}
DAOs are commonly used in governance projects to allow communities to manage decentralized protocols, allocate funds, and make collective decisions. A well-known example of this is the MakerDAO\cite{makerdaoMakerDAOUnbiased}, which governs the Maker Protocol, a decentralized lending platform that manages the stablecoin DAI\cite{Kozhan2021DecentralizedSA}.

In MakerDAO, decisions regarding protocol upgrades, parameter adjustments, and fund allocation are made by the community through token-weighted voting. Any token holder can propose changes, and the community votes to approve or reject them. This decentralized governance model allows for more transparency and community involvement compared to traditional organizations.

Other notable examples include:

\begin{itemize}
    \item \textbf{Uniswap DAO:} Governs the Uniswap decentralized exchange protocol\cite{uniswapUniswapInterface}. The community votes on proposals to adjust fees, add new features, or upgrade the protocol.
    \item \textbf{Compound DAO:} Manages the Compound lending protocol\cite{compoundCompound}, allowing the community to propose and vote on protocol changes.
    \item \textbf{Gitcoin DAO:} Focuses on funding open-source projects through decentralized governance and community-driven funding decisions\cite{gitcoinGitcoinGovernance}.
\end{itemize}

\subsection{The Role of DAOs in DeFi}
DAOs play a crucial role in the decentralized finance (DeFi) ecosystem. Many DeFi protocols are governed by DAOs, which manage everything from protocol upgrades to the allocation of treasury funds\cite{Ushida2021RegulatoryCO}. By allowing token holders to vote on key decisions, DAOs enhance the transparency and decentralization of these platforms.

For instance, in the Aave protocol\cite{aaveAave}, a DAO governs the allocation of its funds, decisions regarding new token listings, and the parameters of the lending pools. This decentralized governance ensures that no single entity controls the platform, and decisions are made by the community in a transparent manner.

DAOs also enable innovation in DeFi by allowing developers to propose new features, upgrades, or experiments, which the community can vote on and implement if approved. This open governance structure leads to a dynamic and evolving ecosystem where the community has direct control over the future of the platform.

\section{Challenges and the Future of DAOs}
Despite their potential, DAOs face several challenges, particularly in terms of legal recognition and scalability. In this section, we will explore the challenges and future directions of DAO development.

\subsection{Legal Status and Regulatory Challenges of DAOs}
One of the biggest challenges facing DAOs is their legal status\cite{Guillaume2023DecentralizedAO}. Since DAOs operate without a centralized entity, it is often unclear who is liable for decisions made by the organization. In traditional organizations, legal entities like corporations or LLCs provide a structure for accountability and liability. However, in DAOs, the decentralized nature makes it difficult to assign responsibility.

Some DAOs have attempted to address this by incorporating as legal entities, such as Wyoming's DAO LLC law\cite{wyolegLegislativeService}, which provides a legal framework for DAOs to operate as LLCs. However, this is still a developing area, and there are many uncertainties regarding how DAOs will be treated by regulators in different jurisdictions.

\subsection{Future Directions of DAO Development}
The future of DAOs is full of potential. As blockchain technology matures, DAOs are likely to become more sophisticated, with improvements in governance mechanisms, scalability, and integration with other blockchain technologies.

One of the key areas of development is improving governance models\cite{Ding2023ASO,Gilson2024EnhancingTD}. While token-weighted voting is the most common system today, new models like quadratic voting and reputation-based voting are being explored to create fairer and more inclusive decision-making processes.

Scalability is another challenge that DAOs need to address. As DAOs grow, the complexity of decision-making increases, and it becomes harder to reach consensus\cite{Sheikh2023BuiltTL}. Solutions such as layer-2 scaling solutions\cite{Mandal2023InvestigatingLS} and cross-chain governance mechanisms\cite{DiRose2018ComparisonAA} are being developed to address these issues.

Finally, DAOs have the potential to disrupt traditional organizational structures. As more industries explore the benefits of decentralized governance, we may see DAOs being used in everything from corporate governance to public service management, offering a more transparent and inclusive way to make decisions.

\part{Blockchain Research in Academia}
\chapter{Current State of Academic Research on Blockchain}
\section{Research Hotspots in Blockchain Technology}
\subsection{Distributed Ledger Technology}
Distributed Ledger Technology (DLT)\cite{Gutlapalli2016CommercialAO} is the core innovation behind blockchain, providing a decentralized method for storing data across multiple nodes in a network. The key features of DLT include security, decentralization, and scalability, which are crucial for the success of blockchain systems. 

\subsubsection{Scalability Challenges:} As blockchain systems grow, one of the main challenges is ensuring scalability—the ability to handle an increasing number of transactions without degrading performance. For example, in Bitcoin’s blockchain, every node must validate each transaction, which limits throughput to around 7 transactions per second (TPS)\cite{Gbel2017IncreasedBS}. Solutions like \textbf{sharding}\cite{CortesGoicoechea2020ResourceAO} divide the ledger into smaller partitions, allowing parallel processing of transactions.

\subsubsection{Security:} Security in DLT involves protecting the ledger from malicious actors who might attempt to alter transaction history. The immutability of blockchain, where data once written cannot be changed, provides strong protection\cite{N2023TheII}. However, researchers continue to explore enhanced cryptographic techniques like \textbf{quantum-resistant algorithms}\cite{Nisha2024LatticeBasedCA}, which aim to safeguard against future threats posed by quantum computing.

\subsubsection{Decentralization:} Decentralization ensures that no single entity controls the ledger. This is a defining feature of blockchain and prevents issues like censorship or single points of failure. However, true decentralization introduces the \textbf{scalability trilemma}\cite{Nakai2023TheBT}—the challenge of balancing decentralization, security, and scalability simultaneously, which continues to be a major research focus.

\subsubsection{Example:} A notable example of a blockchain system with improved DLT architecture is \textbf{Ethereum 2.0}\cite{Asif2023ShapingTF}, which uses a combination of \textbf{Proof of Stake (PoS)} and \textbf{sharding} to increase scalability while maintaining decentralization.

\subsection{Optimization and Improvement of Consensus Algorithms}
The consensus algorithm in a blockchain is the method by which the network reaches agreement on the validity of transactions. Research in this area focuses on optimizing existing algorithms like \textbf{Proof of Work (PoW)} and \textbf{Proof of Stake (PoS)}, as well as developing new ones\cite{ferdous2020blockchainconsensusalgorithmssurvey}.

\subsubsection{Proof of Work (PoW):} PoW\cite{ferdous2020blockchainconsensusalgorithmssurvey}, used in Bitcoin, requires miners to solve computational puzzles to validate transactions. This algorithm is secure but highly energy-intensive. Research on \textbf{ASIC resistance}\cite{Cho2018ASICResistanceOM} and alternative approaches like \textbf{Proof of Space}\cite{Dziembowski2015ProofsOS} aims to mitigate the environmental impact.

\subsubsection{Proof of Stake (PoS):} PoS\cite{Wendl2022TheEI}, used in Ethereum 2.0, selects validators based on the amount of cryptocurrency they hold and are willing to "stake" as collateral. PoS is more energy-efficient than PoW but introduces new challenges such as ensuring \textbf{fairness} and preventing \textbf{centralization of wealth}. 

\subsubsection{New Consensus Mechanisms:} Emerging consensus mechanisms such as \textbf{Delegated Proof of Stake (DPoS)}\cite{Wendl2022TheEI} and \textbf{Byzantine Fault Tolerance (BFT)} aim to improve scalability and security. For example, DPoS allows token holders to vote for delegates who validate transactions, which increases efficiency but requires trust in elected delegates.

\subsubsection{Example:} The \textbf{Tendermint} consensus algorithm\cite{Lei2020ChainedTA}, which is based on BFT, is used by the \textbf{Cosmos Network}\cite{cosmosCosmosInternet}. It ensures fast finality, meaning transactions are confirmed almost instantly, addressing the scalability issue present in traditional PoW systems.

\subsection{Privacy Protection and Anonymity Research}
While blockchain offers transparency, this can sometimes conflict with privacy needs. Research is focusing on enhancing privacy without compromising the integrity of the system.

\subsubsection{Zero-Knowledge Proofs (ZKPs):} ZKPs\cite{ZHOU2024103678} allow a user to prove the validity of a transaction without revealing the transaction details. \textbf{Zcash} \cite{zZcashPrivacyprotecting}is a prominent example of a blockchain that uses \textbf{zk-SNARKs} (Zero-Knowledge Succinct Non-Interactive Arguments of Knowledge\cite{Nitulescu2020zkSNARKsAG}) to ensure privacy.

\subsubsection{Privacy-Preserving Smart Contracts:} Researchers are exploring methods to execute smart contracts without exposing sensitive data. One approach is using \textbf{MPC (Multi-Party Computation\cite{Pei2018SmartCB})}, where multiple parties jointly compute a function while keeping their inputs private.

\subsubsection{Balancing Privacy and Transparency:} While privacy is crucial, blockchain systems must still ensure transparency for regulatory compliance. Researchers are investigating solutions like \textbf{selective disclosure}, where only authorized entities can view certain transaction details\cite{Mukta2023ABI}.

\subsubsection{Example:} In the \textbf{Monero}\cite{getmoneroMoneroProject} blockchain, privacy is enhanced through \textbf{ring signatures} and \textbf{stealth addresses}, which obscure the origin and destination of transactions, making it difficult to trace.

\section{Interdisciplinary Blockchain Research}
\subsection{Integration of Blockchain and Cryptography}
Cryptography is the foundation of blockchain security, ensuring data integrity and protection through techniques like \textbf{hashing}, \textbf{encryption}, and \textbf{digital signatures}.

\subsubsection{Hashing:} Blockchain relies on cryptographic hashing to secure transactions. A hash function takes input data and produces a fixed-size string, which serves as a digital fingerprint of the data. For example, Bitcoin uses the \textbf{SHA-256} hash function to secure its blocks\cite{Courtois2014OptimizingSI}. 

\subsubsection{Digital Signatures:} Digital signatures authenticate the identity of users. In blockchain, each transaction is signed by the sender using their private key. The recipient can verify this signature using the sender’s public key. The \textbf{Elliptic Curve Digital Signature Algorithm (ECDSA)} is widely used for this purpose in Bitcoin\cite{Wang2014SecureIO}.

\subsubsection{Encryption:} Data encryption ensures that only authorized parties can view sensitive information. Research on \textbf{homomorphic encryption}\cite{Zhang2016ARO} allows computations to be performed on encrypted data, which is beneficial for privacy-preserving applications in blockchain.

\subsection{Blockchain and Game Theory}
Game theory provides a framework to analyze how participants behave in a decentralized system, such as a blockchain. Researchers use game theory to study incentive structures and strategies that ensure participants act in a way that benefits the network\cite{Liu2019ASO}.

\subsubsection{Incentive Mechanisms:} In blockchain systems like Bitcoin, miners are incentivized to validate transactions through rewards. However, game theory helps explore how to prevent malicious behaviors, like \textbf{selfish mining}\cite{Li2020RationalPA}, where miners might collude to increase their rewards.

\subsubsection{Cooperation and Security:} Game theory is also used to analyze how participants can cooperate to maintain security\cite{Bhudia2023GameTM}. For instance, in Proof of Stake systems, validators are penalized for dishonest behavior, which deters attacks.

\subsection{Blockchain and Economics}
Blockchain has significant implications for economics, particularly in the field of \textbf{decentralized finance (DeFi)} and \textbf{token economics}.

\subsubsection{Decentralized Finance (DeFi):} DeFi platforms allow users to borrow, lend, and trade assets without intermediaries like banks. Researchers study how these systems can operate efficiently while ensuring security and preventing fraud\cite{Salami2021ChallengesAA,Gupta2022DecentralizedFA}.

\subsubsection{Token Economics:} The design of cryptocurrencies and tokens involves economic principles like \textbf{scarcity} and \textbf{incentive alignment}. Researchers explore how tokenomics can create sustainable ecosystems that incentivize participation and growth\cite{Lamberty2020LeadingDS}.

\section{Academic Conferences and Journals on Blockchain}
\subsection{Major Academic Conferences in the Blockchain Field}
Academic conferences provide a platform for researchers to present their latest findings and collaborate on advancing blockchain technology. Some key conferences include:

\subsubsection{IEEE Blockchain Conference:} This annual conference focuses on blockchain technology, covering topics like consensus algorithms, privacy, and applications in various industries\cite{IEEECon}.

\subsubsection{Crypto Conference:} As one of the premier conferences on cryptography, Crypto often includes sessions on blockchain-related cryptographic research\cite{iacrCrypto2024}.

\subsubsection{International Conference on Blockchain (ICBC):} ICBC is another major event where researchers discuss advancements in blockchain infrastructure, scalability, and real-world applications\cite{ieeeicbcIEEEInternational}.

\subsubsection{High-Impact Journals Related to Blockchain}
Blockchain-related research is often published in high-impact journals, where peer-reviewed articles contribute to the academic understanding of this technology.

\subsubsection{IEEE Transactions on Blockchain:} This journal focuses specifically on blockchain technology, covering a wide range of topics from cryptographic foundations to industry applications\cite{}.

\subsubsection{Ledger:} Ledger\cite{ledgerjournalAboutJournal} is an open-access journal that publishes peer-reviewed research on blockchain and cryptocurrency, with a focus on both technical and economic aspects.

\subsubsection{Cryptography Journals:} Journals like \textbf{Journal of Cryptology}\cite{springerJournalCryptology} and \textbf{Cryptography and Communications}\cite{springerCryptographyCommunications} also publish papers related to the cryptographic techniques used in blockchain systems.

\chapter{Academic Research on Consensus Mechanisms}

\section{Innovations and Challenges in Consensus Algorithms}
Consensus mechanisms are the backbone of blockchain technology\cite{gervais2016security,ferdous2020blockchainconsensusalgorithmssurvey}. They ensure that all participants in a decentralized network agree on the validity of transactions. Over the years, researchers have made significant advances in consensus algorithms to improve blockchain's security, efficiency, and scalability. In this chapter, we explore various research developments and the challenges associated with these mechanisms.

\subsection{In-Depth Research on Byzantine Fault Tolerance Algorithms}
Byzantine Fault Tolerance (BFT) algorithms are designed to reach consensus in a distributed system\cite{Zhang2022ReachingCI}, even when some participants behave maliciously or experience failures. This is essential for maintaining the integrity of decentralized systems like blockchains.

The Byzantine Generals Problem is a famous thought experiment illustrating the difficulties of achieving consensus when participants may send conflicting information. BFT algorithms solve this problem by ensuring that a majority of participants agree on a single version of the truth, even in the presence of malicious actors.

One of the most prominent BFT algorithms is Practical Byzantine Fault Tolerance (PBFT)\cite{Ferenczi2022AnEV}, introduced by Miguel Castro and Barbara Liskov in 1999. PBFT is efficient for smaller systems but becomes increasingly challenging to scale for larger networks, like public blockchains. Therefore, recent research has focused on optimizing BFT for blockchain scalability.

\textbf{Example: Improving BFT for Blockchain Scalability}

For example, research on Tendermint\cite{Buchman2018TheLG}, a popular BFT-based consensus mechanism, has shown promising results in improving scalability by reducing the communication overhead between nodes. In Tendermint, nodes communicate in rounds, and each round has a leader who proposes a block. If two-thirds of the nodes agree, the block is added to the blockchain. This reduces the time and complexity of reaching consensus, especially in networks with many participants.

Research has also focused on improving the security of BFT algorithms by combining them with cryptographic techniques such as threshold signatures, where a certain number of signatures are required to validate transactions. This approach enhances resilience against attacks by making it harder for malicious actors to disrupt the consensus process.

\subsection{Efficiency Optimization of Consensus Algorithms}
Consensus algorithms, particularly in large-scale blockchains, can be energy-intensive and slow. Improving the efficiency of these algorithms is critical for their adoption and sustainability.

\textbf{Reducing Energy Consumption in Proof of Work (PoW)}

Proof of Work (PoW)\cite{gervais2016security,ferdous2020blockchainconsensusalgorithmssurvey} is the consensus algorithm used by Bitcoin. While PoW ensures security by requiring participants to solve complex mathematical puzzles, it consumes vast amounts of energy. Researchers are actively exploring ways to reduce this energy consumption. For instance, projects like Ethereum 2.0 are moving away from PoW and adopting Proof of Stake (PoS), which requires far less computational power.

One research direction involves reducing the complexity of the puzzles in PoW systems. By finding puzzles that still require work but are less energy-intensive, researchers aim to maintain security while lowering environmental impact\cite{Tsabary2019JustES}.

\textbf{Improving Throughput in Proof of Stake (PoS)}

PoS is designed to replace the energy-hungry PoW by selecting validators based on the number of tokens they hold\cite{Wendl2022TheEI}. However, PoS can also suffer from scalability issues. To address this, researchers are investigating ways to improve PoS by reducing latency and increasing transaction throughput.

One approach is the use of sharding\cite{CortesGoicoechea2020ResourceAO}, where the blockchain is divided into smaller partitions, or "shards." Each shard processes transactions in parallel, which greatly increases the number of transactions the network can handle.

\section{Exploration of Novel Consensus Mechanisms}
As blockchain adoption grows, researchers are exploring new consensus mechanisms that offer different trade-offs between security, efficiency, and scalability.

\subsection{Variants and Optimizations of Practical Byzantine Fault Tolerance (PBFT)}
Practical Byzantine Fault Tolerance (PBFT)\cite{Ferenczi2022AnEV} has proven effective for private or permissioned blockchain systems, where the number of nodes is limited and trust is partially established. However, scaling PBFT to public blockchains, which may have thousands of nodes, introduces challenges related to communication overhead and latency.

\textbf{Research Example: Optimizing Communication in PBFT}

Recent research has proposed several optimizations to make PBFT more scalable for large systems. For instance, in traditional PBFT, every node must communicate with every other node, which leads to quadratic communication costs\cite{Yang2018LinBFTLB}. Optimizations like \textit{reducing message complexity}\cite{Bonomi2018MultihopBR} have been proposed, where nodes only need to communicate with a subset of other nodes.

Another optimization focuses on improving the leader election process in PBFT. By ensuring that leaders are selected based on performance and reliability metrics, the system can reduce the chances of leader failures, which would otherwise slow down the consensus process.

\subsection{Research Progress in Zero-Knowledge Proof Consensus}
Zero-knowledge proofs (ZKPs)\cite{ZHOU2024103678} are cryptographic methods that allow one party to prove the validity of a statement without revealing the underlying data. These methods have gained significant attention in the context of blockchain because they can improve both privacy and scalability.

In Zero-Knowledge Proof Consensus (ZKPC), participants verify the correctness of transactions without having to see the actual transaction data. This is particularly useful in privacy-centric applications, where users want to ensure their data is secure and private while still participating in the consensus process.

\textbf{Example: Zero-Knowledge Rollups}

One of the most exciting developments in ZKP is the concept of zero-knowledge rollups\cite{Capko2022StateOT}. In this approach, multiple transactions are bundled together into a single proof, which is then verified on the main blockchain. This reduces the computational burden on the network, as fewer transactions need to be processed individually, thereby improving scalability without sacrificing security.

\section{Research on Hybrid Consensus Algorithms}
Hybrid consensus algorithms\cite{KumarJha2023HybridCM} combine multiple types of consensus mechanisms to leverage their respective strengths while mitigating their weaknesses. These algorithms are particularly useful in blockchain systems that require both high security and scalability.

\subsection{Consensus Mechanisms Combining PoW and PoS}
One promising area of research is the development of hybrid models that combine Proof of Work (PoW) and Proof of Stake (PoS)\cite{Sun2019AnIH}. In these models, PoW is used to initially secure the network and prevent malicious attacks, while PoS is used to validate and approve transactions more efficiently.

\textbf{Example: PoW/PoS Hybrid in Decred}

Decred\cite{decredDecredMoney} is an example of a cryptocurrency that uses a hybrid PoW/PoS consensus mechanism. In Decred, miners perform PoW to propose blocks, but stakeholders (those holding the coin) perform PoS to validate and approve these blocks. This combination ensures both the security benefits of PoW and the energy efficiency of PoS.

\subsection{Research on Hybrid Consensus for Layer 1 and Layer 2}
Layer 1 refers to the base layer of the blockchain, where fundamental operations like transaction processing and block creation take place. Layer 2 solutions are built on top of Layer 1 to improve scalability by offloading certain tasks\cite{Hafid2020ScalingBA}.

Hybrid consensus mechanisms for Layer 1 and Layer 2 aim to combine the security of Layer 1 with the scalability of Layer 2. One popular solution is the Lightning Network\cite{lightningLightningNetwork}, a Layer 2 technology for Bitcoin, where small transactions occur off-chain and are later settled on the main blockchain. Researchers are investigating how to combine Layer 1 and Layer 2 consensus in a way that maximizes both performance and security.

\textbf{Example: Plasma}

Plasma\cite{ethereumPlasmaChains} is another Layer 2 solution that uses a hybrid consensus approach. In Plasma, smart contracts and Merkle trees are used to create "child chains" that run on top of the main blockchain. These child chains handle the majority of transactions, while the main chain is used for dispute resolution. This hybrid approach improves scalability by reducing the load on the main chain while ensuring that the child chains remain secure.

\chapter{Scalability and Performance Research of Blockchain}

\section{Academic Research on Scalability Issues}

\subsection{Comparison of On-Chain and Off-Chain Scaling}
Blockchain scalability has been a major challenge due to the limitations in processing a large number of transactions while maintaining security and decentralization. Researchers have explored two primary methods for scaling blockchains: on-chain scaling and off-chain scaling\cite{Kim2018ASO}. 

\subsubsection{On-Chain Scaling}
On-chain scaling involves making changes directly to the blockchain's base layer\cite{Pawar2020ASO}. A common approach is increasing the block size, which allows more transactions to be processed within each block. For instance, Bitcoin's block size is capped at 1MB, which limits the number of transactions per second. By increasing this block size, we can allow more transactions to be processed at once.

However, increasing block size has trade-offs. While it may enhance throughput, it also increases the amount of data that each node must store and process, potentially leading to centralization as smaller nodes may be unable to keep up. Moreover, larger blocks take longer to propagate through the network, increasing the chance of forks and reducing overall security.

\subsubsection{Off-Chain Scaling (Layer 2)}
Off-chain scaling moves part of the transaction load off the main chain. Layer 2 solutions, such as state channels and rollups, allow users to execute multiple transactions off-chain, with only a final aggregated transaction recorded on-chain. This reduces the load on the main blockchain and allows for much higher transaction throughput.

One well-known example is the Lightning Network for Bitcoin\cite{Seres2019TopologicalAO}, which enables users to open payment channels and conduct numerous transactions without interacting directly with the Bitcoin blockchain. Off-chain solutions can significantly improve performance but often come with complexity in terms of security and usability.

\subsubsection{Trade-offs: Decentralization, Security, and Performance}
Both on-chain and off-chain scaling methods present trade-offs. On-chain scaling can compromise decentralization, as larger blocks may require more powerful nodes, leading to centralization\cite{Kim2018ASO}. Off-chain scaling, while maintaining decentralization, can introduce complexity and security risks, such as the need for participants to trust off-chain systems or mechanisms for dispute resolution\cite{Dong2018CelerNB}.

A detailed comparison is necessary when deciding on the best approach, as no single method can perfectly balance scalability, security, and decentralization.

\subsection{The Theory and Implementation of Sharding}
Sharding \cite{CortesGoicoechea2020ResourceAO}is a scaling technique that aims to improve blockchain throughput by splitting the network into smaller pieces, called "shards." Each shard processes a subset of the total transactions in parallel, allowing multiple transactions to be processed simultaneously, thus increasing throughput.

\subsubsection{Theory of Sharding}
In traditional blockchains, every node must process every transaction, which creates a bottleneck. Sharding proposes a different approach where nodes are split into different shards, and each shard is responsible for only a portion of the transaction load. The key is that no single shard handles all the network data, but the overall security of the network is maintained through coordination between shards.

For example, in Ethereum's sharding implementation\cite{CortesGoicoechea2020ResourceAO}, the network is divided into multiple shards, each functioning as a mini-blockchain. A central component known as the "beacon chain" coordinates the activity of all shards to ensure security and integrity.

\subsubsection{Practical Implementation of Sharding}
The implementation of sharding presents several challenges, especially regarding security. Since each shard only validates its own transactions, attackers might try to target individual shards. One solution is to assign validators randomly to shards, reducing the risk of any single shard being compromised.

Moreover, the communication between shards, known as "cross-shard communication,"\cite{Das2020EfficientCT,Liu2024CHERUBIMAS} is a critical aspect of sharding. Ensuring that transactions between different shards are processed correctly is a key focus of current research. For instance, in Ethereum's roadmap, sharding is planned to be implemented gradually to address these challenges and improve the network's scalability.

\section{Research on Layer 2 Scaling Solutions}

\subsection{Theoretical Foundations of State Channels}
State channels are a Layer 2 solution designed to improve blockchain scalability by allowing participants to transact off-chain. State channels enable multiple transactions to take place between parties off-chain, only recording the final result on the blockchain.

\subsubsection{How State Channels Work}
A state channel is established between two parties by locking some funds in a multi-signature wallet on the blockchain\cite{Dziembowski2019MultipartyVS}. The participants then transact freely off-chain by exchanging signed messages that represent updates to the state of their agreement. When they decide to close the channel, they submit the final state of the channel to the blockchain for settlement.

\begin{lstlisting}[style=python]
# Example Python pseudo-code for a state channel transaction
class StateChannel:
    def __init__(self, balance_party1, balance_party2):
        self.balance_party1 = balance_party1
        self.balance_party2 = balance_party2

    def update_balance(self, party, amount):
        if party == 1:
            self.balance_party1 += amount
            self.balance_party2 -= amount
        else:
            self.balance_party2 += amount
            self.balance_party1 -= amount

    def finalize(self):
        return (self.balance_party1, self.balance_party2)
# Initialize a state channel with two parties
channel = StateChannel(100, 100)
# Party 1 sends 10 to Party 2
channel.update_balance(1, -10)
# Final state after multiple transactions
final_state = channel.finalize()
\end{lstlisting}

The challenge with state channels is ensuring that both parties act honestly. Mechanisms such as time locks and challenge periods are implemented to ensure that if one party tries to submit an outdated state to the blockchain, the other can provide proof and penalize them.

\subsection{Research Progress on Rollups}
Rollups\cite{Thibault2022BlockchainSU} are another Layer 2 solution gaining significant traction in blockchain research. Rollups bundle multiple transactions off-chain, compress them, and submit a summary to the main blockchain. There are two main types of rollups: Optimistic Rollups and Zero-Knowledge (ZK) Rollups.

\subsubsection{Optimistic Rollups}
Optimistic Rollups\cite{9862815} assume transactions are valid by default, only verifying them when challenged. This assumption allows transactions to be processed quickly and at a low cost. However, a dispute mechanism must be in place to verify any potentially fraudulent transactions.

\subsubsection{ZK-Rollups}
ZK-Rollups\cite{Fernando2023PosterWA}, on the other hand, generate cryptographic proofs (known as ZK-SNARKs) to verify the validity of every transaction. These proofs are submitted to the blockchain alongside the transaction summary, providing faster and more secure transaction verification. While ZK-Rollups provide more immediate security assurances, they require more complex computation to generate the cryptographic proofs.

\section{Research on Cross-Chain Technology}

\subsection{Challenges of Cross-Chain Interoperability}
As more blockchain networks emerge, the need for interoperability between different blockchains becomes critical. However, achieving seamless communication between blockchains poses numerous challenges.

\subsubsection{Consensus Differences}
Different blockchains often employ different consensus mechanisms, such as Proof of Work (PoW)\cite{gervais2016security,ferdous2020blockchainconsensusalgorithmssurvey}, Proof of Stake (PoS)\cite{Wendl2022TheEI}, or others. These mechanisms can have different security assumptions and finality guarantees, making it difficult to synchronize state across chains.

\subsubsection{Security Risks}
Another key challenge in cross-chain communication is maintaining security. Cross-chain bridges, which connect different blockchains, must ensure that assets and data are transferred securely between chains. Attacks on these bridges could lead to loss of funds or compromised data.

\subsection{Design of Decentralized Cross-Chain Bridges}
Decentralized cross-chain bridges aim to solve the interoperability problem without relying on centralized intermediaries. These bridges typically employ cryptographic techniques and multi-signature schemes to transfer assets and data across chains securely.

One popular example is the "Polkadot" network\cite{pokladotPolkadotInteroperable}, which enables various blockchains to interoperate through a central relay chain. This allows different blockchains, known as "parachains,"\cite{Scott2023BringingTA} to communicate with each other securely while maintaining their own consensus mechanisms.

\section{Performance Optimization of Blockchain}

\subsection{Strategies for Improving Blockchain Throughput}
Several strategies have been proposed to enhance blockchain throughput, including increasing block size, introducing faster consensus algorithms (e.g., Proof of Stake), and leveraging Layer 2 scaling solutions.

For instance, Ethereum's switch from Proof of Work to Proof of Stake in Ethereum 2.0\cite{Asif2023ShapingTF} is expected to significantly increase transaction throughput by introducing more efficient block validation.

\subsection{Research on Latency and Transaction Confirmation Time}
Reducing latency and improving transaction confirmation times are critical for improving user experience in blockchain networks. Research focuses on optimizing network propagation, reducing block times, and implementing faster finality mechanisms to ensure that transactions are confirmed more quickly without compromising security.

\chapter{Privacy Protection and Data Security}
    \section{Privacy Protection Mechanisms in Blockchain}
    
        \subsection{Academic Research on Zero-Knowledge Proofs}
        
        Zero-Knowledge Proofs (ZKPs)\cite{ZHOU2024103678} are cryptographic protocols that enable one party (the prover) to convince another party (the verifier) that a statement is true without revealing any additional information about the statement. This concept, first introduced by Goldwasser, Micali, and Rackoff in the 1980s\cite{goldreich1994definitions}, has become essential in enhancing privacy in blockchain networks. 

        \subsubsection{How ZKPs Work:} 
        In a ZKP, the prover provides evidence of knowledge without disclosing the underlying data. For example, imagine proving you know the password to a system without revealing the actual password. This is achieved through clever cryptographic techniques.

        \subsubsection{Application in Blockchain:} 
        Blockchains, by nature, are transparent. This is an excellent feature for security, but it raises privacy concerns since all transaction data is visible. ZKPs help solve this problem by allowing users to prove that they have the correct credentials or that they are authorized to perform an action without exposing sensitive details. For instance, in privacy-centric cryptocurrencies like Zcash\cite{zZcashPrivacyprotecting}, ZKPs (specifically zk-SNARKs, a variant of ZKPs\cite{Nitulescu2020zkSNARKsAG}) are used to shield transaction amounts and addresses while still ensuring that the transaction is valid.

        \subsubsection{Example:} 
        Consider Alice sending Bob 10 coins. In a traditional blockchain system like Bitcoin, anyone can see Alice's address, Bob's address, and the transaction amount. However, using a ZKP system like zk-SNARKs, Alice can prove she has enough coins to send Bob 10 coins without revealing her balance, Bob’s address, or the amount sent.
        
        \subsection{In-Depth Study of the MimbleWimble Protocol}
        
        MimbleWimble\cite{cointelegraphWhatMimblewimble} is a blockchain protocol that enhances privacy and scalability by obfuscating transaction details and compressing the overall data. Introduced in 2016, MimbleWimble fundamentally restructures how transaction data is stored and validated.

        \subsubsection{Core Concept:}
        MimbleWimble removes unnecessary data from the blockchain, like the public keys and signatures of each transaction, which are common in most blockchain systems. Instead, it only stores the input and output amounts and a cryptographic commitment that proves that no coins were created or destroyed in the process.

        \subsubsection{Confidential Transactions:} 
        One of the key features of MimbleWimble is Confidential Transactions, which hide transaction amounts using cryptographic blinding factors. This ensures that only the parties involved in the transaction know the amount being transacted, while the rest of the network can still verify that the transaction is valid without seeing any sensitive details.

        \subsubsection{Data Compression:} 
        Another benefit of MimbleWimble is its efficiency in reducing blockchain data size. Since transaction history is compacted, it reduces the need for large amounts of storage, making it easier for nodes to validate the chain.

        \subsubsection{Example:}
        Imagine Alice sends Bob some coins using MimbleWimble. Instead of the blockchain storing the full transaction details, the protocol hides the transaction amounts and other specifics while still allowing validators to confirm that the transaction is valid. Additionally, when Bob spends the coins later, much of the old transaction data is discarded, improving scalability.

        \subsection{Ring Signatures and Applications in Privacy Coins}
        
        Ring signatures\cite{Mercer2016PrivacyOT} are cryptographic methods that allow a user to sign a message on behalf of a group, without revealing which specific member of the group signed the message. This technique is widely used in privacy-focused cryptocurrencies like Monero\cite{getmoneroMoneroProject} to obfuscate transaction details.

        \subsubsection{How Ring Signatures Work:}
        When a user signs a transaction with a ring signature, their signature is mixed with other users’ signatures. This makes it computationally infeasible for anyone to determine who made the transaction. 

        \subsubsection{Applications in Monero:} 
        Monero\cite{getmoneroMoneroProject} uses ring signatures to make transactions anonymous. When a user sends a transaction, their signature is grouped with several other signatures, creating a "ring" of possible signers. As a result, it becomes impossible for an observer to tell which member of the ring made the transaction.

        \subsubsection{Example:}
        Suppose Alice sends 5 Monero coins to Bob. Using ring signatures, Alice's signature will be combined with several other signatures from unrelated transactions. An observer can see that one of the signatures belongs to Alice, but they cannot determine which one with certainty. This provides plausible deniability and strong privacy guarantees.

    \section{Data Security and Privacy-Preserving Computation}
    
        \subsection{Applications of Multi-Party Computation (MPC) in Blockchain}
        
        Multi-Party Computation (MPC)\cite{Evans2019API} is a cryptographic protocol that allows multiple parties to jointly compute a function without revealing their private inputs. This ensures that even though computations are carried out, no party learns anything beyond the result of the computation.

        \subsubsection{How MPC Works:}
        MPC involves splitting the input data into several encrypted shares, with each participant holding only one share. The computation is performed across all parties, and the result is combined in such a way that no party learns the other parties’ inputs.

        \subsubsection{Applications in Blockchain:} 
        In blockchain, MPC can be used for secure voting\cite{Sharma2019ScalableOS}, where participants can vote without revealing their choice, or for private transactions, where multiple parties can transact without revealing their individual transaction data.

        \subsubsection{Example:}
        Imagine a decentralized voting system where voters want to tally votes without revealing who voted for whom. MPC allows this computation by ensuring that no voter learns any other voters' choices, but the result (the final vote count) is correct and verifiable.

        \subsection{Combining Fully Homomorphic Encryption with Blockchain}
        
        Fully Homomorphic Encryption (FHE)\cite{Viand2021SoKFH} allows computations to be performed on encrypted data without needing to decrypt it first. This means sensitive data can remain confidential while still being used in meaningful computations.

        \subsubsection{How FHE Works:}
        FHE enables operations (such as addition or multiplication) to be performed on ciphertexts\cite{Yi2014}. The result is still encrypted, and when decrypted, it matches the result as if the operations had been performed on the plaintext.

        \subsubsection{Applications in Blockchain:} 
        In blockchain, FHE can be used to perform calculations on encrypted transactions, enabling privacy-preserving smart contracts\cite{Solomon2023smartFHEPS}. Users can participate in a blockchain network, submit transactions, or interact with contracts without revealing any sensitive details.

        \subsubsection{Example:}
        Suppose a healthcare blockchain wants to perform statistical analysis on patient data without compromising privacy. With FHE, computations like averages or sums can be performed on encrypted data, and the decrypted result will reveal the correct statistical values without disclosing individual patient information.

        \subsection{Trusted Execution Environments (TEE) and Privacy Protection}
        
        Trusted Execution Environments (TEEs)\cite{Asokan2013TheUP} provide a secure area in a computer’s hardware that ensures sensitive computations are executed in an isolated environment. This helps prevent unauthorized access or tampering.

        \subsubsection{How TEEs Work:}
        TEEs rely on secure hardware to isolate computations and data. Even if the main operating system is compromised, the data inside the TEE remains secure. TEEs are often used for executing sensitive smart contracts in a privacy-preserving manner\cite{li2022sok}.

        \subsubsection{Applications in Blockchain:} 
        TEEs can enhance blockchain security by allowing nodes to perform private computations without revealing the data\cite{Solomon2023smartFHEPS}. For example, sensitive data like private keys or confidential transactions can be processed in the TEE, ensuring that they are never exposed to the broader network.

        \subsubsection{Example:}
        Consider a blockchain that supports private auctions. A TEE can be used to keep all bids confidential while still allowing the smart contract to determine the winner. The contract runs inside the TEE, so no one else, including the network validators, can see the bid amounts.

    \section{Research on Data Immutability}
    
        \subsection{Mechanisms for Detecting Data Tampering}
        
        Blockchain is often praised for its immutability, meaning that once data is written to the blockchain, it cannot be changed. However, detecting attempts at data tampering is crucial for ensuring the integrity of the system.

        \subsubsection{Tamper Detection Mechanisms:}
        Research has focused on using cryptographic proofs and auditing trails to detect tampering\cite{Itkis2003CryptographicTE}. One common method involves storing cryptographic hashes of the data, which can be verified later to ensure the data has not been altered.

        \subsubsection{Example:}
        In a blockchain-based document storage system, every document is hashed, and the hash is stored on the blockchain. If someone attempts to change the document, its hash will no longer match the hash stored on the blockchain, immediately indicating tampering.

        \subsection{Applications of Hash Chains and Merkle Trees}
        
        Hash chains\cite{Fischlin2004FastVO} and Merkle trees\cite{Kuznetsov2024MerkleTI} are essential cryptographic structures used to ensure data integrity and immutability in blockchain systems.

        \subsubsection{How Hash Chains Work:}
        A hash chain is a series of cryptographic hashes where each hash depends on the previous one. This means that altering any part of the chain breaks the entire structure, making it evident if tampering has occurred.

        \subsubsection{Merkle Trees:} 
        A Merkle tree is a binary tree structure where each leaf node is a hash of a block of data, and each non-leaf node is a hash of its children. This allows efficient and secure verification of the data, as only a small part of the tree needs to be checked to verify the integrity of the entire dataset.

        \subsubsection{Example:}
        In blockchain, Merkle trees are used to verify transactions in a block. If a user wants to verify a specific transaction, they do not need to check the entire block; they only need to verify a few hashes up the Merkle tree, making the process highly efficient and secure.

\chapter{Decentralized Storage and Distributed Systems}

\section{Academic Research on Decentralized Storage}

\subsection{Technical Foundations of IPFS and Filecoin}
Decentralized storage systems aim to break the reliance on centralized servers for storing and retrieving data, making it more resilient, transparent, and censorship-resistant. Two prominent technologies in this field are the InterPlanetary File System (IPFS)\cite{cloudflareInterplanetaryFile} and Filecoin\cite{filecoinDecentralizedStorage}. Let’s explore the technical foundation behind how these systems work.

\textbf{IPFS} (InterPlanetary File System)\cite{Benet2014IPFSC} is a protocol designed to create a peer-to-peer network for storing and sharing data. Rather than relying on traditional URL-based addressing (where files are located on specific servers), IPFS uses \textbf{content-based addressing}\cite{Fong2022SecureSS}. This means files are addressed by their content rather than their location. Each file or piece of data is cryptographically hashed, and this hash acts as the file’s unique identifier. When you request a file from the network, IPFS looks for peers that store the file with that particular hash.

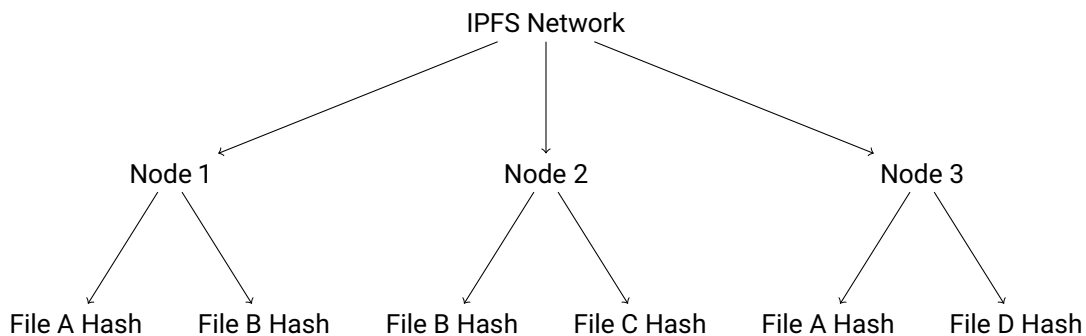
\begin{center}
\begin{tikzpicture}
  [->,level/.style={sibling distance = 5cm/#1, level distance = 2cm}]
  \node {IPFS Network}
    child {node {Node 1}
      child {node {File A Hash}}
      child {node {File B Hash}}
    }
    child {node {Node 2}
      child {node {File B Hash}}
      child {node {File C Hash}}
    }
    child {node {Node 3}
      child {node {File A Hash}}
      child {node {File D Hash}}
    };
\end{tikzpicture}
\end{center}

In this tree diagram, each node in the IPFS network stores a collection of files identified by their hashes. If you want to retrieve \textit{File A}, IPFS uses the hash to search the distributed network for the node(s) containing that file, rather than relying on a centralized server.

\textbf{Filecoin} builds on IPFS by adding a decentralized, incentivized layer for storage. While IPFS is focused on addressing and sharing content, Filecoin adds a marketplace where users can buy and sell storage\cite{filecoinDecentralizedStorage}. Filecoin’s blockchain ensures transparency, and participants are incentivized to store data by earning FIL tokens. This happens through a proof-of-storage mechanism where storage providers prove they are maintaining copies of the files.

In Filecoin, the proof system works in two parts:
\begin{enumerate}
    \item \textbf{Proof of Replication (PoRep):} Ensures that a storage provider has created a unique copy of the data\cite{Fisch2018ScalingPF}.
    \item \textbf{Proof of Space-Time (PoST):} Proves that the provider continues to store the data over a specified time period\cite{Moran2016ProofsOS}.
\end{enumerate}

\subsection{Challenges and Optimization of Decentralized Storage}
Decentralized storage, while promising, faces several challenges that are the focus of ongoing research:

\textbf{1. Data Retrieval Speed:} Unlike centralized systems, where data is retrieved from one server, decentralized systems must locate the nearest peer holding the requested data. This can introduce latency, especially if peers are far apart geographically or if the data is replicated unevenly across the network\cite{Sripanidkulchai2002EnablingEC,Matri2017KeepingUW,Acharya2024OnEO}.

\textbf{2. Redundancy:} To ensure data availability, decentralized systems replicate data across multiple nodes. However, managing this replication efficiently is challenging, as storing too many copies increases costs, while storing too few reduces reliability\cite{Ranganathan2002ImprovingDA}.

\textbf{3. Cost:} Storing data in decentralized systems like Filecoin involves incentives, which can make it more expensive than traditional cloud storage. Research is focusing on balancing storage costs with incentives to ensure a fair and efficient marketplace.

To optimize decentralized storage, researchers are exploring:
\begin{itemize}
    \item \textbf{Data Sharding:} Splitting large datasets into smaller, manageable pieces that are distributed across nodes. This allows for parallel retrieval and can improve speed\cite{Thakur2024SelfhealingNW}.
    \item \textbf{Dynamic Replication:} Adjusting replication levels dynamically based on data access patterns, ensuring that popular data is more readily available while rarely accessed data remains stored at minimal cost\cite{Ranganathan2002ImprovingDA}.
    \item \textbf{Advanced Compression Techniques:} Using compression to reduce the size of the data before storing it on the network, lowering storage and transmission costs.
\end{itemize}

\section{Consistency Issues in Distributed Systems}

\subsection{CAP Theorem and Consistency in Blockchain}
The CAP theorem\cite{Nygaard2023CosteffectiveDU}, proposed by Eric Brewer, is a fundamental principle in distributed systems. It states that a distributed system can only guarantee two of the following three properties at any given time:
\begin{itemize}
    \item \textbf{Consistency (C):} Every read receives the most recent write or an error.
    \item \textbf{Availability (A):} Every request receives a response, without a guarantee that it contains the latest write.
    \item \textbf{Partition Tolerance (P):} The system continues to operate even if communication between nodes is lost.
\end{itemize}

Blockchain systems, such as Bitcoin and Ethereum, often favor \textbf{partition tolerance} and \textbf{availability} over strict consistency\cite{Gilbert2002BrewersCA}. This is because the decentralized nature of blockchain involves many nodes (or miners) that may not always have the latest data. The consensus mechanism ensures eventual consistency—once a block is added to the blockchain, it is eventually propagated to all nodes. However, during network partitions or delays, different nodes might temporarily disagree on the state of the blockchain\cite{Cachin2017BlockchainCP}.

\textbf{Example:} In a blockchain network, suppose two transactions (T1 and T2) occur at nearly the same time, and the network is partitioned. Some nodes may confirm T1 first, while others may confirm T2. Once the partition is resolved, the blockchain’s consensus mechanism will determine which transaction to finalize, ensuring eventual consistency.

\subsection{Reliability Research in Decentralized Networks}
Reliability is critical for decentralized networks. Here are some research areas:

\textbf{1. Fault Tolerance:} Decentralized systems must handle node failures without compromising data availability. Techniques such as data replication, redundancy, and peer-to-peer recovery mechanisms\cite{Sari2015FaultTM} help maintain system reliability even if individual nodes go offline.

\textbf{2. Consensus Mechanisms:} Decentralized networks rely on consensus algorithms to agree on the state of the network. Common algorithms include:
\begin{itemize}
    \item \textbf{Proof of Work (PoW)\cite{gervais2016security,ferdous2020blockchainconsensusalgorithmssurvey}:} Used by Bitcoin, it requires nodes (miners) to solve complex puzzles to validate transactions.
    \item \textbf{Proof of Stake (PoS)\cite{Wendl2022TheEI}:} Validators are selected to confirm transactions based on the amount of cryptocurrency they hold.
\end{itemize}

These mechanisms ensure that even in the presence of malicious or faulty nodes, the network can reach agreement on valid transactions.

\section{Research on Data Storage and Availability}

\subsection{Combining Blockchain with Cloud Computing}
Blockchain can complement traditional cloud storage by providing decentralized control and transparency\cite{Sharma2021BlockchainbasedDA}. In this hybrid model, cloud storage systems are used for scalability and efficiency, while blockchain handles auditing, access control, and verifying data integrity.

For example, companies could store encrypted data in a cloud service while using blockchain smart contracts to manage who has access to that data. This creates an additional layer of security and transparency, preventing unauthorized access and ensuring that any changes to the data are recorded immutably.

\subsection{Exploration of Efficient Data Storage Mechanisms}
Efficient storage mechanisms are crucial for reducing costs and improving data availability in decentralized systems. Key areas of research include:

\textbf{1. Compression Techniques:} Advanced compression algorithms are being explored to reduce the amount of data that needs to be stored and transmitted in decentralized networks\cite{Chen2019BitcoinBC}. This can significantly reduce both storage costs and bandwidth requirements.

\textbf{2. Distributed Databases:} Systems like BigchainDB\cite{githubGitHubBigchaindbbigchaindb} combine blockchain principles with distributed databases, allowing for scalable and decentralized data storage. These databases are designed to handle large volumes of transactions while maintaining decentralization and security.

\textbf{3. Blockchain-based Storage Solutions:} Projects like Sia\cite{siaDecentralizedData} and Storj\cite{storjStorjSmarter} are building decentralized storage systems that leverage blockchain to incentivize storage and ensure data integrity. These systems provide alternatives to centralized cloud providers by offering cheaper, decentralized storage.

In summary, research in decentralized storage and distributed systems is rapidly evolving. By addressing the challenges of speed, redundancy, and cost, and integrating blockchain with cloud computing, these technologies have the potential to transform how data is stored and accessed.

\chapter{Research on Smart Contracts and Decentralized Applications (dApps)}

\section{Formal Verification of Smart Contracts}

\subsection{Application of Formal Verification in Smart Contract Security}
Formal verification is a crucial method used in ensuring the correctness of smart contracts. A smart contract is essentially a program that runs on a blockchain, and once deployed, it is immutable, meaning that its behavior cannot be changed\cite{John2023SmartCA}. This immutability makes it critical to verify that the contract will behave as intended under all conditions before deployment.

Formal verification uses mathematical methods to prove or disprove the correctness of algorithms underlying the smart contract with respect to a given formal specification. This approach is especially important for preventing vulnerabilities such as logical errors, which could result in loss of funds or other unintended outcomes.

For example, if a smart contract is used to manage a simple escrow transaction, formal verification can ensure that the contract will always release the funds to the correct party, under the correct conditions, and never to an unauthorized party.

Let's consider the following Solidity smart contract, which performs an escrow function:

\begin{lstlisting}[style=solidity]
pragma solidity ^0.8.0;

contract Escrow {
    address public payer;
    address public payee;
    uint public amount;
    bool public isReleased;

    constructor(address _payee) payable {
        payer = msg.sender;
        payee = _payee;
        amount = msg.value;
        isReleased = false;
    }

    function release() public {
        require(msg.sender == payer, "Only payer can release funds");
        require(!isReleased, "Funds already released");
        isReleased = true;
        payable(payee).transfer(amount);
    }
}
\end{lstlisting}

In this contract, formal verification can be used to mathematically guarantee that:
\begin{itemize}
    \item The funds will only be released by the payer.
    \item The funds will only be released once, preventing double-spending.
    \item The funds will be transferred to the correct recipient (the \texttt{payee}).
\end{itemize}

Formal verification tools such as \texttt{KeY}\cite{keyprojectProgramVerification}, \texttt{Isabelle/HOL}\cite{ethereumFormalVerification}, and \texttt{Z3}\cite{githubGitHubZ3Proverz3} can be applied to analyze such contracts. These tools check properties like invariant conditions and termination, ensuring that the contract behaves correctly in every possible state.

\subsection{Detection and Fixing of Smart Contract Vulnerabilities}
While formal verification helps prevent certain types of errors, many other tools focus on detecting and fixing vulnerabilities in smart contracts. One such tool is static analysis, which inspects the code without executing it, identifying potential vulnerabilities like reentrancy attacks, integer overflows, and unchecked external calls\cite{Chinen2021RAAS}.

For example, the infamous "DAO Hack" on Ethereum in 2016\cite{Dingman2019DefectsAV} was caused by a reentrancy vulnerability. In this attack, a function in a contract made an external call to another contract before updating its internal state, allowing the attacker to recursively call the function and drain funds from the contract.

\textbf{Reentrancy Example:}
\begin{lstlisting}[style=solidity]
pragma solidity ^0.8.0;

contract VulnerableContract {
    mapping(address => uint) public balances;

    function withdraw(uint _amount) public {
        require(balances[msg.sender] >= _amount, "Insufficient balance");
        (bool success, ) = msg.sender.call{value: _amount}("");
        require(success, "Withdrawal failed");
        balances[msg.sender] -= _amount;
    }

    function deposit() public payable {
        balances[msg.sender] += msg.value;
    }
}
\end{lstlisting}

Here, the vulnerability arises because the contract sends funds to the user (\texttt{msg.sender}) before updating the balance. An attacker can reenter the contract by recursively calling the \texttt{withdraw} function, withdrawing more funds than they are entitled to.

Several tools and techniques help to detect such vulnerabilities:
\begin{itemize}
    \item \textbf{Static Analysis:} Tools like \texttt{Mythril}\cite{githubGitHubConsensysmythril}, \texttt{Slither}\cite{githubGitHubCryticslither}, and \texttt{Oyente}\cite{githubGitHubEnzymefinanceoyente} can automatically scan the bytecode or source code of a contract to identify potential vulnerabilities before deployment.
    \item \textbf{Automated Testing:} Frameworks like \texttt{Truffle}\cite{trufflesuiteHomeTruffle} and \texttt{Hardhat}\cite{hardhatHardhatEthereum} allow developers to write tests to simulate different conditions, ensuring that the contract behaves correctly in every scenario.
    \item \textbf{Symbolic Execution:} This technique systematically explores all possible paths of execution in the smart contract to detect logical errors or vulnerabilities.
\end{itemize}

Once vulnerabilities are detected, developers can fix them using techniques such as:
\begin{itemize}
    \item Using the \texttt{Checks-Effects-Interactions} pattern to prevent reentrancy.
    \item Properly checking for integer overflows and underflows using \texttt{SafeMath}\cite{openzeppelinMathOpenZeppelin} libraries.
    \item Avoiding unchecked external calls.
\end{itemize}

\section{Research on Smart Contract Programming Languages}

\subsection{Security and Optimization of Solidity}
Solidity is the most widely-used smart contract programming language on the Ethereum blockchain\cite{soliditylangHomeSolidity}, but it comes with its own set of security challenges. Common vulnerabilities in Solidity contracts include:
\begin{itemize}
    \item Reentrancy attacks.
    \item Integer overflows and underflows.
    \item Gas limit attacks.
    \item Unprotected self-destruct calls.
\end{itemize}

Researchers have focused on improving both the security and performance of Solidity contracts. For example, reentrancy can be mitigated by adopting the \texttt{Checks-Effects-Interactions} pattern\cite{Britten2021UsingCT}, where internal state changes (effects) are made before making external calls (interactions).

Another area of optimization is gas consumption, which directly impacts the cost of executing smart contracts. For example, researchers have worked on reducing the number of storage operations in Solidity\cite{Britten2021UsingCT}, as writing to the Ethereum blockchain storage is significantly more expensive than reading from it. Techniques such as storing frequently used data in memory instead of storage can lead to substantial gas savings\cite{Junaid2023AnalyzingTP}.

\subsection{Exploration of New Smart Contract Programming Languages}
New programming languages for smart contracts are being developed to overcome the limitations of Solidity in terms of security, readability, and performance. Two such languages are:
\begin{itemize}
    \item \textbf{Vyper} Designed to improve security by being more restrictive and easier to read\cite{githubGitHubVyperlangvyper}:. It eliminates some of the complex features of Solidity that are prone to misuse, such as recursive calls and inline assembly.
    \item \textbf{Pact:} Used primarily on the Kadena blockchain\cite{kadenaStartedWith,kadenaKadenaBlockchain}, Pact focuses on human-readable syntax and built-in formal verification features, making it easier to create secure smart contracts.
\end{itemize}

Vyper\cite{githubGitHubVyperlangvyper}, for example, avoids features that are known to introduce security vulnerabilities, such as the use of infinite loops or the ability to modify low-level assembly code. This makes it a safer alternative to Solidity in situations where simplicity and security are prioritized over performance.

\section{Performance and Security Research on Decentralized Applications}

\subsection{Optimization of User Experience in dApps}
Decentralized applications (dApps) often face performance issues that negatively impact user experience\cite{Zheng2023BlockchainBasedDA}. Some of the key factors affecting dApp usability include:
\begin{itemize}
    \item \textbf{Transaction Speed:} On-chain transactions can be slow, as they require confirmation by the blockchain network. Layer-2 scaling solutions like Optimistic Rollups and zk-Rollups aim to speed up transaction processing without compromising security\cite{9862815,Fernando2023PosterWA}.
    \item \textbf{Interface Design:} Many dApps suffer from poor user interfaces, which can make them difficult to use for non-technical users\cite{Teruel2020EasingI}. Research into improving dApp design focuses on making interfaces more intuitive and easy to navigate.
    \item \textbf{Onboarding Processes:} Setting up a wallet, managing private keys, and understanding gas fees are significant barriers for new users\cite{Durn2020AnAF}. Solutions such as social recovery wallets and gas abstraction techniques help simplify the onboarding process.
\end{itemize}

\subsection{Security Challenges in Decentralized Applications}
Security remains one of the most significant challenges in the development of dApps. Some common threats include:
\begin{itemize}
    \item \textbf{Smart Contract Vulnerabilities:} As mentioned earlier, flaws in smart contracts can lead to severe consequences, such as loss of funds.
    \item \textbf{Network Attacks:} dApps that rely on peer-to-peer networks can be susceptible to Sybil attacks, in which an attacker creates multiple fake identities to gain control of the network.
    \item \textbf{User Privacy:} dApps often operate in a public environment, where user data and transaction details are visible on the blockchain. Solutions like zero-knowledge proofs (ZKPs)\cite{Berentsen2023AnIT} and confidential transactions are being researched to enhance privacy.
\end{itemize}

By addressing these challenges, researchers aim to make decentralized applications more secure and user-friendly, thereby increasing their adoption.

\chapter{Interdisciplinary Research on Blockchain and Socioeconomics}

\section{Game Theory Models in Blockchain}

\subsection{Application of Game Theory in Blockchain Systems}
Game theory provides a crucial lens through which the behavior of participants in blockchain systems can be analyzed\cite{Liu2019ASO}. At its core, game theory studies the strategic interactions between rational decision-makers, which can be applied to participants in blockchain networks, such as miners, validators, or users. 

For example, blockchain systems often use consensus mechanisms like Proof of Work (PoW)\cite{gervais2016security,ferdous2020blockchainconsensusalgorithmssurvey} or Proof of Stake (PoS)\cite{Wendl2022TheEI}. In these settings, participants must decide how much computational power (in PoW) or how much stake (in PoS) to dedicate to validating transactions. A central concept from game theory, the *Nash equilibrium*, can help explain the optimal strategies of these participants, where no one gains by unilaterally changing their strategy.

\textbf{Nash Equilibrium in Proof of Work (PoW):}
In PoW blockchains like Bitcoin, miners must decide how much computational power to allocate to mining blocks\cite{gervais2016security,ferdous2020blockchainconsensusalgorithmssurvey}. If one miner increases their mining power, they might be more likely to solve the cryptographic puzzle and earn rewards. However, if all miners increase their mining power proportionally, the difficulty of mining also increases. The Nash equilibrium occurs when all miners allocate just enough computational power such that no single miner can increase their profits by changing their strategy.

Similarly, cooperative strategies can also be studied. In some blockchain networks, validators or miners might choose to collude to manipulate outcomes, such as reordering transactions. Game theory allows researchers to model these cooperative behaviors and design mechanisms that discourage such behavior, thereby enhancing the security and fairness of the network\cite{Li2020RationalPA}.

\subsection{Academic Research on Incentive Mechanism Design}
The design of incentive mechanisms in blockchain is vital to ensuring the sustainability and security of decentralized systems. Blockchain participants, such as miners or validators, need to be incentivized to maintain the integrity of the network, but these incentives must also balance decentralization and fairness. 

\textbf{Example: Bitcoin's Incentive Structure.}
In Bitcoin, miners are rewarded with newly minted bitcoins and transaction fees for validating transactions and adding them to the blockchain\cite{Azzolini2019StudyingTF}. This reward structure not only motivates miners to maintain the network, but also ensures that new blocks are continually added to the chain, thus keeping the system secure.

Academic research often focuses on creating models that optimize these incentive structures\cite{Schrijvers2016IncentiveCO}. A common challenge in decentralized systems is the risk of *free-riders*—participants who benefit from the network without contributing to it\cite{Yahaya2015FreeRI}. Research in this area aims to design incentive mechanisms that prevent such behaviors and ensure that rewards are distributed fairly based on participation.

One area of particular interest is the concept of *block rewards* diminishing over time\cite{Weinberg2016OnTI}, as in Bitcoin, where the total supply is capped at 21 million coins. Academics explore how systems can continue to incentivize participants even after block rewards significantly decrease, potentially relying more on transaction fees to sustain the network.

\section{Tokenomics and Token Design}

\subsection{Theoretical Foundations of Tokenomics Models}
Tokenomics\cite{Lamberty2020LeadingDS} refers to the economic systems that govern the use and distribution of tokens within a blockchain ecosystem. These tokens are often used to incentivize participants, facilitate governance, or represent value in decentralized networks.

\textbf{Example: Utility Tokens vs. Governance Tokens.}
A common tokenomic model involves two primary types of tokens:
\begin{itemize}
    \item \textbf{Utility Tokens}: These tokens provide users with access to a product or service within the blockchain ecosystem. For example, in the Filecoin network\cite{filecoinDecentralizedStorage}, tokens are used to pay for data storage services.
    \item \textbf{Governance Tokens}: These tokens grant holders voting rights on decisions related to the governance of the blockchain protocol. For instance, holders of tokens in the MakerDAO protocol\cite{makerdaoMakerDAOUnbiased} can vote on key decisions related to the stability and functioning of the network.
\end{itemize}

\subsection{Research on Token Issuance and Distribution Mechanisms}
Academic research on token issuance and distribution focuses on how to effectively launch and distribute tokens within a network to ensure long-term sustainability and growth. One popular distribution method is the Initial Coin Offering (ICO)\cite{Fisch2018InitialCO}, where tokens are sold to early investors to raise funds for a project.

\textbf{Example: ICOs vs. Airdrops.}
\begin{itemize}
    \item \textbf{ICOs}: In an ICO, new projects sell tokens to raise capital\cite{Fisch2018InitialCO}. Investors typically buy these tokens with cryptocurrencies like Ethereum or Bitcoin. The success of an ICO often depends on the perceived value and utility of the token in the network.
    \item \textbf{Airdrops}: In contrast, airdrops distribute tokens for free to a wide group of participants\cite{Allen2022WhyAC}. Airdrops are used to promote network adoption by distributing tokens to existing users, incentivizing them to become active participants in the ecosystem.
\end{itemize}

\section{Research on Decentralized Governance Mechanisms}

\subsection{Academic Exploration of Decentralized Autonomous Organizations (DAOs)}
A Decentralized Autonomous Organization (DAO) is a governance model where decision-making is decentralized and carried out by token holders through voting\cite{Ding2023ASO}. DAOs operate without a central authority, and their rules are encoded in smart contracts.

\textbf{Example: The DAO (2016).}
One of the earliest examples of a DAO was "The DAO,"\cite{Mehar2017UnderstandingAR} which was launched on the Ethereum network in 2016. The idea was to create a decentralized venture capital fund where participants could vote on how to allocate the funds. However, due to vulnerabilities in the code, it was hacked, leading to a loss of \$60 million worth of ether\cite{Mehar2017UnderstandingAR}. This incident sparked extensive academic research into the security and robustness of DAO models.

\subsection{Research on Security and Fairness in Voting Mechanisms}
Voting in decentralized systems comes with several challenges, including security risks like sybil attacks\cite{Yu2006SybilGuardDA}, where an attacker creates multiple fake identities to influence the vote. Ensuring fairness in voting mechanisms is essential to the legitimacy of decentralized governance.

Academic research explores different methods to make voting mechanisms more secure and fair. Some proposals include quadratic voting\cite{Posner2014VotingSQ}, where the cost of each additional vote increases quadratically, preventing wealthy participants from having disproportionate influence.

\section{Impact of Blockchain on the Global Economy}

\subsection{Interaction Between Blockchain and International Financial Systems}
Blockchain technology is transforming international financial systems by reducing the need for intermediaries in cross-border payments and remittances. Traditionally, these transactions involve several intermediaries, leading to high fees and slow transaction times.

\textbf{Example: Ripple’s Cross-Border Payment System.}
Ripple\cite{Islam2022AnalysisOB} uses blockchain to facilitate near-instantaneous cross-border transactions. By using its native token (XRP), Ripple allows banks and financial institutions to transfer money globally in seconds, compared to the several days required by traditional banking systems. This dramatically reduces friction and costs, which is especially significant for industries like remittances, where high fees can represent a significant burden on low-income households.

\subsection{Economic Impact of Blockchain on Trade and Supply Chains}
Blockchain enhances the transparency, security, and efficiency of global trade and supply chains by enabling real-time tracking of goods and services. One of the main benefits of blockchain in supply chains is the creation of an immutable record of transactions, which ensures accountability and reduces the likelihood of fraud or counterfeiting.

\textbf{Example: IBM’s Food Trust Blockchain.}
IBM’s Food Trust blockchain\cite{ibmSupplyChain} is used by companies like Walmart to track the journey of food products from farms to grocery stores. By using blockchain technology, participants in the supply chain can verify the origin and authenticity of products, ensuring that food is safe and ethically sourced. This level of transparency not only improves efficiency but also builds trust with consumers.

\chapter{Future Research Directions in Blockchain}

\section{Blockchain and Quantum Computing}

\subsection{Threats of Quantum Computing to Blockchain Security}
One of the most critical future challenges to blockchain technology is the potential impact of quantum computing\cite{Shrestha2021ARO}. Quantum computers, by leveraging the principles of quantum mechanics, could possess computational power exponentially greater than that of classical computers. This immense capability poses a significant threat to the cryptographic algorithms that secure blockchain networks today.

Current blockchain systems rely heavily on cryptographic algorithms such as RSA and elliptic curve cryptography (ECC) to ensure security\cite{Chandel2019AMA}. For example, Bitcoin and Ethereum use elliptic curve cryptography to generate private and public key pairs. The security of ECC relies on the difficulty of the Elliptic Curve Discrete Logarithm Problem (ECDLP)\cite{Chavan2016ARO}, which is computationally infeasible for classical computers to solve within a reasonable time frame. However, with quantum computing, Shor’s algorithm could be used to efficiently solve both the RSA and ECDLP\cite{Kearney2021VulnerabilityOB}, rendering the encryption methods used in most blockchain systems vulnerable to attack.

To better understand this, imagine a scenario where a powerful quantum computer could break the cryptographic key that secures a blockchain wallet. This would allow a malicious actor to forge digital signatures or manipulate blockchain transactions without authorization, undermining the core trust model of the entire system. As quantum computing continues to evolve, this threat becomes more imminent, and addressing it is crucial for the future security of blockchain technologies.

\subsection{Research on Quantum-Resistant Cryptographic Algorithms}
In response to the threat posed by quantum computing, researchers are actively developing quantum-resistant (or post-quantum) cryptographic algorithms\cite{Nisha2024LatticeBasedCA,Allende2021QuantumresistanceIB}. These algorithms are designed to be secure even against quantum computers, ensuring that blockchain systems can remain robust in a post-quantum world.

One such area of research focuses on lattice-based cryptography, which relies on the hardness of mathematical problems like the Shortest Vector Problem (SVP\cite{Nisha2024LatticeBasedCA}) and the Learning With Errors (LWE) problem\cite{Peikert2016ADO}. These problems are believed to be resistant to both classical and quantum attacks. Another promising approach involves hash-based signatures, which rely on cryptographic hash functions rather than algebraic structures like those used in RSA or ECC\cite{Kermani2015ReliableHT}.

The integration of quantum-resistant cryptographic algorithms into blockchain systems is essential for future-proofing these networks. Researchers are already working on protocols such as Bitcoin's potential upgrade to include quantum-resistant algorithms like Lamport signatures\cite{Zentai2020OnTE}. Although these quantum-resistant methods may increase the size of cryptographic keys and signatures, and therefore impact performance, their adoption is a necessary step in maintaining blockchain security in the quantum era.

\section{Synergies Between Blockchain and the Internet of Things (IoT)}

\subsection{Research on Blockchain Applications in IoT Devices}
The Internet of Things (IoT) refers to the network of interconnected devices that communicate and share data autonomously\cite{Sivaraman2018SmartID}. As the number of IoT devices grows, securing the vast amounts of data they generate becomes increasingly important. Blockchain technology is emerging as a solution for enhancing the security, transparency, and integrity of IoT data.

For example, consider a smart home system where various devices, such as thermostats, security cameras, and appliances, are connected. By integrating blockchain, each device can securely authenticate itself to the network without relying on a centralized authority\cite{Sivaraman2018SmartID}. Blockchain can store immutable records of device interactions, ensuring that data cannot be tampered with after it has been logged.

One specific application of blockchain in IoT is secure device authentication\cite{Li2018ABA}. In a traditional system, a centralized server verifies the identity of each device. However, this creates a single point of failure, which could be exploited by hackers. By using blockchain, a decentralized network can authenticate devices in a distributed manner, eliminating the risk of server compromise.

Another example is the decentralized management of IoT data\cite{Thakker2020SecureDM}. IoT devices generate massive amounts of data that need to be processed, analyzed, and stored. Using blockchain, data can be stored across a distributed ledger, ensuring transparency and data integrity. This decentralized approach also allows for real-time auditability, as anyone with access to the blockchain can verify the authenticity of the data.

\subsection{Designing Decentralized IoT Systems}
A key research area at the intersection of blockchain and IoT involves designing fully decentralized IoT systems, where devices can operate autonomously without reliance on centralized infrastructure\cite{Yu2018IoTChainET}. This decentralization is particularly valuable in scenarios such as smart cities, where a large number of devices need to coordinate securely and efficiently.

One of the major challenges in designing decentralized IoT systems is scalability\cite{Arellanes2019DecentralizedDF}. IoT networks consist of potentially millions of devices, each generating and transmitting data. In a blockchain-based IoT system, this data must be stored and validated across the network, which could lead to significant performance bottlenecks. Therefore, research into scalable consensus mechanisms is crucial for the development of IoT-blockchain integration.

Consensus mechanisms such as Proof of Stake (PoS)\cite{Wendl2022TheEI} or delegated Proof of Stake (dPoS)\cite{Saad2021ComparativeAO} are being explored to reduce the computational overhead in decentralized IoT systems. Additionally, novel solutions such as sharding, where the blockchain is split into smaller partitions (shards) that can process transactions in parallel, are being investigated to improve scalability.

Security is another concern in IoT-blockchain systems\cite{K2020SecuringIA}. IoT devices are often resource-constrained, with limited processing power and battery life. Research is focused on developing lightweight cryptographic protocols\cite{khan2020lightweight} that can provide robust security without draining the resources of these devices. Moreover, protocols for ensuring the privacy of IoT data on the blockchain are essential to prevent sensitive information from being exposed to unauthorized entities.

\section{Integration of Blockchain and Artificial Intelligence}

\subsection{AI-Driven Blockchain System Design}
Artificial Intelligence (AI) is increasingly being integrated into blockchain systems to optimize and enhance various functions. AI can help automate complex processes, improve the efficiency of consensus mechanisms, and provide advanced analytics for blockchain performance.

For instance, AI-driven consensus mechanisms are being explored to replace traditional methods like Proof of Work (PoW)\cite{gervais2016security,ferdous2020blockchainconsensusalgorithmssurvey}, which require significant computational resources. By using machine learning algorithms, blockchain systems can predict the optimal way to validate transactions based on network conditions\cite{safana2020improving}, reducing energy consumption and speeding up transaction processing.

Another area of research involves AI in the management of smart contracts\cite{krichen2023strengthening}. Smart contracts are self-executing contracts with terms encoded directly into code. AI can be used to analyze contract performance, predict potential conflicts, and even autonomously manage contract execution. For example, an AI system could detect anomalous behavior in a blockchain network and automatically adjust the contract terms to mitigate risk.

AI can also improve blockchain performance through predictive analytics\cite{ressi2024ai}. By analyzing patterns in transaction data, AI systems can anticipate network congestion, enabling proactive adjustments to prevent delays. This type of optimization is particularly valuable in decentralized finance (DeFi) platforms, where rapid transaction processing is critical.

\subsection{Research on Blockchain and AI Data Sharing}
The intersection of blockchain and AI presents exciting possibilities for secure, transparent, and decentralized data sharing. AI systems rely heavily on large datasets for training and decision-making. However, sharing sensitive data for AI purposes often raises concerns about privacy, data ownership, and security. Blockchain can provide a solution by creating decentralized data marketplaces where data can be shared securely and transparently.

For example, in healthcare, AI models require access to vast amounts of patient data to improve diagnostics and treatment recommendations\cite{tagde2021blockchain}. Blockchain can enable patients to control and monetize their data by sharing it with AI researchers through a decentralized marketplace. Patients can grant access to their data while maintaining privacy and control, thanks to blockchain's transparency and immutability features.

In collaborative AI development, blockchain can also play a critical role. Imagine multiple AI developers working on a shared project. Blockchain can be used to record and verify each developer’s contributions to ensure that intellectual property rights are respected. This decentralized approach reduces the need for intermediaries and fosters trust among collaborators.

Blockchain's ability to create auditable records of data transactions makes it a powerful tool for addressing the challenges of AI data sharing, ensuring that AI models are trained on accurate, secure, and transparent data.

\part{Blockchain Regulation and Future Development}

\chapter{Legal and Regulatory Challenges of Blockchain}

\section{Legal Definitions of Blockchain}

Blockchain technology, though relatively new, has quickly gained attention from regulators, lawmakers, and legal experts. At its core, blockchain is a distributed ledger system that records transactions in a secure, transparent, and immutable manner. However, despite its technical aspects being well-understood by developers, its legal definition remains a complex and evolving issue. This section explores the different ways in which blockchain is being defined and categorized within legal frameworks.

\subsection{Legal Validity of Smart Contracts}

Smart contracts\cite{Ene2020SmartC} are one of the most revolutionary features enabled by blockchain technology. These are self-executing contracts where the terms of the agreement between buyer and seller are directly written into lines of code. Smart contracts automatically enforce and execute the terms of the contract once certain conditions are met.

The legal recognition of smart contracts presents several challenges:

1. \textbf{Traditional Contract Law}: Traditional contracts usually require offer, acceptance, consideration (value exchanged), and the intention to create legal relations\cite{MacDonald20186IT}. In a blockchain context, smart contracts meet many of these criteria but differ in their form\cite{Jaswant2021SmartCA}, as they are written in code rather than natural language.

    For example, consider a simple transaction:
    
    \begin{lstlisting}[style=solidity]
    pragma solidity ^0.8.0;
    
    contract SimpleContract {
        address payable buyer;
        address payable seller;
        
        constructor(address payable _seller) {
            buyer = msg.sender;
            seller = _seller;
        }

        function purchase() external payable {
            require(msg.value > 0, "Payment must be greater than 0");
            seller.transfer(msg.value);
        }
    }
    \end{lstlisting}
    
    In this contract, the agreement to transfer funds upon purchase is executed automatically when a buyer sends a payment. There is no need for intermediaries, but how is such a contract viewed under traditional legal systems?

2. \textbf{Jurisdictional Issues}: Another challenge is determining which jurisdiction's laws apply to smart contracts\cite{SanzBayn2019KeyLI}. Since blockchain is decentralized and transactions can occur across borders, disputes could arise over which legal framework governs the contract's enforcement.

3. \textbf{Consumer Protection}: Legal systems often have provisions that protect consumers from unfair contract terms or situations where they may not fully understand the implications of the contract. With smart contracts being code-based, individuals without technical knowledge may face difficulties understanding the terms they are agreeing to.

\subsection{Regulatory Challenges of Blockchain Technology}

Blockchain's decentralized nature and its ability to support anonymous transactions pose significant challenges to existing regulatory frameworks. Governments and regulatory bodies face the dilemma of how to regulate a technology that operates across borders\cite{alayanAksoy2022SmartCT}, often outside of traditional legal structures.

\textbf{Key Challenges}:
\begin{itemize}
    \item \textbf{Anonymity and Pseudonymity}: One of blockchain's unique characteristics is that it allows users to engage in transactions without revealing their real-world identities. While this can provide privacy, it also presents challenges for regulators who are tasked with enforcing anti-money laundering (AML) and Know Your Customer (KYC) rules\cite{Barsan2019PublicBT}.
    
    \item \textbf{Jurisdictional Ambiguity}: Since blockchain operates on a global scale, determining which country's laws apply to specific transactions is a difficult issue\cite{Mohsin2021BlockchainLA}. For instance, a user in Europe could enter into a smart contract with another user in Asia, raising questions about which jurisdiction governs the transaction in case of a dispute.
    
    \item \textbf{Cross-Border Transactions}: Traditional regulatory frameworks are often based on geographical borders. However, blockchain transactions occur globally and instantaneously, making it difficult for regulators to track and enforce local laws.
    
    \item \textbf{Regulating Decentralized Systems}: Blockchain networks often operate without a central entity, which means there is no single point of control. This raises questions about who is responsible for compliance with regulations.
\end{itemize}

\section{Global Blockchain Regulatory Policies}

The regulatory landscape for blockchain varies significantly across countries, with some nations embracing the technology and others taking a more cautious or restrictive approach.

\subsection{Current Regulatory Landscape in Different Countries}

Different countries have adopted different approaches to regulating blockchain technology and cryptocurrencies. Below are some notable examples:

\begin{itemize}
    \item \textbf{United States}: In the U.S., blockchain and cryptocurrencies are subject to a patchwork of federal and state regulations. The Securities and Exchange Commission (SEC) classifies certain digital assets as securities, which means they must comply with existing securities laws\cite{DiMatteo2019BlockchainBasedFS}. Meanwhile, the Commodity Futures Trading Commission (CFTC) has jurisdiction over cryptocurrencies like Bitcoin when they are used as commodities\cite{Nathan2020InOO}.

    \item \textbf{European Union}: The EU has been working on creating a unified regulatory framework for blockchain and cryptocurrencies. The Markets in Crypto-Assets (MiCA) regulation\cite{europaMarketsCryptoAssets}, for example, seeks to create a clear legal framework for digital assets across all member states, providing more clarity for businesses operating within the EU.

    \item \textbf{China}: China has taken a more restrictive approach, banning cryptocurrency trading and Initial Coin Offerings (ICOs). However, the Chinese government has shown strong support for blockchain technology itself, particularly in its potential applications for state-run digital currencies\cite{cnnChinaMakes}.

    \item \textbf{Singapore}: Singapore is considered a blockchain-friendly jurisdiction, with clear regulatory frameworks that support innovation while ensuring compliance with AML and KYC regulations\cite{Alekseenko2022LegalRO}. The Monetary Authority of Singapore (MAS) has developed guidelines that allow blockchain startups to operate in a relatively low-risk environment.
\end{itemize}

\subsection{Major Points of Regulatory Contention}

Regulating blockchain presents several key points of contention, as policymakers try to balance the need for innovation with the necessity of protecting consumers and maintaining financial stability.

\begin{itemize}
    \item \textbf{Privacy vs. Transparency}: While blockchain can provide greater transparency in financial systems, it also raises privacy concerns. Regulatory bodies are challenged with ensuring transparency, especially in financial transactions, without infringing on individuals' privacy rights.
    
    \item \textbf{AML and Consumer Protection}: Regulators often focus on ensuring compliance with Anti-Money Laundering (AML) laws, as blockchain's ability to facilitate anonymous transactions can be exploited for illicit activities. Striking the right balance between enforcing AML and enabling legitimate users to maintain their privacy remains a contentious issue.
    
    \item \textbf{Stifling Innovation vs. Financial Stability}: Regulatory overreach can potentially stifle innovation. Governments are tasked with maintaining financial stability and preventing systemic risks while ensuring that regulation does not hinder the growth of blockchain technology.
\end{itemize}

\section{Cryptocurrency and Securities Regulations}

\subsection{Legal Boundaries Between Securities and Cryptocurrencies}

A critical issue in blockchain regulation is the distinction between cryptocurrencies and securities. This distinction is important because securities are heavily regulated, and companies issuing them must comply with a variety of legal requirements.

\textbf{How Regulators Classify Digital Assets}:
\begin{itemize}
    \item In the U.S., the SEC uses the \textit{Howey Test}\cite{secSECgovFramework} to determine whether a digital asset qualifies as a security. The test asks whether an investment of money is made in a common enterprise with the expectation of profit primarily from the efforts of others.
    \item Cryptocurrencies like Bitcoin, which are decentralized and do not rely on a central entity, are generally not considered securities. However, many Initial Coin Offerings (ICOs) and token sales involve digital assets that may meet the definition of a security\cite{Senderowicz2018SECFO}.
\end{itemize}

\subsection{Legal Norms for Token Offerings and Fundraising}

Initial Coin Offerings (ICOs) and Security Token Offerings (STOs) are common methods for blockchain startups to raise funds\cite{Kondova2019BlockchainIS} . However, these fundraising mechanisms often fall under the purview of securities law.

\textbf{Key Legal Requirements}:
\begin{itemize}
    \item \textbf{Compliance with Securities Law}: ICOs that issue tokens resembling securities must comply with the relevant securities regulations. This often involves registering the offering with the SEC or seeking an exemption.
    
    \item \textbf{Investor Protections}: To protect investors, regulators often require issuers of securities to provide clear and accurate information about the investment opportunity. This may include the requirement to file disclosures, such as a prospectus.
    
    \item \textbf{Cross-border Fundraising}: Given the global nature of blockchain, token offerings often involve investors from multiple jurisdictions. This introduces additional complexity, as issuers must comply with the securities laws of each jurisdiction where investors reside.
\end{itemize}

\chapter{Future Development of Blockchain}
    
    \section{Technical Challenges of Blockchain Technology}
        \subsection{Scalability and Performance Bottlenecks}
        One of the most significant challenges facing blockchain technology today is scalability\cite{Khan2021SystematicLR}. Blockchain networks, especially popular ones like Bitcoin and Ethereum, face limitations in transaction throughput and latency. As more users and applications adopt these networks, the speed at which transactions are processed becomes slower, and the system becomes congested.

        For example, Bitcoin can only process about 7 transactions per second (TPS), and Ethereum around 15-30 TPS. In contrast, centralized payment systems like Visa can handle upwards of 24,000 TPS\cite{}. This comparison highlights a clear gap in performance, which must be addressed for blockchain to achieve mass adoption.

        The reason for these bottlenecks lies in the consensus mechanisms and the decentralized nature of blockchain. Every transaction needs to be validated by nodes (computers) in the network, which introduces latency. Moreover, every node must store a copy of the entire blockchain, limiting the system's ability to scale.

        \subsubsection{Research and Development Solutions:} Several solutions are being developed to tackle scalability issues:
        \begin{itemize}
            \item \textbf{Layer 2 Solutions:} These are built on top of the main blockchain (Layer 1) to handle a large volume of transactions off-chain, reducing the load on the base layer. The Lightning Network\cite{Seres2019TopologicalAO} for Bitcoin and Optimistic Rollups\cite{9862815} for Ethereum are prime examples. These solutions allow transactions to be settled quickly, only using the base layer for final settlement.
            \item \textbf{Sharding:} Sharding involves splitting the blockchain into smaller, manageable "shards,"\cite{CortesGoicoechea2020ResourceAO} each capable of processing its own transactions and smart contracts. By distributing the workload across multiple shards, the system can achieve higher throughput. Ethereum 2.0\cite{Asif2023ShapingTF} is expected to implement sharding as part of its scalability upgrade.
        \end{itemize}
        
        These innovations are critical to improving blockchain performance, making it more viable for widespread use.

        \subsection{Solutions to the Energy Consumption Problem}
        Another major challenge is the high energy consumption associated with blockchain, particularly those using Proof of Work (PoW)\cite{gervais2016security,ferdous2020blockchainconsensusalgorithmssurvey} consensus mechanisms. Bitcoin, for example, uses an immense amount of electricity to secure its network—comparable to the energy consumption of some small countries.

        PoW works by requiring miners to solve complex mathematical puzzles to add new blocks to the blockchain, which consumes significant computational power and, by extension, energy. This has led to criticism of blockchain’s environmental impact.

        \subsubsection{Alternative Approaches:} To address this issue, several energy-efficient consensus mechanisms have been developed:
        \begin{itemize}
            \item \textbf{Proof of Stake (PoS)} Unlike PoW, where miners compete to solve puzzles, PoS selects validators based on the number of tokens they hold and are willing to "stake" as collateral\cite{Wendl2022TheEI}:. This significantly reduces the energy consumption since the competition and heavy computations are removed. Ethereum is transitioning from PoW to PoS with Ethereum 2.0 to address these concerns.
            \item \textbf{Other Energy-Efficient Mechanisms:} Several other mechanisms, such as Delegated Proof of Stake (DPoS)\cite{article}, Proof of Authority (PoA)\cite{ethereumProofofauthorityPoA,Islam2022ACA}, and Proof of Space-Time (PoST)\cite{Moran2016ProofsOS}, have been developed to further minimize the energy requirements while maintaining security.
        \end{itemize}

        As blockchain continues to evolve, energy efficiency will be a key consideration for its broader acceptance.

    \section{Integration of Blockchain and the Internet of Things (IoT)}
        \subsection{Use Cases of Blockchain in IoT}
        Blockchain and the Internet of Things (IoT) are two emerging technologies that can complement each other, particularly in areas like security, data integrity, and automation. IoT devices generate a massive amount of data, and securing that data while ensuring efficient communication is crucial.

        \subsubsection{Examples of Blockchain in IoT:}
        \begin{itemize}
            \item \textbf{Secure Device Communication:} IoT devices often face security vulnerabilities, making them prime targets for cyberattacks. By using blockchain, devices can communicate with each other securely in a decentralized and tamper-proof environment\cite{Sivaraman2018SmartID}. For instance, each device can have a unique blockchain-based identity, preventing unauthorized access.
            \item \textbf{Decentralized Data Storage:} IoT generates enormous amounts of data, and blockchain can be used to store this data securely across a distributed network\cite{Thakker2020SecureDM}. For example, IoT data from a smart city can be stored in a blockchain network to ensure transparency and prevent tampering.
            \item \textbf{Automated Machine-to-Machine (M2M)\cite{Chanthong2020BlockchainAS} Transactions:} Blockchain enables smart contracts, which can automate transactions between IoT devices. For example, electric vehicles could automatically pay for charging services using a blockchain-based system without human intervention.
        \end{itemize}

        These use cases demonstrate the potential for blockchain to enhance the security and efficiency of IoT ecosystems.

        \subsection{Combining Smart Devices and Blockchain}
        Integrating smart devices with blockchain technology offers several benefits, especially in enhancing security, transparency, and automation\cite{Hmoud2023IoTAB}. Smart devices can interact in real-time, and blockchain can ensure that data transferred between these devices remains secure and verifiable.

        \subsubsection{Potential Benefits:}
        \begin{itemize}
            \item \textbf{Improved Security:} Each smart device can be assigned a unique digital identity on the blockchain, making it nearly impossible for hackers to impersonate or tamper with them.
            \item \textbf{Transparency:} Blockchain’s immutable ledger ensures that any interactions between smart devices are recorded transparently, allowing users to track activities and ensure trust.
            \item \textbf{Interoperability:} Different manufacturers produce smart devices, often leading to compatibility issues. Blockchain can standardize the communication protocols, allowing devices from various brands to interact smoothly within the same network.
        \end{itemize}

        These integrations provide a promising future where blockchain can enhance the capabilities of smart devices in IoT ecosystems.

    \section{Blockchain and Artificial Intelligence (AI)}
        \subsection{Convergence Points Between AI and Blockchain}
        Artificial Intelligence (AI) and blockchain are both powerful technologies, and their convergence offers immense potential. AI algorithms can be used to enhance the efficiency, security, and automation of blockchain networks.

        \subsubsection{How AI Enhances Blockchain:}
        \begin{itemize}
            \item \textbf{Optimizing Consensus Mechanisms:} AI \cite{feng2024deeplearningmachinelearning} can be used to optimize how consensus mechanisms work in blockchain networks. For example, AI can predict and adjust the difficulty of PoW algorithms dynamically, making the network more efficient\cite{Chakraborty2022APD}.
            \item \textbf{Data Analysis:} Blockchain networks generate vast amounts of data. AI can help analyze this data, finding patterns, identifying trends, and even predicting network performance issues, which can improve decision-making processes.
            \item \textbf{Automation:} AI can further enhance the automation capabilities of blockchain-based systems, such as smart contracts\cite{Krichen2023StrengtheningTS}. For instance, AI-driven smart contracts could make more complex and adaptive decisions based on real-time data.
        \end{itemize}
        
        The synergy between AI and blockchain can significantly improve the overall performance of decentralized networks.

        \subsection{AI-Driven Decentralized Networks}
        One of the most intriguing possibilities is the development of AI-driven decentralized networks. In such systems, AI agents operate autonomously within blockchain-based ecosystems, making decisions, processing transactions, and interacting with smart contracts without human intervention.

        \subsubsection{Applications:}
        \begin{itemize}
            \item \textbf{Smart Contracts:} AI can make smart contracts more intelligent, allowing them to adapt to unforeseen situations or changes in external conditions\cite{Taherdoost2022BlockchainTA}. For example, AI could adjust contract terms dynamically based on market conditions.
            \item \textbf{Decentralized Autonomous Organizations (DAOs):} DAOs are organizations that operate on blockchain, with rules encoded in smart contracts\cite{Ding2023ASO}. AI could take these a step further by allowing the organization to learn and evolve based on data patterns, making more complex decisions autonomously.
        \end{itemize}
        
        This combination of AI and blockchain holds great promise for the future of decentralized systems.

    \section{Impact of Blockchain on Society and the Economy}
        \subsection{Future Outlook of Decentralized Economies}
        The concept of decentralized economies is becoming increasingly popular with the rise of blockchain technology. These economies allow individuals to engage in peer-to-peer transactions without intermediaries like banks or governments, creating more freedom and lowering transaction costs.

        \subsubsection{Decentralized Finance (DeFi):} One of the most impactful applications of blockchain in decentralized economies is Decentralized Finance (DeFi)\cite{Jensen2021AnIT}. DeFi platforms allow users to lend, borrow, and trade financial assets directly with each other using smart contracts, eliminating the need for traditional financial institutions.
        
        \subsubsection{Governance Models:} Blockchain also enables new forms of governance. Decentralized Autonomous Organizations (DAOs)\cite{Beck2018GovernanceIT} allow stakeholders to vote on decisions, distributing power more evenly across a community.

        \subsection{Impact of Blockchain on the Global Financial System}
        Blockchain is poised to reshape the global financial system by introducing more efficient, transparent, and inclusive solutions. Cross-border payments, for instance, are slow and expensive using traditional systems. Blockchain can streamline these payments, making them faster and more affordable.

        \subsubsection{Examples:}
        \begin{itemize}
            \item \textbf{Cross-Border Payments:} With blockchain, cross-border payments can be completed in minutes rather than days, with significantly lower fees. Ripple’s XRP and Stellar’s XLM are two examples of blockchain platforms focusing on this use case.
            \item \textbf{Central Bank Digital Currencies (CBDCs):} Governments are exploring blockchain for issuing digital currencies that operate on blockchain. CBDCs\cite{Svoboda2024CentralBD} could make the financial system more accessible, especially for the unbanked population, creating a more inclusive global economy.
        \end{itemize}
        
        As blockchain continues to develop, its impact on the global financial system could be transformative, creating a more connected and efficient economy.

\bibliographystyle{ieeetr}
\bibliography{reference}

\end{document}